\documentclass[aps,prb,twocolumn,longbibliography,superscriptaddress,amsmath,amssymb,floatfix,showpacs]{revtex4-2}
\usepackage{amsmath}
\usepackage{amssymb}
\usepackage{bm}
\usepackage{color}
\usepackage[utf8]{inputenc}
\usepackage{pifont}
\usepackage[colorlinks,linkcolor={blue},citecolor={blue},urlcolor={blue}]{hyperref}
\usepackage{mathtools}
\usepackage{booktabs}
\usepackage{multirow}
\usepackage{physics}
\usepackage{relsize}
\usepackage{mathrsfs}
\usepackage{mathdots}

\begin{document}

\title{Floquet Superlattices and Edge States in Graphene Nanoribbons}

\author{Siam Sarower}
\thanks{These authors contributed equally to this work.}
%\email{ssarower@students.kennesaw.edu}
\affiliation{Department of Physics, Kennesaw State University, Marietta, Georgia 30060, USA}
\author{Jonathon Dvorscak}
\thanks{These authors contributed equally to this work.}
%\email{jd277822@ohio.edu}
\affiliation{Department of Physics and Astronomy and Nanoscale and Quantum Phenomena Institute, Ohio University, Athens, Ohio 45701, USA}
\author{Nancy P. Sandler}
%\email{sandler@ohio.edu}
\affiliation{Department of Physics and Astronomy and Nanoscale and Quantum Phenomena Institute, Ohio University, Athens, Ohio 45701, USA}
\author{Mahmoud M. Asmar}
\email{masmar@kennesaw.edu}
\affiliation{Department of Physics, Kennesaw State University, Marietta, Georgia 30060, USA}

\begin{abstract}
Structured light provides a route to imprint spatially patterned Floquet potentials onto quantum materials. As a particular example, we study a zigzag graphene nanoribbon driven by two coherent tilted beams, whose interference creates a periodic polarization pattern that gives rise to a photo-induced superlattice. The matching between the periodicity of the optical field and the nanoribbon width leads to two regimes in the quasienergy spectrum: matched profiles preserve degenerate edge branches, while mismatched profiles yield a boundary-induced gap that survives in wide ribbons. We propose a two-edge model that captures this splitting through residual hybridization and the boundary-sampled optical field. The quasienergy gap reverses between valleys, leading to a valley-selective boundary response. Our results establish light-induced superlattices as a flexible method for valley selectivity in finite-size Dirac-like materials through tunable edge-state quasienergy splitting.
\end{abstract}
\maketitle

%%%%%%%%%%%%%%%%%%%%%%%%%%%%%%%%%%%%%%%%%%%%%%%%%%%%%%%%%%%%%%%%%%%%%%%%%%%%%%%%%%%%%%%%%%%%%%%%%%%%%
\section{Introduction}
Periodic driving provides a versatile dynamical route for reshaping electronic spectra, symmetries, and topology in quantum materials~\cite{Oka_RMP,flreview3,flreview1,flreview2,flreview4,flreview5}. In irradiated solids, electronic states hybridize with their photon replicas, producing Floquet bands and dressed quasiparticles with gaps, velocities, and topological character distinct from those of the equilibrium system~\cite{FloqTIReview}. This framework has motivated predictions of light-induced topological phases, anomalous transport, magnetic responses, and correlated nonequilibrium behavior~\cite{ftrans1,FloquetTI,ftrans3,Virtual-ph,FloquetTI2,graphene-top-ins,mitraandoka1,martin3,Asmar2024}. Experimental progress has followed several complementary directions: time-resolved photoemission has resolved Floquet--Bloch bands in topological insulators, black phosphorus, and graphene~\cite{FloqExp2,FloqExp1,BlackPhosphorus,FloqExp4,TrARPSGraph}; optical probes have revealed light-induced Stark shifts in WS$_2$, large modulation of nonlinear response in MnPS$_3$, and strong Floquet dressing of excitons~\cite{Floq_WS2,Floquet_Modualation,Floq_WS22}; and transport measurements have observed the light-induced anomalous Hall effect, steady Floquet--Andreev states, and signatures of long-lived Floquet steady states in graphene~\cite{FloqExp3,Park2022FloquetAndreev,FloqTraspGraphen}. 

While most realizations employ spatially uniform illumination, structured light adds a further layer of control by allowing the drive to carry spatial information~\cite{lightSt1,lightSt}. A particularly well-known example is provided by vortex light beams, which carry orbital angular momentum and therefore expand the set of optical degrees of freedom available for driving matter~\cite{lightV4,lightV3,lightV1}. More generally, spatially structured drives imprint real-space dependencies directly onto electronic systems~\cite{MD1,babakVLB,Lauren2025}, opening a route to space-time Floquet engineering.

Once the drive carries spatial structure, the geometry of the sample becomes an essential part of the Floquet problem. Graphene superlattices provide a useful precedent: models including scalar periodic potentials have predicted the modification of the Dirac spectra due to anisotropic velocity renormalization, minibands, gap generation, and additional Dirac points~\cite{Park2008NP,Park2008PRL,BreyFertig2009,Arovas2010,Snyman2009GappedState}. Similarly, models based on periodic mass or gap modulations have shown 
generation of interface states and anisotropic Dirac bands~\cite{Semenoff2008,Maksimova2012,DeMartino2023}.
Structured irradiation extends these ideas to dynamic potentials by allowing the superlattice profile to be imposed optically rather than through static gates. In finite graphene nanoribbons, the length scale introduced by the optical superlattice period must be considered together with the geometric scale set by the ribbon width. 
Zigzag-terminated graphene nanoribbons are especially interesting in this context because they host edge-localized states near charge neutrality~\cite{Fujita1996,Nakada1996,BreyFertig2006,Son2006}. The ubiquitous occurrence of boundary states has been observed experimentally by scanning-probe methods in chiral nanoribbons and nanoribbons with atomically engineered zigzag and chiral terminations~\cite{Tao2011,Zhang2013}. Static boundary potentials can substantially modify the edge spectra, including reshaping edge dispersions and opening gaps~\cite{Yao2009EdgeStates,Apel2011GNRPotentials}. 
 
Here we study the interplay between the system size and the length scale introduced by irradiation patterns in a minimal setting: a zigzag-terminated graphene nanoribbon irradiated by two coherent laser beams tilted from the same axis. This irradiation geometry was recently shown to generate a spatially periodic light--matter coupling in armchair nanoribbons, producing quasi-one-dimensional Floquet supercells with bulk topological states and scalable photocurrents, as shown in Ref.~\cite{Torres1}. In that case, the optical modulation defines a commensurate supercell with the armchair ribbon width, and the central physics arises from polarization interfaces within the irradiated region. Here, by contrast, the intrinsic zigzag edge states provide direct probes of whether the optical modulation is edge matched or edge mismatched with the ribbon size. This distinction also reflects the broader structure of graphene boundary conditions: the ideal armchair edge mixes the two valleys and lacks the zero-energy edge band characteristic of zigzag ribbons~\cite{BreyFertig2006,AkhmerovBeenakker2008}, whereas more general, {\it e.g.} mixed, chiral, or rational terminations can support localized or dispersive boundary states whose properties depend on the microscopic edge structure~\cite{Jaskolski2011,Fefferman2022}. The zigzag ribbon therefore provides a simple representative geometry for a broader class of finite-size Dirac systems in which pre-existing boundary modes couple to light. 

%%%%%%%%%%%%%%%%%%%%%%%%%%%%%%%%%%%%%%%%%%%%%%%%%%%%%%%%%%%%%%%%%%%%%%%%%%%%%%%%%%%%%%%%%%%%%%%%%%%%%
\section{Space--time modulated system}
To make the competition between the structured light and material length scales explicit, we consider a zigzag-terminated graphene nanoribbon that is translationally invariant along the $x$ direction and finite along $y$. The ribbon is driven by two coherent monochromatic beams whose wave vectors lie in the $yz$ plane. Because the beams are tilted with respect to the surface normal ($z$), their optical phases acquire a $y$-dependence across the ribbon width. Their coherent superposition therefore produces both intensity and polarization profiles that vary with position across the ribbon. This structured drive, illustrated schematically in Fig.~\ref{Fig1}(a), provides the optical origin of the Floquet-engineered superlattice.

%%%%%%%%%%%%%%%%%%%%%%%%%%%%%%%%%%%%%%%%%%%%%%%%
\subsection{Light Field}\label{light}
\begin{figure}[ht!]
  \centering
  \includegraphics[width=0.48\textwidth]{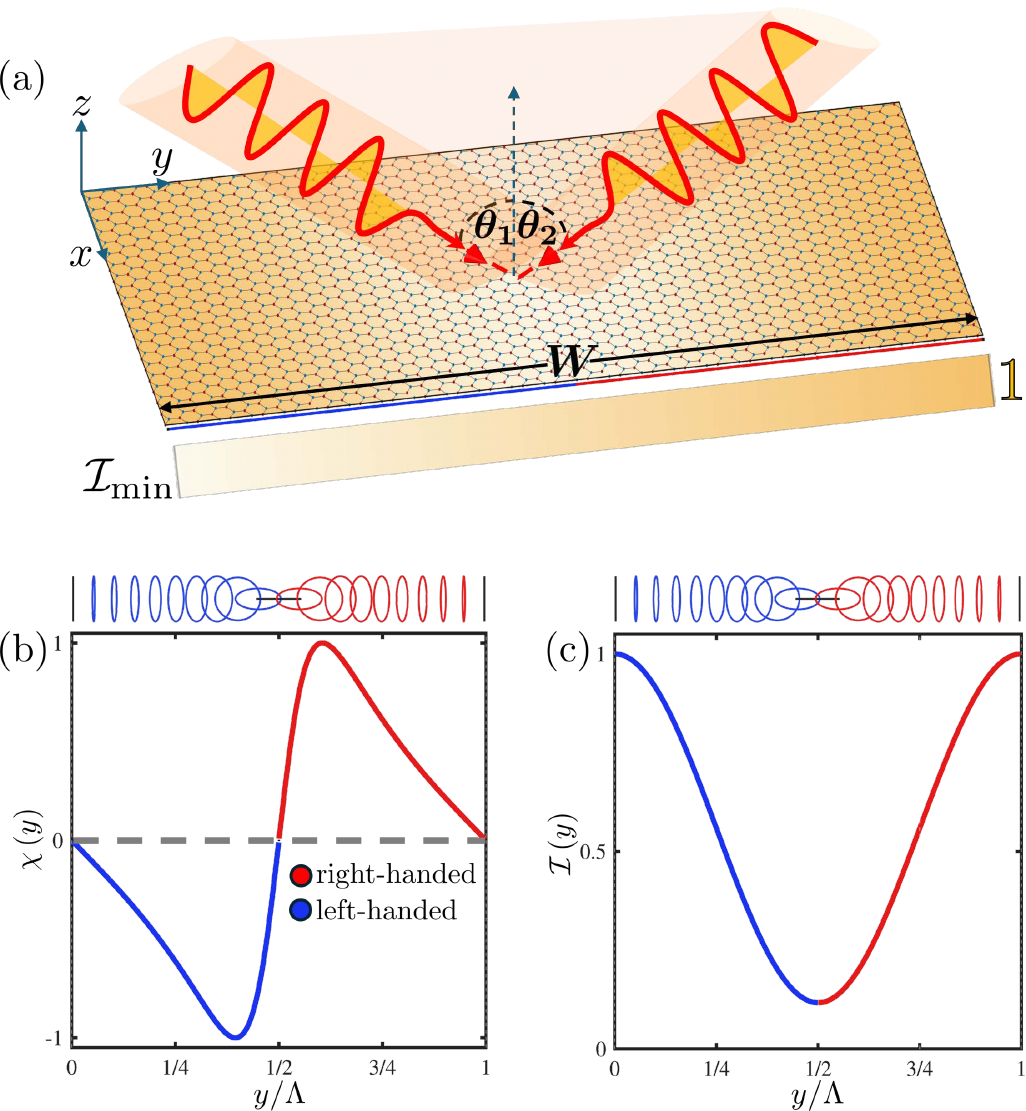}
\caption{Optical profiles produced by two coherent tilted beams incident on a zigzag graphene nanoribbon for $\theta_1=-\theta_2=70^\circ$ and $\phi_1=\phi_2-\pi=0$. 
(a) Schematic of the driven ribbon, showing the spatially modulated in-plane intensity $\mathcal I(y)$ and the beam geometry. 
(b) Normalized handedness $\chi(y)$; the dashed line marks locally linear polarization, $\chi=0$, while blue and red indicate opposite handedness. 
(c) Normalized in-plane intensity profile $\mathcal I(y)$. 
The polarization ellipses above (b) and (c) illustrate the local polarization state across one optical period.}
\label{Fig1}
\end{figure}

To see how this spatial structure arises, we first consider a single beam of frequency $\Omega$, wave number $q=\Omega/c$, and incidence angle $\theta_i$ measured from the surface normal, as shown in Fig.~\ref{Fig1}(a). 
The phase $\phi_i$ fixes the polarization state of beam $i$: $\phi_i=0$ or $\pi$ correspond to linear polarization, while $\phi_i=\pm\pi/2$ to circular polarization, with the sign determining the handedness. Intermediate values of $\phi_i$ correspond to elliptical polarization.

The propagation phase of the tilted beam is ${\bm q}_i\cdot\bm r-\Omega t$, with
${\bm q}_i=q(0,\sin\theta_i,-\cos\theta_i)$.
After evaluating the field on the graphene plane, $z=0$, this phase becomes
$
\vartheta_i(y,t)=qy\sin\theta_i-\Omega t
$. 
%%%
%%%
Thus, on the ribbon, the tilt gives the plane wave a $y$-dependent phase. The corresponding vector-potential components at the graphene plane are
%%%
\begin{equation}\label{Ai_z}
\bm{\mathcal A}_{i}
=
\frac{A_0}{\sqrt{2}}
\left[
\cos\vartheta_i,\,
\cos(\vartheta_i+\phi_i)\cos\theta_i,\,
\cos(\vartheta_i+\phi_i)\sin\theta_i
\right],
\end{equation}
%%%
where $A_0=E_0/\Omega$. 
Since the dynamics of electrons in graphene are confined to the $xy$ plane, the $z$ component does not couple to the low-energy electronic motion. 
For the two beam system the coupling to graphene is therefore determined only by
%%
%% 
%%%
\begin{equation}\label{A_parallel_sum}
\bm{\mathcal A}_{\parallel}(y,t)
=
\sum_{i=1}^{2}
\left[
\mathcal A_{i,x}(y,t)\hat{\bm x}
+
\mathcal A_{i,y}(y,t)\hat{\bm y}
\right].
\end{equation}
%%%

Having identified the in-plane field that couples to graphene, we now express it in a form that separates the rapid time-dependent oscillation from the spatial profile of the drive. The field remains monochromatic at frequency $\Omega$, but the coherent superposition of the two tilted beams makes the local field amplitude and the polarization profile depend on the transverse coordinate $y$. We write
%%%
\begin{equation}\label{jones_vector}
\bm{\mathcal A}_{\parallel}(y,t)
=
{\rm Re}\left[\bm a(y)e^{-i\Omega t}\right],
\;
\bm a(y)=\frac{A_{0}}{\sqrt{2}}[a_x(y)\hat{\bm x}+a_y(y)\hat{\bm y}].
\end{equation}
%%%
The Jones vector $\bm a(y)$ with complex components specifies the local in-plane optical field: its magnitude gives the field strength, while the relative phase between $a_x$ and $a_y$ determines the local polarization. For the two-beam drive, it is given by
%%%
\begin{equation}\label{jones_ay}
a_x(y)=%\frac{A_0}{\sqrt{2}}
\sum_{j=1}^{2} e^{iqy\sin\theta_j},\; a_y(y)= \sum_{j=1}^{2}[a_x(y)]_j\cos\theta_j\,e^{i\phi_j} .
\end{equation}
%%%
Here $[a_x(y)]_j$ denotes the contribution of beam $j$ to the $x$ component. The phase $
\alpha_j(y)=qy\sin\theta_j
$, appears in both in-plane components, while the $y$ component contains the projection factor $\cos\theta_j$ and the polarization phase $\phi_j$. As a result, the coherent sum of the two beams produces position-dependent in-plane intensity and polarization patterns across the ribbon.

At each position $y$, the time-dependent vector $\bm{\mathcal A}_{\parallel}(y,t)$ traces a full polarization modulation over one optical cycle. We describe it through the Stokes parameters normalized to $A^2_0$
%%%
\begin{align}\label{stokes_parameters}
S_0(y) &= \frac{1}{2}\left(|a_x|^2+|a_y|^2\right),
&
S_1(y) &= \frac{1}{2}\left(|a_x|^2-|a_y|^2\right),
\nonumber\\
S_2(y) &= {\rm Re}(a_xa_y^*),
&
S_3(y)&= {\rm Im}(a_xa_y^*) .
\end{align}
%%%
The quantity $S_0(y)\equiv\mathcal I(y)$ defines the in-plane field-intensity profile of the drive and is proportional to the time-averaged optical intensity at the location of the graphene membrane, as shown in Fig.~\ref{Fig1}. The parameters $S_1$ and $S_2$ describe the linear-polarization content and fix the orientation and shape of the polarization ellipse, while $S_3$ measures the circular-polarization component. We also define the normalized handedness
%%%
\begin{equation}\label{handedness}
\chi(y)=S_3(y)/S_0(y),
\end{equation}
%%%
so that $\chi(y)=\pm1$ corresponds to circular polarization of opposite handedness, while $\chi=0$ corresponds to linear polarization.

For the two-beam drive, the spatial dependence of the intensity and handedness is controlled 
by the phase difference 
$
\Delta\alpha(y)=\alpha_1(y)-\alpha_2(y)
=qy(\sin\theta_1-\sin\theta_2).
$
To write the resulting expressions compactly, we define
$
\delta(y)=\Delta\alpha(y)/2
$
and
$
c_j=\cos\theta_j .
$
Substituting Eq.~\eqref{jones_ay} into Eq.~\eqref{stokes_parameters} gives the in-plane field-intensity profile
%%%
\begin{equation}
\mathcal I(y)
=
2\cos^2\delta
+
\frac{c_1^2+c_2^2}{2}
+
c_1c_2\cos(2\delta+\phi_1-\phi_2),
\label{intensity_general}
\end{equation}
%%%
and the normalized handedness
%%%
\begin{equation}
\chi(y)
=
\frac{
-2\cos\delta
\left[
c_1\sin(\phi_1+\delta)
+
c_2\sin(\phi_2-\delta)
\right]
}
{
\mathcal I(y)
}.
\label{chi_general}
\end{equation}
%%%
These expressions show that $\mathcal I(y)$ and $\chi(y)$ are both periodic functions of the phase difference $\Delta\alpha(y)$, exhibiting the common spatial period
%%%
\begin{equation}\label{period_general}
\Lambda
=
\frac{2\pi}
{q|\sin\theta_1-\sin\theta_2|},
\end{equation}
%%%
that repeats across the width of the ribbon.  
We note that although $\mathcal I(y)$ and $\chi(y)$ repeat over the same length, they describe different aspects of the drive: $\mathcal I(y)$ gives the local intensity of the in-plane field, while $\chi(y)$ gives the local polarization. Thus, within one period, $\chi(y)$ changes sign when the field becomes linearly polarized.

For the special symmetric case shown in Fig.~\ref{Fig1},
$
\theta_1=-\theta_2=\theta$,
$\phi_1=0$,
$\phi_2=\pi,
$
we have
$
\delta(y)=qy\sin\theta
$
and
$
c_1=c_2=\cos\theta .
$
Eqs.~\eqref{intensity_general} and \eqref{chi_general} reduce to
%%%
\begin{equation}
{\mathcal I(y)}
=
1+\cos^2\theta
+
\sin^2\theta\,
\cos\!\left(2qy\sin\theta\right),
\label{intensity_symmetric}
\end{equation}
%%%
and
%%%
\begin{equation}
\chi(y)
=
-\frac{
2\cos\theta\,
\sin\!\left(2qy\sin\theta\right)
}
{
1+\cos^2\theta
+
\sin^2\theta\,
\cos\!\left(2qy\sin\theta\right)
}.
\label{chi_symmetric}
\end{equation}
%%%
In the symmetric geometry, the relative phase varies as
$
\Delta\alpha(y)=2qy\sin\theta,
$
and the common period of $\mathcal I(y)$ and $\chi(y)$ is
$
\Lambda=\pi/(q\sin\theta).
$
The optical structure generated by the coherent two-beam drive is illustrated in Fig.~\ref{Fig1}. The tilted beams interfere on the graphene plane, producing a standing-wave-like modulation of the in-plane intensity $\mathcal I(y)$ across the ribbon [Fig.~\ref{Fig1}(a)]. At the same time, the phase difference varies with position, so the local polarization evolves continuously across the period. This is represented by the polarization ellipses in Fig.~\ref{Fig1}(b) and (c): blue and red indicate opposite handedness, while straight lines mark the points where the handedness changes sign. This change is explicitly shown in Fig.~\ref{Fig1}(b). The normalized intensity profile $\mathcal I(y)$, which weights the local strength of the light--matter coupling is shown in Fig.~\ref{Fig1}(c).

%%%%%%%%%%%%%%%%%%%%%%%%%%%%%%%%%%%%%%%%%%%%%%%%
\subsection{Light--matter Coupling and Effective Floquet Mass}

Having identified the in-plane optical field in Eq.~\eqref{A_parallel_sum}, we now couple it to the low-energy Dirac quasiparticles of graphene. Near the two inequivalent valleys $K$ and $K'$, the continuum Hamiltonian is~\cite{graphrev}
%%%
\begin{equation}
H^0_\tau
=
v_{\rm F}\left(\tau\sigma_x p_x+\sigma_y p_y\right),
\end{equation}
%%%
where $v_{\rm F}$ is the Fermi velocity, $\bm p$ is measured from the corresponding valley, $\sigma_i$ acts on the sublattice pseudospin, and $\tau=\pm1$ denotes the $K$ and $K'$ valleys. The optical field enters through the minimal-coupling substitution
$
\bm p\rightarrow \bm p+e\bm{\mathcal A}_{\parallel}(y,t).
$

Since $\bm{\mathcal A}_{\parallel}$ varies only along the finite direction $y$, translational invariance along the ribbon direction is preserved. We therefore use $k_x$ as a good quantum number. Using the complex-amplitude representation introduced in Eq.~\eqref{jones_vector}, the driven Hamiltonian contains only the static term and the first harmonics at $\pm\Omega$:
%%%
\begin{equation}
\label{time_periodic_H}
H_\tau(k_x,y,t)
=
H^0_\tau(k_x,y)
+
V_{\tau,-}(y)e^{-i\Omega t}
+
V_{\tau,+}(y)e^{i\Omega t},
\end{equation}
%%%
where
%%%
$
H^0_\tau(k_x,y)
=
v_{\rm F}
\left(
\tau\hbar k_x\sigma_x
-
i\hbar\sigma_y\partial_y
\right)
$,
%%%
and
%%%
\begin{align}
V_{\tau,-}(y)
&=
\frac{e v_{\rm F}A_0}{2\sqrt{2}}
\left[
\tau a_x(y)\sigma_x
+
a_y(y)\sigma_y
\right],
\nonumber\\
V_{\tau,+}(y)
&=
\frac{e v_{\rm F}A_0}{2\sqrt{2}}
\left[
\tau a_x^*(y)\sigma_x
+
a_y^*(y)\sigma_y
\right].
\label{eq:Vpm}
\end{align}
%%%
Here $V_{\tau,-}$ and $V_{\tau,+}$ describe the coupling between neighboring Floquet modes. Physically, these terms encode virtual absorption and emission processes associated with the monochromatic drive.

Because Eq.~\eqref{time_periodic_H} is periodic in time with period $T=2\pi/\Omega$, the solutions can be written in Floquet form as
$\psi_{\tau,n}(k_x,y,t)=e^{-i\epsilon_n t/\hbar}\phi_{\tau,n}(k_x,y,t)$, where $\epsilon_n$ is the quasienergy defined modulo $\hbar\Omega$, and $\phi_{\tau,n}(k_x,y,t+T)=\phi_{\tau,n}(k_x,y,t)$. Expanding the periodic mode as
$
\phi_{\tau,n}(k_x,y,t)
=
\sum_m e^{im\Omega t}\phi_{\tau,n}^{m}(k_x,y)
$
maps the time-dependent Schr\"odinger equation onto an eigenvalue problem in Floquet--Sambe space~\cite{Floq-Shirley,Floq-Sambe}
%%%
\begin{equation}
\sum_m
\left(
H^F_\tau
\right)_{m'm}
\phi_{\tau,n}^{m}(k_x,y)
=
\epsilon_n
\phi_{\tau,n}^{m'}(k_x,y).
\end{equation}
%%%
The Floquet Hamiltonian matrix elements are
%%%
\begin{eqnarray}
\left(
H^F_\tau
\right)_{m'm}
&=&
\left(
H_\tau
\right)_{m'm}
+
m\hbar\Omega\,\delta_{m'm},
\nonumber\\
\left(
H_\tau
\right)_{m'm}
&=&
\frac{1}{T}
\int_0^T
H_\tau(k_x,y,t)
e^{i(m-m')\Omega t}\,dt .
\end{eqnarray}
%%%
Using Eq.~\eqref{time_periodic_H}, this gives the block-tridiagonal structure
%%%
\begin{equation}
\begin{aligned}
\left(
H^F_\tau
\right)_{m'm}
=
&
\left[
H^0_\tau(k_x,y)
+
m\hbar\Omega
\right]\delta_{m'm}
\\
&+
V_{\tau,+}(y)\delta_{m',m+1}
+
V_{\tau,-}(y)\delta_{m',m-1}.
\end{aligned}
\label{floquet_block}
\end{equation}
%%%
This representation organizes the driven problem in terms of sectors for different Floquet modes separated by $\hbar\Omega$. In the off-resonant regime the drive admixes neighboring Floquet modes and produces an effective static correction to the electronic Hamiltonian. Following the van Vleck construction~\cite{VanVleck,flreview1,flreview5,Lauren2025}, this correction is obtained perturbatively in inverse powers of $\Omega$,
%%%
\begin{equation}
H^{\rm eff}_\tau(k_x,y)
=
H^0_\tau(k_x,y)
+
\frac{[V_{\tau,+}(y),V_{\tau,-}(y)]}{\hbar\Omega}
+
O(\Omega^{-2}).
\end{equation}
%%%
The commutator describes the difference between the two virtual processes in which the electron absorbs and emits Floquet modes in opposite order. 
Using 
$
[V_{\tau,+},V_{\tau,-}]
=
\tau(e v_{\rm F}A_0)^2
{\rm Im}\!\left[a_x(y)a_y^*(y)\right]\sigma_z/2 ,
$
the effective Hamiltonian becomes
%%%
\begin{equation}
H^{\rm eff}_\tau(k_x,y)
=
H^0_\tau(k_x,y)
+
\tau\Delta(y)\sigma_z,
\label{effective_mass_H}
\end{equation}
%%%
where the light-induced Floquet mass is
$
\Delta(y)
=
(e v_{\rm F}A_0)^2
{\rm Im}\left[a_x(y)a_y^*(y)\right]/{(2\hbar\Omega)}$ and originates from the local circular polarization of the in-plane drive. This circular component is encoded in the Stokes parameter $S_3$, giving
%%%
\begin{equation}
\Delta(y)
=
\frac{(e v_{\rm F}A_0)^2}{2\hbar\Omega}S_3(y)
=
\frac{\hbar\Omega}{2}g^2\;\mathcal{I}(y)\chi(y),
\label{mass_stokes}
\end{equation}
%%%
$\mathcal I(y)\chi(y)$ modulates the amplitude and sign of the local light--matter coupling $g=ev_{\rm F} A_0/(\hbar\Omega)$. Clearly the mass vanishes when the circular component vanishes at linear polarization points where $\chi(y)=0$.

Using the optical profiles derived in Eqs.~\eqref{intensity_general} and \eqref{chi_general}, the mass can now be obtained directly from Eq.~\eqref{mass_stokes}. The light-matter coupling introduces the characteristic mass scale
$
\Delta_0=\hbar\Omega g^2
$,
that can be used to normalize the Floquet mass, rendering 
%%%
\begin{equation}
\frac{\Delta(y)}{\Delta_0}
=
-\frac{\mathcal C}{2}
-
\frac{1}{2}
\left[
\mathcal C\cos\Delta\alpha(y)
+
\mathcal D\sin\Delta\alpha(y)
\right],
\label{mass_offset_osc}
\end{equation}
%%%
where
$
\mathcal C=c_1\sin\phi_1+c_2\sin\phi_2
$
and
$
\mathcal D=c_1\cos\phi_1-c_2\cos\phi_2
$.

%%%
\begin{figure}[ht]
  \centering
  \includegraphics[width=0.48\textwidth]{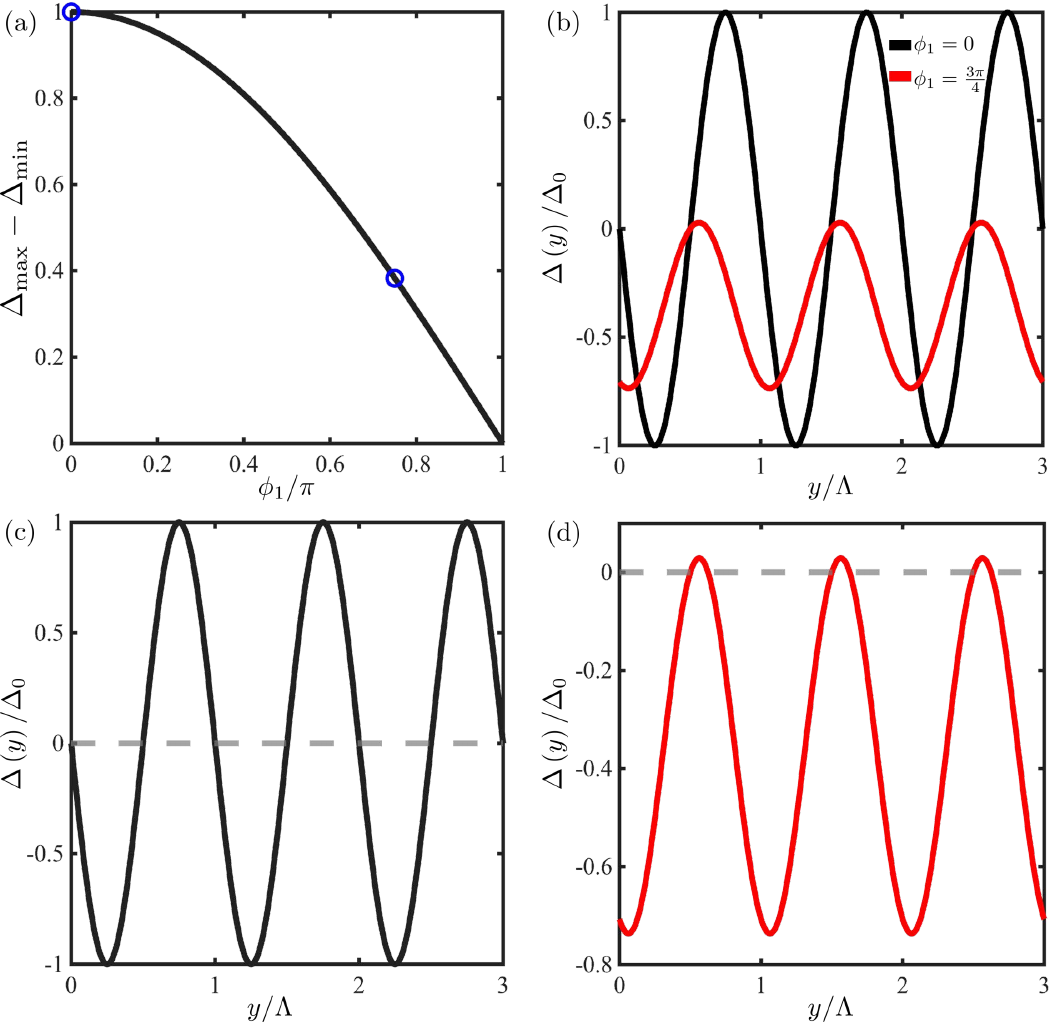}
\caption{(a) Mass contrast $\Delta_{\max}-\Delta_{\min}$ as a function of $\phi_1$, with $\phi_2=\pi$ kept fixed; blue dots indicate the phases used in (b)--(d). 
(b) Normalized mass profiles $\Delta(y)/\Delta_0$ for $\phi_1=0$ and $\phi_1=3\pi/4$. 
(c),(d) Corresponding real-space mass domains.} %All other parameters are as in Fig.~\ref{Fig1}.}
\label{Fig2}
\end{figure}
%%%
The light-induced mass has two contributions with distinct physical origins. The spatially uniform term, $-\mathcal C/2$, comes from the total circular polarization carried by the beams after projection onto the graphene plane. For beam $j$, this projected circular weight is proportional to
$
c_j\sin\phi_j=\cos\theta_j\sin\phi_j .
$
The average mass is nonzero when the projected circular contributions of the two beams do not cancel,
$
c_1\sin\phi_1+c_2\sin\phi_2\neq 0
$.
The remaining terms in Eq.~\eqref{mass_offset_osc} 
form an oscillating mass superlattice whose period is fixed by $\Delta\alpha(y)$. 
Thus, the beam polarization phases and incidence angles control the average mass and the spatial distribution of positive and negative mass domains, while the superlattice period is fixed by Eq.~\eqref{period_general}.

For the special symmetric configuration the two beams have equal in-plane projection factors
$
c_1=c_2=\cos\theta,
$
and both are linearly polarized. The uniform offset vanishes because
$
\mathcal C=c_1\sin\phi_1+c_2\sin\phi_2=0.
$
The oscillating part is controlled by
$
\mathcal D=c_1\cos\phi_1-c_2\cos\phi_2=2\cos\theta,
$
while
$
\Delta\alpha(y)=2qy\sin\theta .
$
Equivalently, using Eqs.~\eqref{intensity_symmetric} and \eqref{chi_symmetric},
%%%
$
\mathcal I(y)\chi(y)
=
-2\cos\theta\,
\sin\!\left(2qy\sin\theta\right).
$
%%%
Thus, Eq.~\eqref{mass_offset_osc} reduces to
%%%
$
\Delta(y)
=
-\Delta_0\cos\theta\,
\sin\!\left(2qy\sin\theta\right).
$
%%%
The corresponding effective Hamiltonian is
%%%
\begin{equation}
H^{\rm eff}_\tau(k_x,y)
=
H^0_\tau(k_x,y)
-
\tau \Delta_0\cos\theta
\sin\!\left(2qy\sin\theta\right)\sigma_z .
\end{equation}
%%%
Although each beam is linearly polarized, their coherent superposition produces a spatially varying polarization profile that includes circular, elliptical, and linear regimes across the ribbon, Fig.~\ref{Fig1}(b). The resulting handedness changes sign periodically, producing alternating positive and negative mass domains with zero spatial average in a full period. The induced mass therefore forms a one-dimensional optical superlattice with the period $\Lambda$ determined by the handedness profile discussed in Eq.~\eqref{period_general}. For the symmetric geometry,
$
\Lambda=
\pi/{(q\sin\theta)}.
$
This optical length scale controls the commensuration between the light-induced spatial period and the ribbon width.

Fig.~\ref{Fig2} shows how the optical beams phases controls the mass profile. Fig.~\ref{Fig2}(a) shows the mass contrast, $\Delta_{\max}-\Delta_{\min}$, as $\phi_1$ changes, while all other beam parameters are kept fixed. In terms of Eq.~\eqref{mass_offset_osc}, this contrast is controlled by the amplitude of the oscillating space-dependant component, while the vertical shift is controlled by the offset $-\mathcal C/2$. The blue dots identify the two representative phases shown in Figs.~\ref{Fig2}(b)--(d). In Fig.~\ref{Fig2}(b) we compare the corresponding normalized profiles $\Delta(y)/\Delta_0$. For $\phi_1=0$, the symmetric linear-polarization configuration discussed above has zero offset, so the mass is purely oscillatory and alternates between positive and negative domains, as shown in Fig.~\ref{Fig2}(c). For $\phi_1=3\pi/4$, with $\phi_2$ kept fixed at $\pi$, the first beam carries a finite projected circular component. This makes $\mathcal C>0$ and produces a negative offset $-\mathcal C/2$, shifting the mass profile downward. As a result, the positive-mass regions become narrow while the negative-mass regions extend over most of the period, as shown in Fig.~\ref{Fig2}(d). 

%%%%%%%%%%%%%%%%%%%%%%%%%%%%%%%%%%%%%%%%%%%%%%%%%%%%%%%%%%%%%%%%%%%%%%%%%%%%%%%%%%%%%%%%%%%%%%%%%%%%%
\section{Quasienergy spectrum and states of a space-time-modulated ribbon}
\label{quasienergy_states}

Having derived the effective Floquet Hamiltonian in Eq.~\eqref{effective_mass_H}, we now determine the quasienergy spectrum and eigenstates of a finite size zigzag graphene nanoribbon. We focus on the interplay between the light-induced superlattice period $\Lambda$ and the ribbon width $W$, which controls the spatial arrangement of the mass domains.

%%%%%%%%%%%%%%%%%%%%%%%%%%%%%%%%%%%%%%%%%%%%%%%%
\subsection{Numerical Quasienergy Spectrum and States}
\label{numerical_spectrum}

%%%
%%%
\begin{figure}[ht!]
  \centering
  \includegraphics[width=0.48\textwidth]{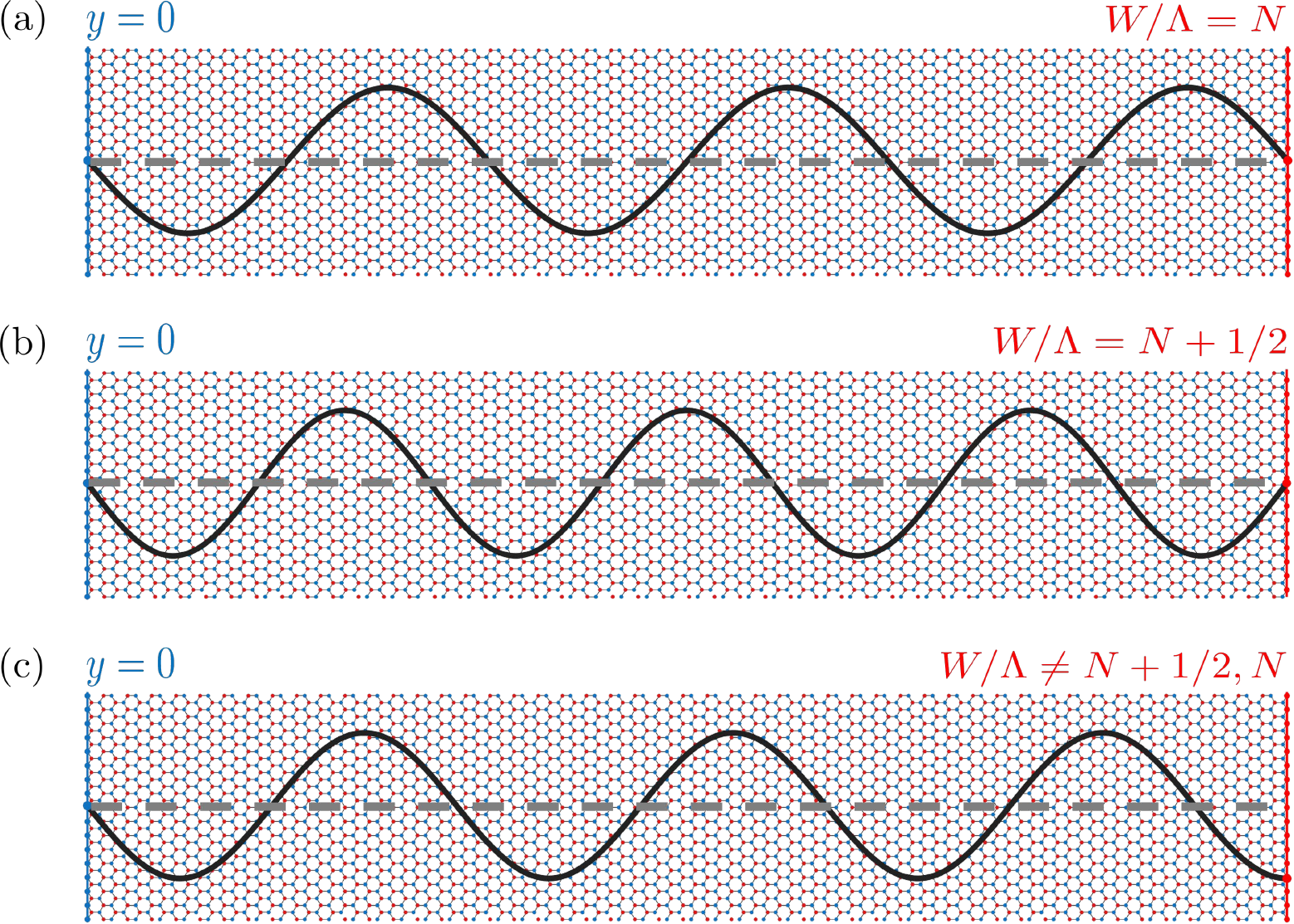}
\caption{Boundary matching of the Floquet mass profile $\Delta(y)$ in a finite zigzag graphene nanoribbon. The black curve shows $\Delta(y)$, the gray dashed line marks $\Delta=0$, and the boundaries are at $y=0$ and $y=W$.  (a) Integer edge matching, $W/\Lambda=N$, with both boundaries at equivalent mass nodes.  (b) Half-integer edge matching, $W/\Lambda=N+1/2$, where the right boundary also lies at a mass node.  (c) Edge mismatch, $W/\Lambda\neq N$ and $W/\Lambda\neq N+1/2$, where the right boundary samples a finite local mass. Here $N\in\mathbb{Z}^{+}$.}
\label{Fig3new}
\end{figure}
%%%
%%%
To resolve the quasienergy spectrum and the space dependence of the light-induced states, we discretize the effective Floquet Hamiltonian in Eq.~\eqref{effective_mass_H} on a uniform mesh across the finite ribbon width, $0\leq y\leq W$. The spatially varying Floquet mass $\Delta(y)$ is evaluated directly on this mesh, while the first-order Dirac derivatives require a discretization that preserves the physical zigzag boundaries and avoids spurious lattice states. A conventional central approximation,
$
\partial_y f_i
\approx
{(f_{i+1}-f_{i-1})}/{(2h)}
+
O(h^2)
$,
is unsuitable because it generates the fermion-doubling problem characteristic of discretized Dirac Hamiltonians~\cite{Stacey1982,szafran2019fd,PhysRevC.106.L051303}. 

We instead represent the derivatives by complementary forward and backward operators~\cite{PhysRevC.106.L051303}. For instance for the three-point difference, the forward derivative is 
%%%
$
D_{+}f_i
=
(-3f_i+4f_{i+1}-f_{i+2})/{(2h)}
+O(h^2)
$, and the backward one is 
$
D_{-}f_i
=
(3f_i-4f_{i-1}+f_{i-2})/{(2h)}
+O(h^2)
$. The numerical results presented in this work were obtained using $4$-point finite differences with $O(h^3)$. The asymmetric discretization removes the even--odd mesh symmetry responsible for fermion doubling without introducing a Wilson-type mass term or otherwise modifying the continuum Hamiltonian~\cite{Stacey1982,Beenakker2023TangentFermions,walsh2026}.
%%%
Applying them to the off-diagonal elements of the Dirac Hamiltonian gives
%%%
\begin{equation}
H^{\rm eff}_{\tau,h}
=
\begin{pmatrix}
\tau\Delta
&
\hbar v_{\rm F}(\tau k_x-D_{+})
\\
\hbar v_{\rm F}(\tau k_x+D_{-})
&
-\tau\Delta
\end{pmatrix},
\label{discrete_effective_H}
\end{equation}
%%%
where $D_{+}^{\dagger}=-D_{-}$, resulting in a Hermitian Hamiltonian. 

The same forward--backward structure is naturally compatible with the zigzag termination of the ribbon. For the sublattice convention used here, the left zigzag edge terminates on the $A$ sublattice, so that $\psi_{B,\tau}(0)=0$, while the right edge terminates on the $B$ sublattice, giving $\psi_{A,\tau}(W)=0$. The forward derivative samples the wave function from the left boundary toward the ribbon interior, whereas the backward derivative samples it from the right boundary toward the interior. Their complementary use therefore follows the opposite sublattice terminations of the two zigzag edges. 

%%%
\begin{figure}[ht!]
  \centering
  \includegraphics[width=0.48\textwidth]{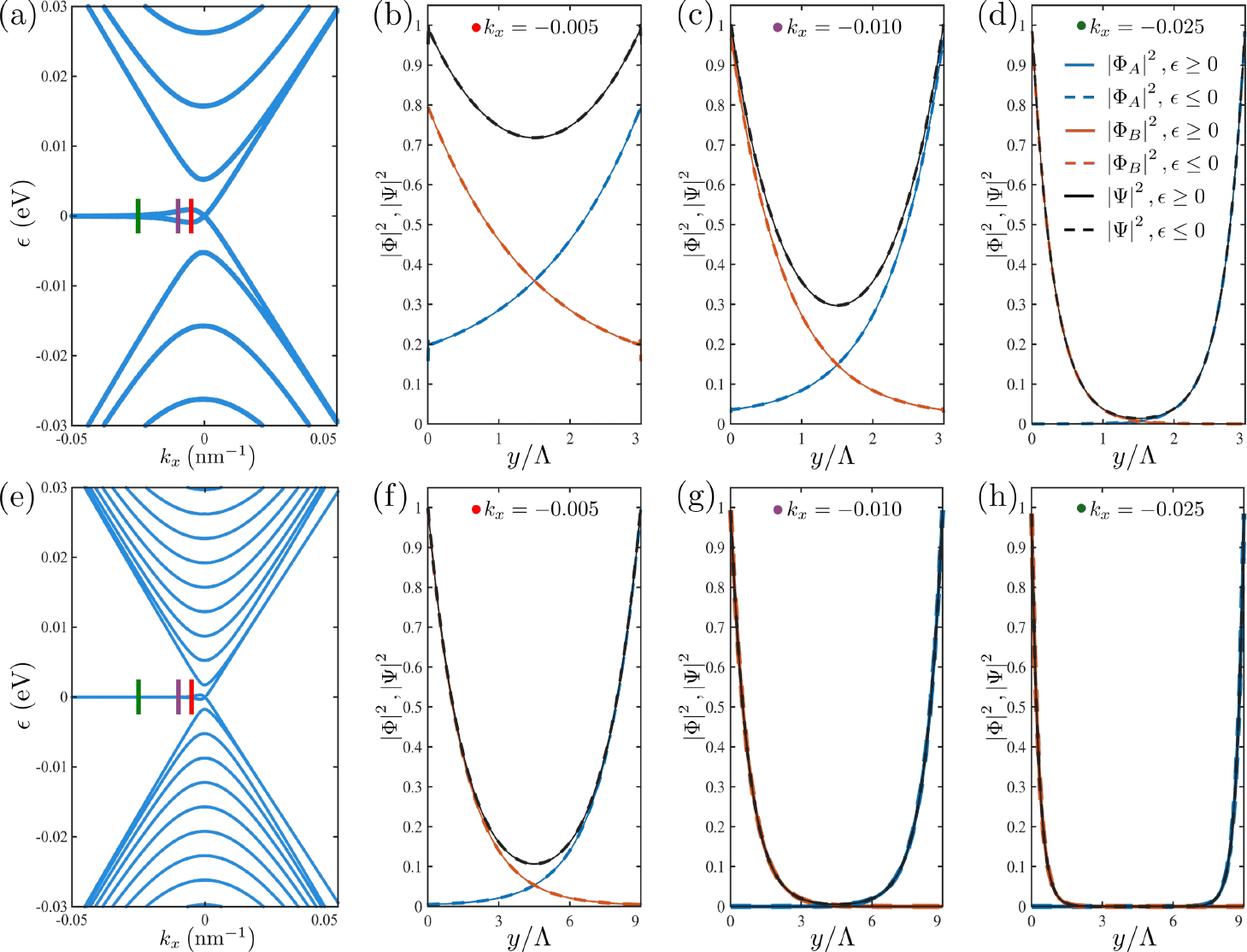}
\caption{Equilibrium energy spectra and edge-state profiles of undriven zigzag graphene nanoribbons with $W=3\Lambda$ (a)--(d) and $W=9\Lambda$ (e)--(h), where $\Lambda=66~{\rm nm}$ is used as a reference length for comparison with the irradiated system. (a),(e) Energy spectra at the $K$ valley ($\tau=+1$); the colored vertical markers indicate $k_x=-0.005$, $-0.010$, and $-0.025~{\rm nm}^{-1}$. (b)--(d),(f)--(h) Corresponding sublattice-resolved densities $|\Phi_A|^2$ and $|\Phi_B|^2$ and total density $|\Psi|^2$. 
}
\label{Fig3}
\end{figure}
%%%
%%%%%%%%%%%%%%%%%%%%%%%%%%%HERE%%%%%%%%%%%%%%%%%%%%
Before presenting the spectra for the driven system, it is useful to distinguish two boundary classes of the Floquet mass profile. For the symmetric two-beam geometry used below,
$
\Delta(y)
=
-\Delta_0\cos\theta\sin(2\pi y/\Lambda),
$
the left boundary at $y=0$ is always a mass node. The right boundary at $y=W$ is also a mass node if
$
\Delta(W)=0,
$
or equivalently when
$
W/\Lambda=\ell/2$, $\ell\in\mathbb{Z}^{+}$.
We refer to these configurations as {\it``edge matched.''} They include integer-period ribbons, $W=N\Lambda$ ($N\in\mathbb{Z}^{+}$), where both boundaries lie at equivalent mass nodes [Fig.~\ref{Fig3new}(a)], and half-integer-period ribbons, $W=(N+1/2)\Lambda$, where the right boundary also lies at a mass node [Fig.~\ref{Fig3new}(b)]. In contrast, when
$
W/\Lambda\ne \ell/{2}$,
the right boundary samples a finite local mass $\Delta(W)\neq0$. We refer to the configuration as {\it``edge mismatched''} [Fig.~\ref{Fig3new}(c)]. This distinction will play a crucial role in the low-energy edge spectrum.

%%%%%%%%%%%%%%%%%%%%%%%%%%%%%%%%%%%%%%%%%%%%%%%%
\subsection{Edge Matched Floquet Mass Modulation} 
\label{commensurate_edge_states}

Figs.~\ref{Fig3}(a) and \ref{Fig3}(e) show the equilibrium spectra for an undriven zigzag nanoribbon of width $W=3\Lambda$ and $W=9\Lambda$, respectively. The results are shown for the $K$ valley, where the edge branches occur at negative $k_x$; at $K'$, they are obtained by time-reversal $k_x\rightarrow -k_x$. 
As expected, the spectra contain the characteristic zigzag flat bands near $\epsilon=0$. In the semi-infinite width limit, the edge-state amplitude decays as
%%%
$
|\Psi_{\rm edge}(d)|
\propto
e^{-d/\xi_{\rm edge}}$, $\xi_{\rm edge}\simeq |k_x|^{-1},
$
%%%
so states closer to the $K$ valley penetrate farther into the ribbon and can hybridize across a finite width~\cite{Nakada1996,Fujita1996,BreyFertig2006,Wakabayashi2010}.

%%%
\begin{figure}[ht!]
  \centering
  \includegraphics[width=0.48\textwidth]{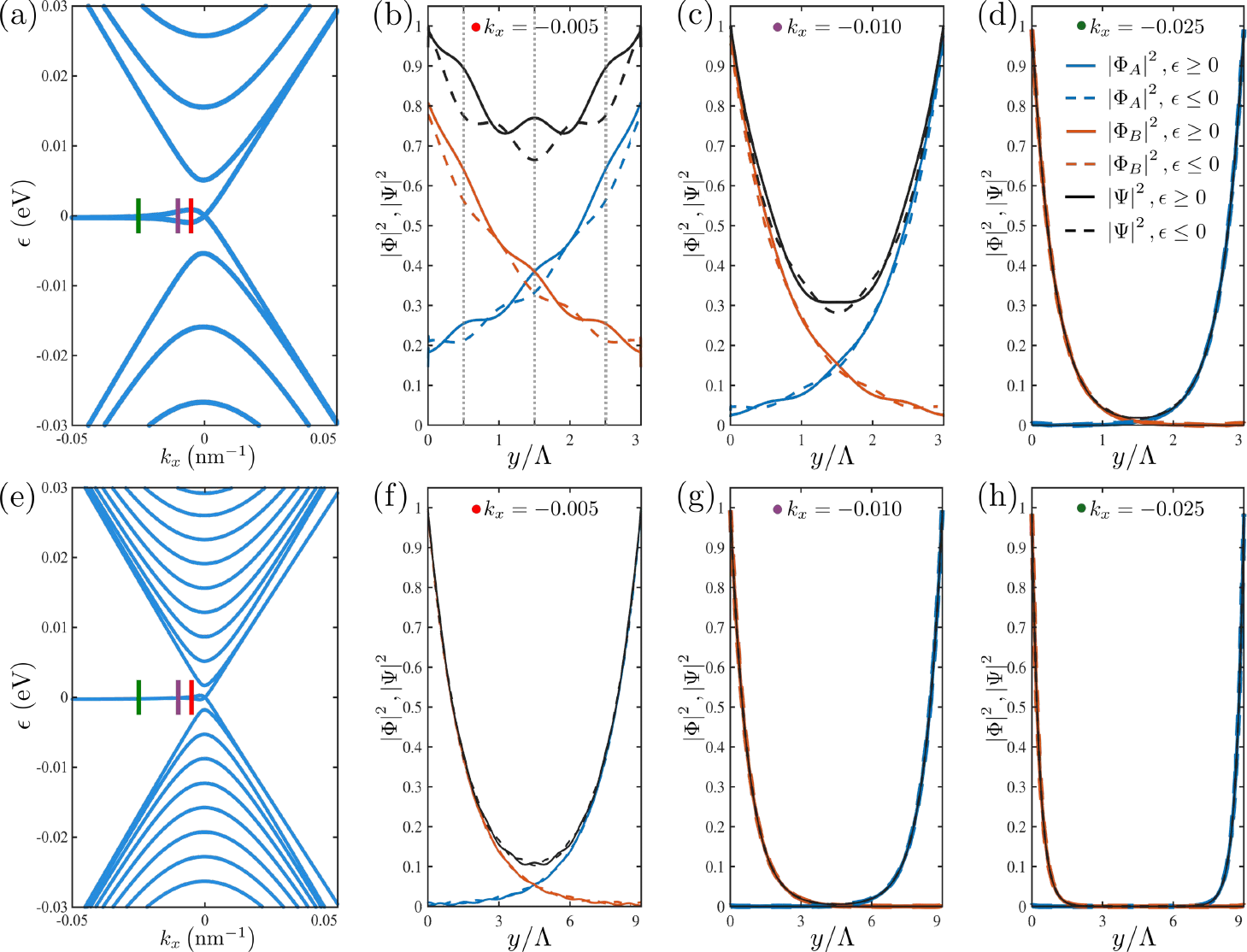}
\caption{Floquet spectra and edge-state profiles for irradiated zigzag graphene nanoribbons with $W=3\Lambda$ (a)--(d) and $W=9\Lambda$ (e)--(h). The results are for the $K$ valley ($\tau=+1$) and the symmetric two-beam configuration $\theta_1=-\theta_2=70^\circ$, $\phi_1=0$, and $\phi_2=\pi$, with $\hbar v_{\rm F}=0.66~{\rm eV\,nm}$, $\Lambda=66~{\rm nm}$, and $\Delta_0=4.0~{\rm meV}$, as defined in Eq.~\eqref{mass_offset_osc}. 
(a),(e) Quasienergy spectra; the colored vertical markers indicate $k_x=-0.005$, $-0.010$, and $-0.025~{\rm nm}^{-1}$. (b)--(d),(f)--(h) Sublattice-resolved and total probability densities, $|\Phi_A|^2$, $|\Phi_B|^2$, and $|\Psi|^2$.
}
\label{Fig4}
\end{figure}
%%%
This behavior is illustrated by the densities in Figs.~\ref{Fig3}(b)--(d) and (f)--(h). For the narrower ribbon, $W=3\Lambda$, the low-$|k_x|$ states have long penetration depths and strongly overlap across the ribbon, producing the finite-size quasienergy splitting near $k_x=0$ in Fig.~\ref{Fig3}(a). As $|k_x|$ increases from $0.005$ to $0.025~{\rm nm}^{-1}$ [Figs.~\ref{Fig3}(b)--\ref{Fig3}(d)], the penetration depth decreases, the overlap is suppressed, and the states become localized on opposite zigzag edges with sublattice polarization, as expected~\cite{Nakada1996}. Increasing the width to $W=9\Lambda$ further reduces this overlap [Figs.~\ref{Fig3}(e)--\ref{Fig3}(h)]. Thus, increasing either $|k_x|$ or $W$ drives the system toward the decoupled-edge limit, where the left- and right-localized edge states become degenerate up to exponentially small finite-size corrections.

We next consider the symmetric two-beam configuration described by the effective Floquet Hamiltonian in Eq.~\eqref{effective_mass_H}. We first focus on edge matched widths, where both physical boundaries lie at mass nodes, as illustrated in Figs.~\ref{Fig3new}(a) and \ref{Fig3new}(b). 
For $W=3\Lambda$ [Fig.~\ref{Fig4}(a)], the low-energy branches retain the characteristic zigzag-edge dispersion and exhibit a small finite-size quasienergy splitting near $k_x=0$, where the large penetration depth allows states localized at opposite boundaries to hybridize. Increasing the width to $W=9\Lambda$ [Fig.~\ref{Fig4}(e)] strongly reduces this overlap and correspondingly suppresses the low-quasienergy splitting. Thus, for both edge matched widths, the quasienergy spectrum remains close to that of the corresponding undriven ribbon in Fig.~\ref{Fig3}. However, this apparent spectral similarity conceals a qualitatively different real-space structure of the eigenstates.

The Floquet-induced reconstruction of the wave functions is most pronounced for $W=3\Lambda$ [Figs.~\ref{Fig4}(b)--\ref{Fig4}(d)]. At $k_x=-0.005~{\rm nm}^{-1}$ [Fig.~\ref{Fig4}(b)], the edge-state wave functions retain appreciable weight throughout the ribbon interior and therefore probe several mass domains. Then, in contrast to the smooth decay of the corresponding equilibrium states in Fig.~\ref{Fig3}(b), their extended tails acquire an oscillatory modulation locked to the periodic Floquet mass. Away from the physical boundaries, the positive-quasienergy state develops density maxima near
$
y/\Lambda=n+1/2
$,
where $\Delta(y)$ changes from negative to positive, and minima near
$
y/\Lambda=n
$,
where $\Delta(y)$ changes from positive to negative, with $n\in\mathbb{Z}$. The negative-quasienergy state exhibits the complementary pattern. The physical edges with maxima at $y=0$ and $y=W$ remain dominated by the zigzag boundary condition and should be distinguished from the interior modulations generated by the mass domain walls.

At $k_x=-0.010~{\rm nm}^{-1}$ [Fig.~\ref{Fig4}(c)], the reduced penetration depth means fewer mass domains sampling and weakened oscillations. For $k_x=-0.025~{\rm nm}^{-1}$ [Fig.~\ref{Fig4}(d)], the wave functions are confined almost entirely to the physical edges and no longer appreciably probe the internal mass sign reversals. 
A similar suppression occurs for the wider ribbon, $W=9\Lambda$. The increased edge separation strongly reduces finite-size hybridization [Fig.~\ref{Fig4}(e)]. At fixed $k_x$, the intrinsic penetration depth is unchanged, but the wave functions decay before reaching the opposite edge. Consequently, the interior weight is weak in Fig.~\ref{Fig4}(f) and becomes negligible in Figs.~\ref{Fig4}(g) and \ref{Fig4}(h). 

These trends reflect the competition between localization at the physical zigzag boundaries and localization at the light-induced mass interfaces. Each sign reversal of the Dirac mass defines a kink or antikink state that supports a Jackiw--Rebbi-type localization~\cite{JackiwRebbi1976}. States with long penetration depths sample several such interfaces and acquire a pronounced oscillatory structure, whereas increasing $|k_x|$ strengthens physical-edge localization and increasing $W$ suppresses opposite-edge hybridization. 

Similarly at $K'$, the spectrum is reversed under $k_x\rightarrow -k_x$, while the Floquet mass term changes sign because it is proportional to $\tau\Delta(y)\sigma_z$. As a result, the real-space density modulation is interchanged between the positive- and negative-quasienergy branches.

This valley contrast should therefore be understood as an edge-state readout of the Floquet mass superlattice. The states remain anchored to the same physical zigzag boundaries, but their interior density maxima follow opposite mass domains at $K$ and $K'$ for a fixed quasienergy branch. This distinguishes the present zigzag geometry from the armchair-ribbon superlattice of Ref.~\cite{Torres1}, where the absence of the characteristic zero-energy zigzag edge band makes the periodic modulation primarily a bulk superlattice effect. The pre-existing zigzag edge states convert the periodic Floquet mass into a valley-dependent real-space density modulation for states with $\epsilon\simeq0$.

%%%%%%%%%%%%%%%%%%%%%%%%%%%%%%%%%%%%%%%%%%%%%%%%
\subsection{Edge Mismatched Floquet Mass Modulation}
% and edge-state degeneracy lifting}
\label{incommensurate_edge_states}

\begin{figure}[ht!]
  \centering
  \includegraphics[width=0.48\textwidth]{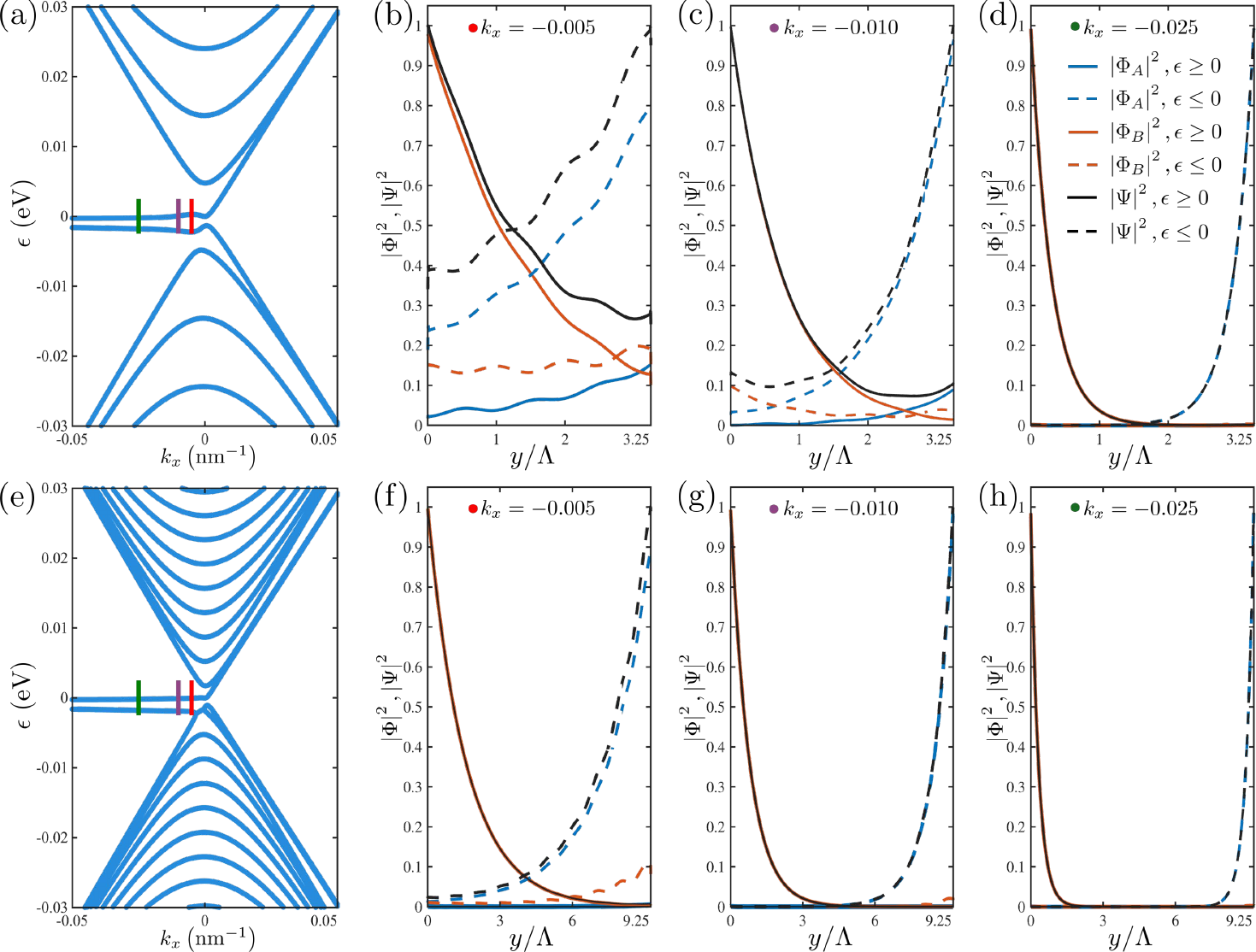}
\caption{Floquet spectra and edge-state profiles for edge mismatched zigzag graphene nanoribbons with $W=3.25\Lambda$ (a)--(d) and $W=9.25\Lambda$ (e)--(h), where $\Lambda=64.2$ nm for the same symmetric two-beam configuration in Fig.~\ref{Fig4}. 
(a),(e) Quasienergy spectra at the $K$ valley; colored markers indicate $k_x=-0.005$, $-0.010$, and $-0.025~{\rm nm}^{-1}$. 
(b)--(d),(f)--(h) Corresponding sublattice-resolved and total densities. 
}
\label{Fig5}
\end{figure}

We now turn to edge mismatched irradiation, where the Floquet mass profile terminates inside a finite-mass domain at the right boundary instead of at a mass node [Fig.~\ref{Fig3new}(c)]. 
To isolate the effect of this boundary mismatch, we fix the physical width and mass scale and vary only the modulation period $\Lambda$. Fig.~\ref{Fig5} shows two representative edge mismatched cases.

The main difference from the edge matched case is that the two physical edges no longer terminate at equivalent points of the Floquet mass profile, as shown in Fig.~\ref{Fig3new}(c). 
In the edge mismatched cases shown in Fig.~\ref{Fig5}, the two edges sample different boundary values and nearby domains of $\Delta(y)$. The mismatch between the mass superlattice period and the sample width breaks the degeneracy between states that would be degenerate in the case of matched edges.  
The spectrum is therefore controlled not only by edge-state overlap, but also by how the Floquet mass profile terminates at the boundaries.

The corresponding wave functions in Fig.~\ref{Fig5} reveal how this boundary mismatch modifies the edge states. For $W=3.25\Lambda$ [Figs.~\ref{Fig5}(b)--\ref{Fig5}(d)], the states still show oscillations associated with the internal mass domains, but the pattern is no longer symmetric between the states at the two boundaries. 
For the wider ribbon, $W=9.25\Lambda$ [Figs.~\ref{Fig5}(f)--\ref{Fig5}(h)], the states become strongly localized near the physical edges, as in the edge matched case. However, because the two edges terminate in different local mass environments, the left- and right-localized states stay spectrally split even after their spatial overlap is suppressed.

\begin{figure}[ht!]
  \centering
  \includegraphics[width=0.5\textwidth]{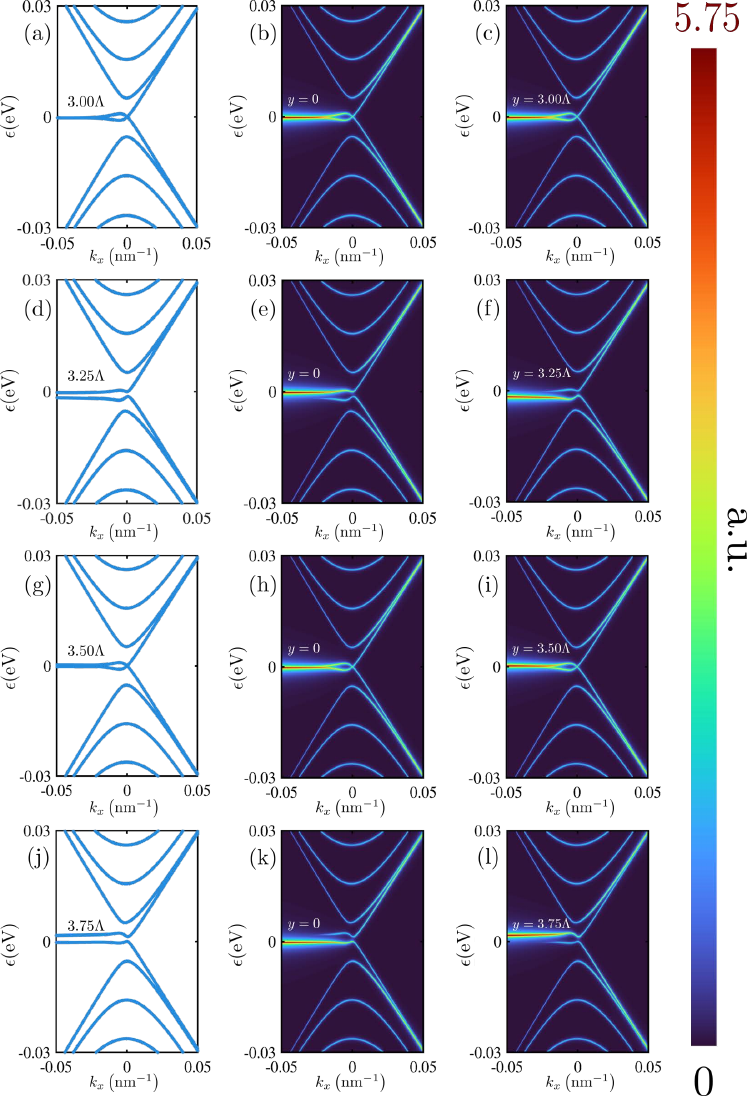}
\caption{Boundary-resolved LDOS for irradiated zigzag graphene nanoribbons with fixed physical width $W=198~{\rm nm}$ and varying mass-period ratio $W/\Lambda=3.00$, $3.25$, $3.50$, and $3.75$ from top to bottom. The physical width is fixed to the $W=3\Lambda$ reference case of Fig.~\ref{Fig4}, while $\Lambda$ is varied; all other parameters are as in Fig.~\ref{Fig4}. 
(a),(d),(g),(j) Quasienergy spectra; (b),(e),(h),(k) LDOS at the left edge, $y=0$; and (c),(f),(i),(l) LDOS at the right edge, $y=W$. 
The LDOS is momentum resolved, $\log[1+\rho(y,E,k_x)]$, in arbitrary units.
}
\label{Fig6}
\end{figure}

The edge selectivity is further resolved by the momentum-resolved boundary local density of states (LDOS) in Fig.~\ref{Fig6}. The spectra in the left column track the evolution of the edge branches as $W/\Lambda$ is varied from the integer edge matched value $3.00$, through the edge mismatched value $3.25$, to the half-integer edge matched value $3.50$, and then to the edge mismatched value $3.75$. The middle and right columns show the LDOS at $y=0$ and $y=W$, respectively. For the $K$ valley,
$
\Delta(y)=-\Delta_0\cos\theta\sin(2\pi y/\Lambda),
$
so the left edge always lies at a mass node, $\Delta(0)=0$, and its edge branch remains pinned near $\epsilon=0$ once the hybridization is beyond resolution. The right edge instead samples $\Delta(W)\propto-\sin(2\pi W/\Lambda)$. Thus, for $N<W/\Lambda<N+1/2$ the right-edge branch shifts to negative quasienergies, while for $N+1/2<W/\Lambda<N+1$ it shifts to positive quasienergies. At the edge matched points $W/\Lambda=N$ and $W/\Lambda=N+1/2$, the right edge also lies at a mass node, so the edge branch is restored toward $\epsilon=0$. 
At the $K'$ valley, this behaviour is reversed.
Consequently, the node-pinned branch remains localized at $y=0$, while the branch shifted away from $\epsilon=0$ is localized at $y=W$. 

%%%
%%%
In this setup, the right edge behaves as a quasienergy-selective valley/chirality filter 
since the effective mass has opposite sign at $K$ and $K'$. 
Therefore, at $y=W$, the sign of the quasienergy selects the valley and the corresponding edge-branch dispersion: for $N<W/\Lambda<N+1/2$, negative-energy right-edge states belong to $K$ and positive-energy right-edge states belong to $K'$, while for $N+1/2<W/\Lambda<N+1$ this assignment is reversed. Thus, the edge-mismatch lifts the edge degeneracy and enables valley selection at the right boundary.

%%%
%%%
%%%%%%%%%%%%%%%%%%%%%%%%%%%%%%%%%%%%%%%%%%%%%%%%
\subsection{Effective Edge-state Model}
\label{effective_edge_model}

To describe the numerical trends found above, we project the Floquet problem onto the low-energy subspace of the two zigzag edge states,
$
\{\ket{\Psi_L},\ket{\Psi_R}\}.
$
The construction follows the adapted tunneling approach of Ref.~\cite{thin4}, applied here to graphene nanoribbons. In this approach, the ribbon is treated as a massless Dirac-like graphene region bounded by massive Dirac regions that generate the zigzag terminations. The finite-mass interface problem, the regularized evaluation of the left--right tunneling matrix element, and the subsequent hard-wall limit are given in Appendix~\ref{appendix_effective_edge_model}.

To leading order in the left--right overlap, the projected edge Hamiltonian is
%%%
\begin{equation}
H_{\rm edge}^{\tau}(k_x)
=
\begin{pmatrix}
\varepsilon_L^\tau(k_x) & t(k_x) \\
t^*(k_x) & \varepsilon_R^\tau(k_x)
\end{pmatrix}_{L,R}.
\label{edge_H_mass}
\end{equation}
%%%
Here $t(k_x)$ is the unirradiated hybridization matrix element between the two edge states, and
$
\varepsilon_{L,R}^{\tau}
=
\bra{\Psi_{L,R}}
\tau\Delta(y)\sigma_z
\ket{\Psi_{L,R}}
$
are the light-induced mass projections. Because the $L$ (left) and $R$ (right) states occupy opposite sublattices in the hard-wall zigzag limit, the mass operator $\sigma_z$ becomes a diagonal, edge-dependent term in the $\{\ket{\Psi_L},\ket{\Psi_R}\}$ basis.

In the unirradiated limit, $\varepsilon_L^\tau=\varepsilon_R^\tau=0$, so the low-quasienergy splitting comes only from the overlap between the two edge states as a manifestation of finite-size effects. As shown in Appendix~\ref{appendix_effective_edge_model}, the tunneling matrix element 
%is evaluated at finite confining mass and only then taken to the hard-wall zigzag limit. Choosing the relative phase of $\ket{\Psi_L}$ and $\ket{\Psi_R}$ so that this matrix element 
is real, positive, and given by
%%%
\begin{figure}[ht!]
  \centering
  \includegraphics[width=0.48\textwidth]{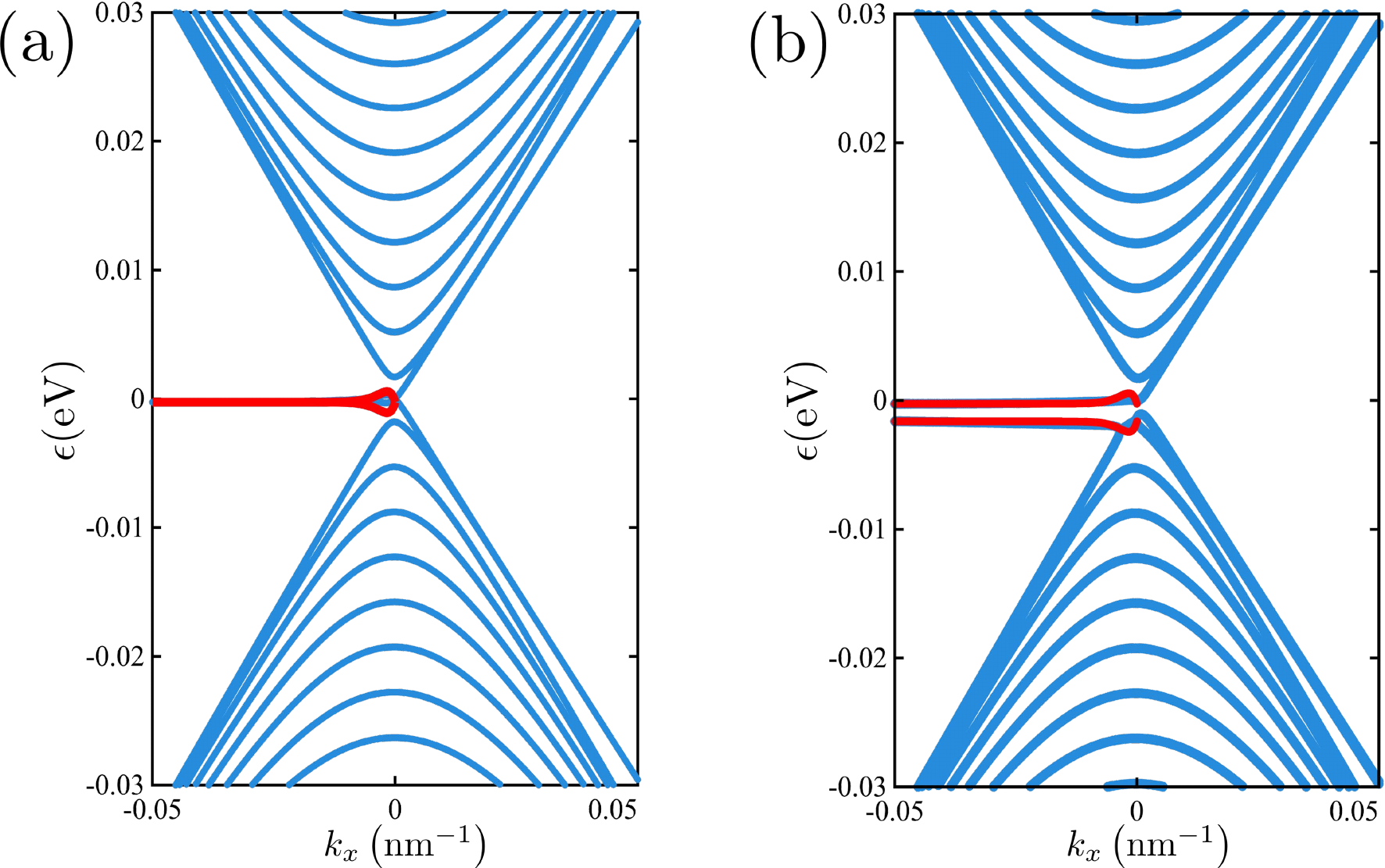}
\caption{Boundary-induced quasienergy splitting from the effective edge-state model. 
(a),(b) Numerical spectra, blue, compared with the analytical two-edge model, red, for $W/\Lambda=9.00$ and $W/\Lambda=9.25$, respectively. The edge matched case remains nearly degenerate, while the edge mismatched case shows a finite right-edge shift. All parameters are the same as in Fig.~\ref{Fig4}.}
\label{Fig7}
\end{figure}
%%%
%%%
\begin{equation}
t(k_x)
=
2\hbar v_{\rm F} |k_x|
e^{-W|k_x|}.
\label{edge_tunneling}
\end{equation}
%%%
Thus, $t(k_x)$ sets the physical hybridization scale between the two edges for $|k_x|>0$. 

As shown in Appendix~\ref{appendix_effective_edge_model}, the interface solutions determine the allowed low-energy momentum sector through normalizability. For the zigzag branch considered here, the normalizable solutions satisfy $\tau k_x<0$, corresponding to $k_x<0$ at $K$ and $k_x>0$ at $K'$. The same derivation fixes the sublattice assignment of the two edge states according to 
$
{\rm sgn}(\tau k_x)=s_L(k_x)$,
$-{\rm sgn}(\tau k_x)=s_R(k_x).
$
Since the Floquet mass enters as $\tau\Delta(y)\sigma_z$, these sublattice signs determine the edge-quasienergy shifts. For strongly localized edge states, the expectation values are controlled mainly by the mass sampled at the corresponding boundary,
%%%
$
\varepsilon_L^\tau(k_x)
\simeq
\tau s_L(k_x)\Delta(0)$,
$\varepsilon_R^\tau(k_x)
\simeq
\tau s_R(k_x)\Delta(W).
$
%%%
For the symmetric irradiation geometry, the left edge is therefore pinned near zero quasienergy, whereas the right edge shifts whenever the mass profile terminates at a finite value. This is precisely the boundary dependence resolved in the LDOS of Fig.~\ref{Fig6}.

The same projected Hamiltonian can also be written in the sublattice-polarized basis that emerges in the free standing nanoribbon limit. For the symmetric irradiation geometry, only the right boundary can sample a finite mass $\Delta(W)\ne0$, and Eq.~\eqref{edge_H_mass} becomes
%%%
\begin{equation}
H_{F,\rm edge}^{\tau}(k_x)
=
t(k_x)\sigma_x
+
\frac{\tau\Delta(W)}{2}
\left[
\sigma_z
-
{\rm sgn}(\tau k_x)\sigma_0
\right].
\label{HF_edge_projected}
\end{equation}
%%%
The solutions of Eq.~\eqref{HF_edge_projected} are understood within the same normalizable sector, $\tau k_x<0$. The first term gives the residual left--right hybridization, while the second term causes the right-edge quasienergy shift. For the symmetric profile the sublattice component at the right-edge branch has $\varepsilon_R^\tau\simeq -\tau{\rm sgn}(\tau k_x)\Delta(W)$ and $\varepsilon_L^\tau=0$.

In general, the quasienergy eigenvalues of the projected two-edge Hamiltonian are
%%%
\begin{equation}
\epsilon_{\pm}^{\tau}(k_x)
=
\frac{\varepsilon_L^\tau+\varepsilon_R^\tau}{2}
\pm
\sqrt{
|t(k_x)|^2
+
\left(
\frac{\varepsilon_L^\tau-\varepsilon_R^\tau}{2}
\right)^2
}.
\label{edge_energies}
\end{equation}
%%%
This expression separates the two contributions to the zero quasienergy splitting: finite-size hybridization through $t(k_x)$ and boundary inequivalence through $\varepsilon_L^\tau-\varepsilon_R^\tau$. In Appendix~\ref{appendix_effective_edge_model} (Fig.~\ref{FigS1}) we show that this effective description captures the numerical quasienergy splitting as the mass cycles through one period at the right boundary. 
In the localized-edge limit, where $t(k_x)$ is exponentially small, the signed quasienergy splitting is controlled mainly by the mass difference at the edge,
%%%
$
\delta_{\rm edge}^{\tau}
\equiv
\varepsilon_R^\tau-\varepsilon_L^\tau
\simeq
-\tau{\rm sgn}(\tau k_x)\Delta(W).
$
%%%
Specializing to the symmetric mass profile gives
%%%
$
\delta_{\rm edge}^{\tau}
\simeq
\tau{\rm sgn}(\tau k_x)\Delta_0\cos\theta
\sin(2\pi W/\Lambda).
$
%%%
For the $K$ valley, where $\tau=+1$ and $k_x<0$, this reduces to
%%%
$
\delta_{\rm edge}^{K}
\simeq
-\Delta_0\cos\theta
\sin(2\pi W/\Lambda).
$
%%%
The quasienergy splitting therefore vanishes at edge matched values.

For edge matched ribbons, 
both boundaries lie at mass nodes
and $\Delta(0)=\Delta(W)=0$, so the edge-state quasienergy splitting is controlled only by the exponentially small tunneling due to finite size effects. The comparison in Fig.~\ref{Fig7}(a) shows that the analytical edge-state dispersion reproduces the near-degenerate low-energy numerical branches. For edge mismatched ribbons, $\Delta(W)\neq0$, 
the right-edge branch remains shifted even when the left and right edge states are spatially separated, as shown in Fig.~\ref{Fig7}(b). 

The effective Hamiltonian also gives a simple interpretation of the valley-selective response. The right-edge quasienergy shift is controlled by the mass term,
%%%
$
\varepsilon_R^\tau
\simeq
-\tau{\rm sgn}(\tau k_x)\Delta(W).
$
%%%
For the allowed edge-state sector, $\tau k_x<0$, the sign of this shift reverses when going from $K$ to $K'$. Thus, the same edge mismatch that shifts the right-edge branch at $K$ shifts the $K'$ branch in the opposite direction, while the left edge remains pinned because $\Delta(0)=0$. This describes the boundary-resolved LDOS in Fig.~\ref{Fig6}: edge matched ribbons show nearly zero-quasienergy boundary modes, whereas edge mismatched ribbons show a right-boundary energy shift that appears at opposite quasienergy and opposite momentum in the complementary valley. Therefore, at a fixed right boundary, the sign of the shifted edge-state energy selects the valley and the corresponding edge-branch direction. Overall, the effective edge-state model reduces the numerical spectra to a boundary-matching mechanism: $t(k_x)$ controls the residual left--right hybridization, while the projected Floquet mass sets the right-edge shift through $\Delta(W)$. 
%edge matched ribbons remain nearly degenerate, whereas edge-mismatched ribbons split the edge branches in a valley-dependent way. 
By tuning how $\Lambda$ matches the physical ribbon edges, the optical superlattice becomes a direct knob for controlling edge-state quasienergy splitting and valley/chirality selectivity.

%%%%%%%%%%%%%%%%%%%%%%%%%%%%%%%%%%%%%%%%%%%%%%%%%%%%%%%%%%%%%%%%%%%%%%%%%%%%%%%%%%%%%%%%%%%%%%%%%%%%%
\section{Conclusion}

We have studied a zigzag graphene nanoribbon driven by two coherent tilted beams and shown that the resulting space--time modulation generates an optically tunable Floquet mass superlattice. The beam geometry controls the period of the optical superlattice, while the local handedness of the polarization pattern controls the sign of the induced Dirac mass. This provides a flexible way to tune how the mass periodicity matches the zigzag boundaries.

The low-quasienergy edge states are controlled by the interplay between the ribbon width, the optical period, and the edge-state localization length. In edge matched configurations, both boundaries lie at mass nodes, so the spectrum remains close to that of the undriven ribbon, apart from exponentially small finite-size splittings. However, states with long penetration depth are reconstructed by the internal mass interfaces. In edge mismatched configurations, the two boundaries sample inequivalent parts of the mass profile, producing a boundary-induced quasienergy splitting that survives in the decoupled-edge limit. Because the Floquet mass changes sign between valleys, this shift is valley dependent.

An effective two-edge Hamiltonian captures these results by separating residual left--right hybridization from the boundary-dependent mass projection. The model reproduces the numerical quasienergy splitting, captures how it vanishes for edge matched profiles, and accounts for the valley reversal of the shifted branch. Thus, the optical superlattice acts as a boundary-matching knob that controls edge-state quasienergy splitting and valley selectivity.

These effects can be probed experimentally using the flexibility of the two-beam geometry. Changing the incidence angles tunes the optical superlattice period, taking a nanoribbon from the edge matched to the edge mismatched regime. Furthermore, the beam phases and intensities control the mass profile and contrast, without requiring a new sample or static patterning. Momentum-resolved signatures could be accessed using time-resolved photoemission, as in Floquet-band measurements in driven solids~\cite{FloqExp2,FloqExp1,TrARPSGraph}, or by micro/nano-ARPES on graphene nanoribbons and related two-dimensional devices~\cite{Karakachian2020GNRARPES,Jiang2023}. Real-space signatures could be tested with scanning tunneling spectroscopy, which has resolved graphene-nanoribbon edge states~\cite{Tao2011,Zhang2013,Ruffieux2016,Wang2016,Brede2023}, or with ultrafast pump--probe scanning probes~\cite{TimeSTM}. Transport measurements with edge contacts, nonlocal geometries, or photocurrent detection under structured illumination could reveal the associated quasienergy- and boundary-selective response~\cite{FloqExp3,FloqTraspGraphen,Gorbachev2014ValleyHall,Sui2015ValleyTransport,Torres1}.
Because the edge-mismatch-induced shift reverses between valleys, the sign and location of the shifted branch provide a direct route to valley-selective boundary response. These results establish optical boundary matching as a flexible control principle for tuning edge-state quasienergy splitting and valley selectivity in finite Dirac materials.

\begin{acknowledgements}
This work was supported by the U.S. Department of Energy, Office of Science, Basic Energy Sciences, under Award \# DE-SC0025703.
\end{acknowledgements}  

%%%%%%%%%%%%%%%%%%%%%%%%%%%%%%%%%%%%%%%%%%%%%%%%%%%%%%%%%%%%%%%%%%%%%%%%%%%%%%%%%%%%%%%%%%%%%%%%%%%%%
\appendix
\section{Derivation of the effective edge-state Hamiltonian}
\label{appendix_effective_edge_model}

In this Appendix we derive the effective two-edge Hamiltonian used in Sec.~\ref{effective_edge_model}. We first keep the confining masses finite, solve the isolated left and right interface problems, and evaluate the residual left--right hybridization. Only after these matrix elements are obtained do we take the hard-wall zigzag limit, $m\rightarrow\infty$. This order of limits is essential: the tunneling matrix element is formally the product of a diverging confining mass and a vanishing wave function tail. In the hard-wall limit, the edge states become fully sublattice polarized, which justifies the pseudospin form of the projected Floquet Hamiltonian.

We start from the valley-resolved continuum Hamiltonian
%%%
\begin{equation}
H_0^\tau(k_x,y)
=
\hbar v_{\rm F}
\left(
\tau k_x\sigma_x
-
i\sigma_y\partial_y
\right),
\label{app_H0}
\end{equation}
%%%
where $\tau=\pm1$ labels the $K$ and $K'$ valleys. The insulating regions outside the ribbon are described by $H_{\pm m}^{\tau}=H_0^\tau\pm m\sigma_z$, with $m>0$. The graphene ribbon occupies $0<y<W$. The left boundary interfaces with a $+m\sigma_z$ region, while the right boundary interfaces with a $-m\sigma_z$ region:
%%%
\begin{align}
H_L^\tau
&=
H_0^\tau\Theta(y)
+
H_{+m}^{\tau}\Theta(-y),
\nonumber\\
H_R^\tau
&=
H_0^\tau\Theta(W-y)
+
H_{-m}^{\tau}\Theta(y-W).
\label{app_HLR}
\end{align}
%%%
The finite strip is constructed as
%%%
\begin{equation}
H_{\rm strip}^{\tau,(0)}
=
H_L^\tau
+
H_R^\tau
-
H_0^\tau ,
\label{app_Hstrip}
\end{equation}
%%%
where the last term avoids double counting the massless graphene region between the two interfaces.

For an interface eigenstate we write
%%%
\begin{equation}
\Psi_\tau(k_x,y)
=
e^{ik_xx}
\begin{pmatrix}
\Phi_{A,\tau}(y)
\\
\Phi_{B,\tau}(y)
\end{pmatrix}.
\label{app_ansatz}
\end{equation}
%%%
Defining $\widetilde E=E/(\hbar v_{\rm F})$ and $\widetilde m=m/(\hbar v_{\rm F})$, the decay constants in the insulating and graphene regions are $\lambda_i=(\widetilde m^2+k_x^2-\widetilde E^2)^{1/2}$ and $\lambda_g=(k_x^2-\widetilde E^2)^{1/2}$. The interface equations have normalizable zero-energy solutions, for which $\widetilde E=0$ and $\lambda_g=|k_x|$. We label the two sectors by $s={\rm sgn}(\tau k_x)$ and use the sublattice spinors $\ket{A}=(1,0)^T$ and $\ket{B}=(0,1)^T$.

To keep the notation compact, we define the finite-mass envelopes $g_L=e^{-\lambda_g y}\Theta(y)$, $i_L=e^{\lambda_i y}\Theta(-y)$, $g_R=e^{-\lambda_g(W-y)}\Theta(W-y)$, and $i_R=e^{-\lambda_i(y-W)}\Theta(y-W)$. We also define $\eta_{\pm}=(\lambda_i\mp\lambda_g)/\widetilde m$ and $N_{\pm}=[1/(2\lambda_i)+\eta_{\pm}^2/(2\lambda_g)]^{-1/2}$.

At finite $m$, the left-interface states localized near $y=0$ are
%%%
\begin{align}
\Psi_L^{+}
&=
N_{+}
\left[
\eta_{+}g_L\ket{A}
+
i_L\ket{B}
\right],
\nonumber\\
\Psi_L^{-}
&=
N_{-}
\left[
\eta_{-}g_L\ket{B}
+
i_L\ket{A}
\right],
\label{app_left_finite}
\end{align}
%%%
where $\Psi_L^{+}$ applies for $\tau k_x>0$, while $\Psi_L^{-}$ applies for $\tau k_x<0$. Similarly, the right-interface states localized near $y=W$ are
%%%
\begin{align}
\Psi_R^{+}
&=
N_{+}
\left[
\eta_{+}g_R\ket{B}
+
i_R\ket{A}
\right],
\nonumber\\
\Psi_R^{-}
&=
N_{-}
\left[
\eta_{-}g_R\ket{A}
+
i_R\ket{B}
\right].
\label{app_right_finite}
\end{align}
%%%
These finite-mass states contain a graphene component and an evanescent component in the insulating region. Keeping these finite-mass tails is essential for obtaining the correct hybridization before the hard-wall limit is taken.

We now take $m\rightarrow\infty$. In this hard-wall zigzag limit, $\lambda_i\simeq\widetilde m$, $\eta_{\pm}\rightarrow1$, and $N_{\pm}\rightarrow\sqrt{2|k_x|}$. The insulating-region weight is pushed out of the low-energy Hilbert space, while the graphene components remain finite. Thus,
%%%
\begin{align}
\Psi_L^{+}
&\rightarrow
f_L(y)\ket{A},
&
\Psi_R^{+}
&\rightarrow
f_R(y)\ket{B},
\nonumber\\
\Psi_L^{-}
&\rightarrow
f_L(y)\ket{B},
&
\Psi_R^{-}
&\rightarrow
f_R(y)\ket{A},
\label{app_edge_limit_states}
\end{align}
%%%
where $f_L(y)=\sqrt{2|k_x|}e^{-|k_x|y}\Theta(y)$ and $f_R(y)=\sqrt{2|k_x|}e^{-|k_x|(W-y)}\Theta(W-y)$. Eq.~\eqref{app_edge_limit_states} makes the sublattice polarization explicit: in the hard-wall limit,
%%%
\begin{equation}
\sigma_z\Psi_L^{s}
=
s\Psi_L^{s},
\qquad
\sigma_z\Psi_R^{s}
=
-s\Psi_R^{s}.
\label{app_sublattice_polarization}
\end{equation}
%%%
The projected low-energy subspace can therefore be viewed either as the left--right edge basis or as a two-state sublattice-polarized basis. In this basis, $\sigma_z$ distinguishes the two decoupled edges, while $\sigma_x$ mixes them through their residual overlap. For the low-energy zigzag branch considered in the main text, normalizability selects $\tau k_x<0$, so $k_x<0$ at $K$ and $k_x>0$ at $K'$. The relevant hard-wall edge states are therefore $\Psi_L^{-}$ and $\Psi_R^{-}$.

We next evaluate the finite-size hybridization. Following the logic of Ref.~\cite{thin4}, this matrix element must be computed before taking the hard-wall limit. If one sets $m\rightarrow\infty$ too early, the interface wave functions vanish in the massive regions while the confining mass operator diverges, making the off-diagonal matrix element formally ill defined. We therefore keep $m$ finite and evaluate the matrix elements using Eqs.~\eqref{app_left_finite} and \eqref{app_right_finite}. The relevant remote mass terms are $H_L^\tau-H_0^\tau=m\sigma_z\Theta(-y)$ and $H_R^\tau-H_0^\tau=-m\sigma_z\Theta(y-W)$. To control the intermediate integrals, we introduce a regulator $R_\gamma(y)=e^{-y^2/\gamma^2}$ and evaluate $H_{ij}^{(0)}(\gamma,m)=\int dy\,R_\gamma(y)\Psi_i^{s\dagger}H_{\rm strip}^{\tau,(0)}\Psi_j^s$, with $i,j=L,R$. The finite-$m$ integrals are evaluated first, the regulator is then removed by taking $\gamma\rightarrow\infty$, and only afterward is the hard-wall zigzag limit $m\rightarrow\infty$ taken. With this order of limits, the diagonal terms vanish, $H_{LL}^{(0)}=H_{RR}^{(0)}=0$, whereas the off-diagonal terms remain finite. Taking $\gamma\rightarrow\infty$ and then $m\rightarrow\infty$ gives
%%%
\begin{equation}
H_{LR}^{(0)}
=
-s\,2\hbar v_{\rm F} |k_x|e^{-W|k_x|},
\qquad
H_{RL}^{(0)}
=
\left[H_{LR}^{(0)}\right]^* .
\label{app_HLR_hybrid}
\end{equation}
%%%
The sign depends on the relative phase convention for the two isolated edge states and has no physical consequence. 
%%%
\begin{figure}[ht!]
  \centering
  \includegraphics[width=0.48\textwidth]{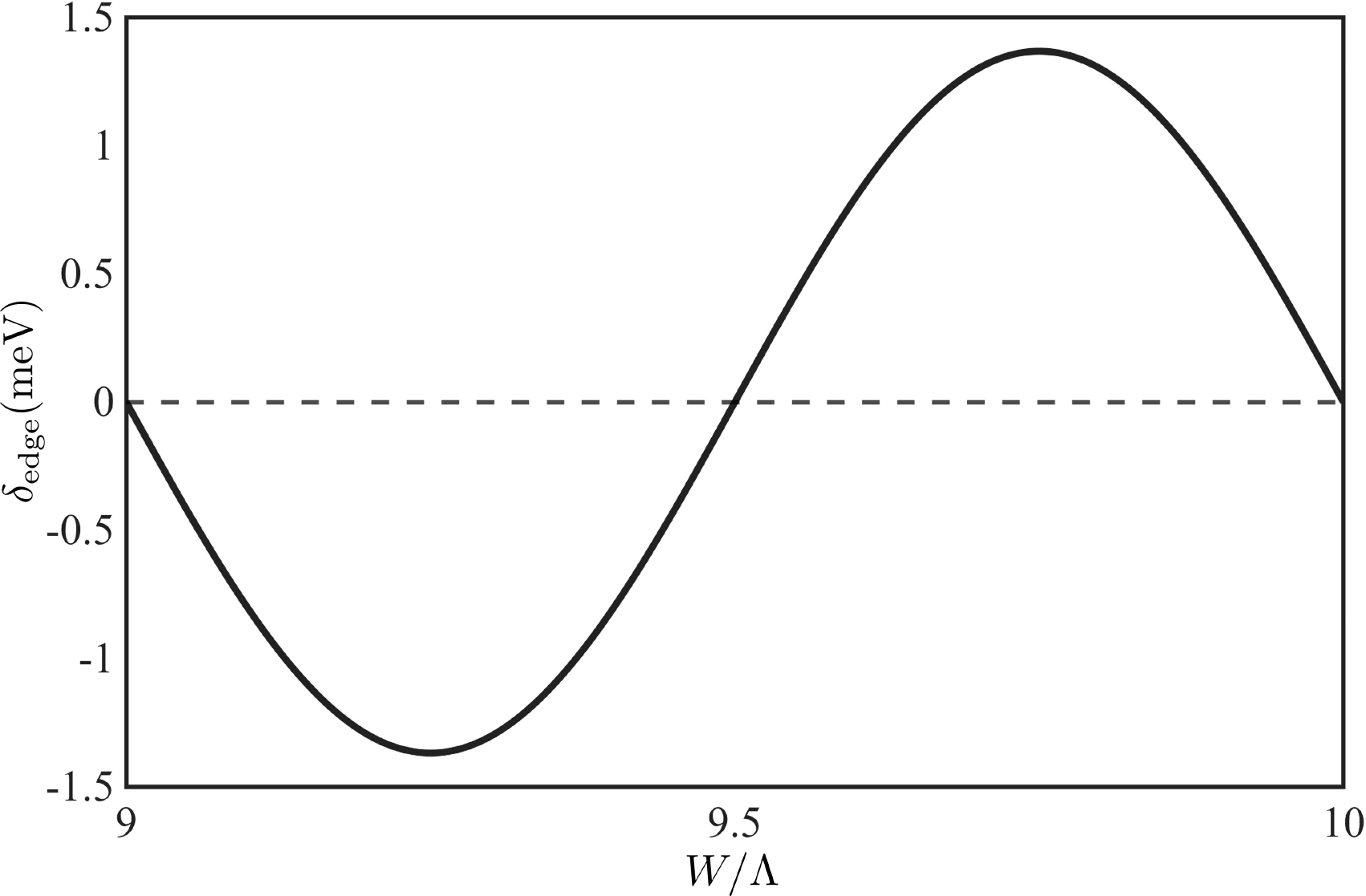}
\caption{Numerical evaluation of the signed quasienergy splitting $\delta_{\rm edge}$ versus $W/\Lambda$ over one mass period. The zeros occur at edge matched widths, where the right boundary lies at a mass node. All parameters are the same as in Fig.~\ref{Fig4}.}
\label{FigS1}
\end{figure}
%%%
%%%
Choosing the phase so that the hybridization is real and positive, we define
%%%
\begin{equation}
t(k_x)
=
2\hbar v_{\rm F} |k_x|e^{-W|k_x|}.
\label{app_tkx}
\end{equation}
%%%
The unirradiated projected Hamiltonian is therefore $H_{\rm edge}^{(0)}(k_x)=t(k_x)\sigma_x$.

We now include the light-induced Floquet mass, $V_F^\tau(y)=\tau\Delta(y)\sigma_z$. Because the limiting edge states are sublattice polarized, this term is diagonal in the projected subspace. Using Eq.~\eqref{app_sublattice_polarization}, the matrix elements are
%%%
\begin{align}
\varepsilon_L^\tau(k_x)
&=
\tau s
\int_0^W dy\, |f_L(y)|^2\Delta(y),
\nonumber\\
\varepsilon_R^\tau(k_x)
&=
-\tau s
\int_0^W dy\, |f_R(y)|^2\Delta(y).
\label{app_eps_integrals}
\end{align}
%%%
These expressions show that the projected mass is a weighted average of the Floquet mass over the finite edge-state envelope. For strongly localized edge states, these integrals are controlled by the boundary values, so $\varepsilon_L^\tau\simeq\tau{\rm sgn}(\tau k_x)\Delta(0)$ and $\varepsilon_R^\tau\simeq-\tau{\rm sgn}(\tau k_x)\Delta(W)$. Combining the hybridization and projected mass terms gives
%%%
\begin{equation}
H_{\rm edge}^{\tau}(k_x)
=
\begin{pmatrix}
\varepsilon_L^\tau & t(k_x)
\\
t(k_x) & \varepsilon_R^\tau
\end{pmatrix}_{L,R},
\label{app_Hedge_full}
\end{equation}
%%%
with eigenvalues
%%%
\begin{equation}
\begin{aligned}
\epsilon_{\pm}^{\tau}
=
\frac{\varepsilon_L^\tau+\varepsilon_R^\tau}{2}
\pm
\sqrt{
t^2(k_x)
+
\left[
\frac{\varepsilon_L^\tau-\varepsilon_R^\tau}{2}
\right]^2
}.
\end{aligned}
\label{app_edge_energies}
\end{equation}
%%%

For the symmetric two-beam geometry, $\Delta(y)=-\Delta_0\cos\theta\sin(2\pi y/\Lambda)$. Thus $\Delta(0)=0$, while $\Delta(W)=-\Delta_0\cos\theta\sin(2\pi W/\Lambda)$. The left edge is pinned at a mass node, while the right edge shifts whenever the mass profile terminates at a finite value. Eq.~\eqref{app_Hedge_full} then becomes
%%%
\begin{equation}
H_{\rm edge}^{\tau}
=
\begin{pmatrix}
0 & t(k_x)
\\
t(k_x) &
-\tau{\rm sgn}(\tau k_x)\Delta(W)
\end{pmatrix}_{L,R}.
\label{app_Hedge_symmetric}
\end{equation}
%%%
Equivalently, in the sublattice-polarized pseudospin basis,
%%%
\begin{equation}
\begin{aligned}
H_{F,\rm edge}^{\tau}
=
t(k_x)\sigma_x
+
\frac{\tau\Delta(W)}{2}
\left[
\sigma_z
-
{\rm sgn}(\tau k_x)\sigma_0
\right],
\end{aligned}
\label{app_HFR}
\end{equation}
%%%
with the solutions restricted to the normalizable sector $\tau k_x<0$. The bracket in Eq.~\eqref{app_HFR} annihilates the left-edge-state and shifts the right-edge-state by $-\tau{\rm sgn}(\tau k_x)\Delta(W)$.

In the localized-edge limit, where $t(k_x)$ is exponentially small, the signed quasienergy splitting is controlled by the boundary mass difference:
%%%
$
\delta_{\rm edge}^{\tau}
\equiv
\varepsilon_R^\tau-\varepsilon_L^\tau
\simeq
-\tau{\rm sgn}(\tau k_x)\Delta(W).
$
%%%
Using the symmetric mass profile,
%%%
\begin{equation}
\delta_{\rm edge}^{\tau}
\simeq
\tau{\rm sgn}(\tau k_x)
\Delta_0\cos\theta
\sin\left(\frac{2\pi W}{\Lambda}\right).
\label{app_delta_edge_symmetric}
\end{equation}
%%%
For the $K$ valley shown in the numerical spectra, $\tau=+1$ and $k_x<0$, so
$
\delta_{\rm edge}^{K}
\simeq
-
\Delta_0\cos\theta
\sin\left(2\pi W/{\Lambda}\right).
$
%%%
The quasienergy splitting vanishes when the right edge also lies at a mass node corresponding to edge matched widths.
Since the Floquet mass enters as $\tau\Delta(y)\sigma_z$, the sign of the right-edge shift reverses between $K$ and $K'$, producing the valley-dependent boundary response discussed in Sec.~\ref{effective_edge_model}.

\bibliography{References.bib}

\begin{thebibliography}{75}%
\makeatletter
\providecommand \@ifxundefined [1]{%
 \@ifx{#1\undefined}
}%
\providecommand \@ifnum [1]{%
 \ifnum #1\expandafter \@firstoftwo
 \else \expandafter \@secondoftwo
 \fi
}%
\providecommand \@ifx [1]{%
 \ifx #1\expandafter \@firstoftwo
 \else \expandafter \@secondoftwo
 \fi
}%
\providecommand \natexlab [1]{#1}%
\providecommand \enquote  [1]{``#1''}%
\providecommand \bibnamefont  [1]{#1}%
\providecommand \bibfnamefont [1]{#1}%
\providecommand \citenamefont [1]{#1}%
\providecommand \href@noop [0]{\@secondoftwo}%
\providecommand \href [0]{\begingroup \@sanitize@url \@href}%
\providecommand \@href[1]{\@@startlink{#1}\@@href}%
\providecommand \@@href[1]{\endgroup#1\@@endlink}%
\providecommand \@sanitize@url [0]{\catcode `\\12\catcode `\$12\catcode
  `\&12\catcode `\#12\catcode `\^12\catcode `\_12\catcode `\%12\relax}%
\providecommand \@@startlink[1]{}%
\providecommand \@@endlink[0]{}%
\providecommand \url  [0]{\begingroup\@sanitize@url \@url }%
\providecommand \@url [1]{\endgroup\@href {#1}{\urlprefix }}%
\providecommand \urlprefix  [0]{URL }%
\providecommand \Eprint [0]{\href }%
\providecommand \doibase [0]{https://doi.org/}%
\providecommand \selectlanguage [0]{\@gobble}%
\providecommand \bibinfo  [0]{\@secondoftwo}%
\providecommand \bibfield  [0]{\@secondoftwo}%
\providecommand \translation [1]{[#1]}%
\providecommand \BibitemOpen [0]{}%
\providecommand \bibitemStop [0]{}%
\providecommand \bibitemNoStop [0]{.\EOS\space}%
\providecommand \EOS [0]{\spacefactor3000\relax}%
\providecommand \BibitemShut  [1]{\csname bibitem#1\endcsname}%
\let\auto@bib@innerbib\@empty
%</preamble>
\bibitem [{\citenamefont {Aoki}\ \emph {et~al.}(2014)\citenamefont {Aoki},
  \citenamefont {Tsuji}, \citenamefont {Eckstein}, \citenamefont {Kollar},
  \citenamefont {Oka},\ and\ \citenamefont {Werner}}]{Oka_RMP}%
  \BibitemOpen
  \bibfield  {author} {\bibinfo {author} {\bibfnamefont {H.}~\bibnamefont
  {Aoki}}, \bibinfo {author} {\bibfnamefont {N.}~\bibnamefont {Tsuji}},
  \bibinfo {author} {\bibfnamefont {M.}~\bibnamefont {Eckstein}}, \bibinfo
  {author} {\bibfnamefont {M.}~\bibnamefont {Kollar}}, \bibinfo {author}
  {\bibfnamefont {T.}~\bibnamefont {Oka}},\ and\ \bibinfo {author}
  {\bibfnamefont {P.}~\bibnamefont {Werner}},\ }\bibfield  {title} {\bibinfo
  {title} {Nonequilibrium dynamical mean-field theory and its applications},\
  }\href {https://doi.org/10.1103/RevModPhys.86.779} {\bibfield  {journal}
  {\bibinfo  {journal} {Reviews of Modern Physics}\ }\textbf {\bibinfo {volume}
  {86}},\ \bibinfo {pages} {779} (\bibinfo {year} {2014})}\BibitemShut
  {NoStop}%
\bibitem [{\citenamefont {Basov}\ \emph {et~al.}(2017)\citenamefont {Basov},
  \citenamefont {Averitt},\ and\ \citenamefont {Hsieh}}]{flreview3}%
  \BibitemOpen
  \bibfield  {author} {\bibinfo {author} {\bibfnamefont {D.}~\bibnamefont
  {Basov}}, \bibinfo {author} {\bibfnamefont {R.}~\bibnamefont {Averitt}},\
  and\ \bibinfo {author} {\bibfnamefont {D.}~\bibnamefont {Hsieh}},\ }\bibfield
   {title} {\bibinfo {title} {{Towards properties on demand in quantum
  materials}},\ }\href {https://doi.org/10.1038/nmat5017} {\bibfield  {journal}
  {\bibinfo  {journal} {Nature Materials}\ }\textbf {\bibinfo {volume} {16}},\
  \bibinfo {pages} {1077} (\bibinfo {year} {2017})}\BibitemShut {NoStop}%
\bibitem [{\citenamefont {Oka}\ and\ \citenamefont
  {Kitamura}(2019)}]{flreview1}%
  \BibitemOpen
  \bibfield  {author} {\bibinfo {author} {\bibfnamefont {T.}~\bibnamefont
  {Oka}}\ and\ \bibinfo {author} {\bibfnamefont {S.}~\bibnamefont {Kitamura}},\
  }\bibfield  {title} {\bibinfo {title} {Floquet engineering of quantum
  materials},\ }\href
  {https://doi.org/10.1146/annurev-conmatphys-031218-013423} {\bibfield
  {journal} {\bibinfo  {journal} {Annual Review of Condensed Matter Physics}\
  }\textbf {\bibinfo {volume} {10}},\ \bibinfo {pages} {387} (\bibinfo {year}
  {2019})}\BibitemShut {NoStop}%
\bibitem [{\citenamefont {Rudner}\ and\ \citenamefont
  {Lindner}(2020)}]{flreview2}%
  \BibitemOpen
  \bibfield  {author} {\bibinfo {author} {\bibfnamefont {M.~S.}\ \bibnamefont
  {Rudner}}\ and\ \bibinfo {author} {\bibfnamefont {N.~H.}\ \bibnamefont
  {Lindner}},\ }\bibfield  {title} {\bibinfo {title} {{Band structure
  engineering and non-equilibrium dynamics in Floquet topological
  insulators}},\ }\href {https://www.nature.com/articles/s42254-020-0170-z}
  {\bibfield  {journal} {\bibinfo  {journal} {Nature Reviews Physics}\ }\textbf
  {\bibinfo {volume} {2}},\ \bibinfo {pages} {229} (\bibinfo {year}
  {2020})}\BibitemShut {NoStop}%
\bibitem [{\citenamefont {Harper}\ \emph {et~al.}(2020)\citenamefont {Harper},
  \citenamefont {Roy}, \citenamefont {Rudner},\ and\ \citenamefont
  {Sondhi}}]{flreview4}%
  \BibitemOpen
  \bibfield  {author} {\bibinfo {author} {\bibfnamefont {F.}~\bibnamefont
  {Harper}}, \bibinfo {author} {\bibfnamefont {R.}~\bibnamefont {Roy}},
  \bibinfo {author} {\bibfnamefont {M.~S.}\ \bibnamefont {Rudner}},\ and\
  \bibinfo {author} {\bibfnamefont {S.}~\bibnamefont {Sondhi}},\ }\bibfield
  {title} {\bibinfo {title} {Topology and broken symmetry in {Floquet}
  systems},\ }\href {https://doi.org/10.1146/annurev-conmatphys-031218-013721}
  {\bibfield  {journal} {\bibinfo  {journal} {Annual Review of Condensed Matter
  Physics}\ }\textbf {\bibinfo {volume} {11}},\ \bibinfo {pages} {345}
  (\bibinfo {year} {2020})},\ \Eprint
  {https://arxiv.org/abs/https://doi.org/10.1146/annurev-conmatphys-031218-013721}
  {https://doi.org/10.1146/annurev-conmatphys-031218-013721} \BibitemShut
  {NoStop}%
\bibitem [{\citenamefont {de~la Torre}\ \emph {et~al.}(2021)\citenamefont
  {de~la Torre}, \citenamefont {Kennes}, \citenamefont {Claassen},
  \citenamefont {Gerber}, \citenamefont {McIver},\ and\ \citenamefont
  {Sentef}}]{flreview5}%
  \BibitemOpen
  \bibfield  {author} {\bibinfo {author} {\bibfnamefont {A.}~\bibnamefont
  {de~la Torre}}, \bibinfo {author} {\bibfnamefont {D.~M.}\ \bibnamefont
  {Kennes}}, \bibinfo {author} {\bibfnamefont {M.}~\bibnamefont {Claassen}},
  \bibinfo {author} {\bibfnamefont {S.}~\bibnamefont {Gerber}}, \bibinfo
  {author} {\bibfnamefont {J.~W.}\ \bibnamefont {McIver}},\ and\ \bibinfo
  {author} {\bibfnamefont {M.~A.}\ \bibnamefont {Sentef}},\ }\bibfield  {title}
  {\bibinfo {title} {Colloquium: Nonthermal pathways to ultrafast control in
  quantum materials},\ }\href {https://doi.org/10.1103/RevModPhys.93.041002}
  {\bibfield  {journal} {\bibinfo  {journal} {Reviews of Modern Physics}\
  }\textbf {\bibinfo {volume} {93}},\ \bibinfo {pages} {041002} (\bibinfo
  {year} {2021})}\BibitemShut {NoStop}%
\bibitem [{\citenamefont {Cayssol}\ \emph {et~al.}(2013)\citenamefont
  {Cayssol}, \citenamefont {D{\'o}ra}, \citenamefont {Simon},\ and\
  \citenamefont {Moessner}}]{FloqTIReview}%
  \BibitemOpen
  \bibfield  {author} {\bibinfo {author} {\bibfnamefont {J.}~\bibnamefont
  {Cayssol}}, \bibinfo {author} {\bibfnamefont {B.}~\bibnamefont {D{\'o}ra}},
  \bibinfo {author} {\bibfnamefont {F.}~\bibnamefont {Simon}},\ and\ \bibinfo
  {author} {\bibfnamefont {R.}~\bibnamefont {Moessner}},\ }\bibfield  {title}
  {\bibinfo {title} {{Floquet topological insulators}},\ }\href
  {https://doi.org/10.1002/pssr.201206451} {\bibfield  {journal} {\bibinfo
  {journal} {Physica Status Solidi (RRL)--Rapid Research Letters}\ }\textbf
  {\bibinfo {volume} {7}},\ \bibinfo {pages} {101} (\bibinfo {year}
  {2013})}\BibitemShut {NoStop}%
\bibitem [{\citenamefont {Oka}\ and\ \citenamefont {Aoki}(2009)}]{ftrans1}%
  \BibitemOpen
  \bibfield  {author} {\bibinfo {author} {\bibfnamefont {T.}~\bibnamefont
  {Oka}}\ and\ \bibinfo {author} {\bibfnamefont {H.}~\bibnamefont {Aoki}},\
  }\bibfield  {title} {\bibinfo {title} {{Photovoltaic Hall effect in
  graphene}},\ }\href {https://doi.org/10.1103/PhysRevB.79.081406} {\bibfield
  {journal} {\bibinfo  {journal} {Physical Review B}\ }\textbf {\bibinfo
  {volume} {79}},\ \bibinfo {pages} {081406} (\bibinfo {year}
  {2009})}\BibitemShut {NoStop}%
\bibitem [{\citenamefont {Lindner}\ \emph {et~al.}(2011)\citenamefont
  {Lindner}, \citenamefont {Refael},\ and\ \citenamefont
  {Galitski}}]{FloquetTI}%
  \BibitemOpen
  \bibfield  {author} {\bibinfo {author} {\bibfnamefont {N.~H.}\ \bibnamefont
  {Lindner}}, \bibinfo {author} {\bibfnamefont {G.}~\bibnamefont {Refael}},\
  and\ \bibinfo {author} {\bibfnamefont {V.}~\bibnamefont {Galitski}},\
  }\bibfield  {title} {\bibinfo {title} {{Floquet topological insulator in
  semiconductor quantum wells}},\ }\href
  {https://www.nature.com/articles/nphys1926} {\bibfield  {journal} {\bibinfo
  {journal} {Nat. Phys.}\ }\textbf {\bibinfo {volume} {7}},\ \bibinfo {pages}
  {490} (\bibinfo {year} {2011})}\BibitemShut {NoStop}%
\bibitem [{\citenamefont {Gu}\ \emph {et~al.}(2011)\citenamefont {Gu},
  \citenamefont {Fertig}, \citenamefont {Arovas},\ and\ \citenamefont
  {Auerbach}}]{ftrans3}%
  \BibitemOpen
  \bibfield  {author} {\bibinfo {author} {\bibfnamefont {Z.}~\bibnamefont
  {Gu}}, \bibinfo {author} {\bibfnamefont {H.~A.}\ \bibnamefont {Fertig}},
  \bibinfo {author} {\bibfnamefont {D.~P.}\ \bibnamefont {Arovas}},\ and\
  \bibinfo {author} {\bibfnamefont {A.}~\bibnamefont {Auerbach}},\ }\bibfield
  {title} {\bibinfo {title} {Floquet spectrum and transport through an
  irradiated graphene ribbon},\ }\href
  {https://doi.org/10.1103/PhysRevLett.107.216601} {\bibfield  {journal}
  {\bibinfo  {journal} {Physical Review Letters}\ }\textbf {\bibinfo {volume}
  {107}},\ \bibinfo {pages} {216601} (\bibinfo {year} {2011})}\BibitemShut
  {NoStop}%
\bibitem [{\citenamefont {Kitagawa}\ \emph {et~al.}(2011)\citenamefont
  {Kitagawa}, \citenamefont {Oka}, \citenamefont {Brataas}, \citenamefont
  {Fu},\ and\ \citenamefont {Demler}}]{Virtual-ph}%
  \BibitemOpen
  \bibfield  {author} {\bibinfo {author} {\bibfnamefont {T.}~\bibnamefont
  {Kitagawa}}, \bibinfo {author} {\bibfnamefont {T.}~\bibnamefont {Oka}},
  \bibinfo {author} {\bibfnamefont {A.}~\bibnamefont {Brataas}}, \bibinfo
  {author} {\bibfnamefont {L.}~\bibnamefont {Fu}},\ and\ \bibinfo {author}
  {\bibfnamefont {E.}~\bibnamefont {Demler}},\ }\bibfield  {title} {\bibinfo
  {title} {{Transport properties of nonequilibrium systems under the
  application of light: Photoinduced quantum Hall insulators without Landau
  levels}},\ }\href {https://doi.org/10.1103/PhysRevB.84.235108} {\bibfield
  {journal} {\bibinfo  {journal} {Phys. Rev. B}\ }\textbf {\bibinfo {volume}
  {84}},\ \bibinfo {pages} {235108} (\bibinfo {year} {2011})}\BibitemShut
  {NoStop}%
\bibitem [{\citenamefont {Rudner}\ \emph {et~al.}(2013)\citenamefont {Rudner},
  \citenamefont {Lindner}, \citenamefont {Berg},\ and\ \citenamefont
  {Levin}}]{FloquetTI2}%
  \BibitemOpen
  \bibfield  {author} {\bibinfo {author} {\bibfnamefont {M.~S.}\ \bibnamefont
  {Rudner}}, \bibinfo {author} {\bibfnamefont {N.~H.}\ \bibnamefont {Lindner}},
  \bibinfo {author} {\bibfnamefont {E.}~\bibnamefont {Berg}},\ and\ \bibinfo
  {author} {\bibfnamefont {M.}~\bibnamefont {Levin}},\ }\bibfield  {title}
  {\bibinfo {title} {Anomalous edge states and the bulk-edge correspondence for
  periodically driven two-dimensional systems},\ }\href
  {https://doi.org/10.1103/PhysRevX.3.031005} {\bibfield  {journal} {\bibinfo
  {journal} {Physical Review X}\ }\textbf {\bibinfo {volume} {3}},\ \bibinfo
  {pages} {031005} (\bibinfo {year} {2013})}\BibitemShut {NoStop}%
\bibitem [{\citenamefont {Usaj}\ \emph {et~al.}(2014)\citenamefont {Usaj},
  \citenamefont {Perez-Piskunow}, \citenamefont {Foa~Torres},\ and\
  \citenamefont {Balseiro}}]{graphene-top-ins}%
  \BibitemOpen
  \bibfield  {author} {\bibinfo {author} {\bibfnamefont {G.}~\bibnamefont
  {Usaj}}, \bibinfo {author} {\bibfnamefont {P.~M.}\ \bibnamefont
  {Perez-Piskunow}}, \bibinfo {author} {\bibfnamefont {L.~E.~F.}\ \bibnamefont
  {Foa~Torres}},\ and\ \bibinfo {author} {\bibfnamefont {C.~A.}\ \bibnamefont
  {Balseiro}},\ }\bibfield  {title} {\bibinfo {title} {{Irradiated graphene as
  a tunable Floquet topological insulator}},\ }\href
  {https://doi.org/10.1103/PhysRevB.90.115423} {\bibfield  {journal} {\bibinfo
  {journal} {Phys. Rev. B}\ }\textbf {\bibinfo {volume} {90}},\ \bibinfo
  {pages} {115423} (\bibinfo {year} {2014})}\BibitemShut {NoStop}%
\bibitem [{\citenamefont {Dehghani}\ \emph {et~al.}(2015)\citenamefont
  {Dehghani}, \citenamefont {Oka},\ and\ \citenamefont {Mitra}}]{mitraandoka1}%
  \BibitemOpen
  \bibfield  {author} {\bibinfo {author} {\bibfnamefont {H.}~\bibnamefont
  {Dehghani}}, \bibinfo {author} {\bibfnamefont {T.}~\bibnamefont {Oka}},\ and\
  \bibinfo {author} {\bibfnamefont {A.}~\bibnamefont {Mitra}},\ }\bibfield
  {title} {\bibinfo {title} {{Out-of-equilibrium electrons and the Hall
  conductance of a Floquet topological insulator}},\ }\href
  {https://doi.org/10.1103/PhysRevB.91.155422} {\bibfield  {journal} {\bibinfo
  {journal} {Physical Review B}\ }\textbf {\bibinfo {volume} {91}},\ \bibinfo
  {pages} {155422} (\bibinfo {year} {2015})}\BibitemShut {NoStop}%
\bibitem [{\citenamefont {Kumar}\ \emph {et~al.}(2020)\citenamefont {Kumar},
  \citenamefont {Rodriguez-Vega}, \citenamefont {Pereg-Barnea},\ and\
  \citenamefont {Seradjeh}}]{martin3}%
  \BibitemOpen
  \bibfield  {author} {\bibinfo {author} {\bibfnamefont {A.}~\bibnamefont
  {Kumar}}, \bibinfo {author} {\bibfnamefont {M.}~\bibnamefont
  {Rodriguez-Vega}}, \bibinfo {author} {\bibfnamefont {T.}~\bibnamefont
  {Pereg-Barnea}},\ and\ \bibinfo {author} {\bibfnamefont {B.}~\bibnamefont
  {Seradjeh}},\ }\bibfield  {title} {\bibinfo {title} {{Linear response theory
  and optical conductivity of Floquet topological insulators}},\ }\href
  {https://doi.org/10.1103/PhysRevB.101.174314} {\bibfield  {journal} {\bibinfo
   {journal} {Physical Review B}\ }\textbf {\bibinfo {volume} {101}},\ \bibinfo
  {pages} {174314} (\bibinfo {year} {2020})}\BibitemShut {NoStop}%
\bibitem [{\citenamefont {Asmar}\ and\ \citenamefont {Tse}(2024)}]{Asmar2024}%
  \BibitemOpen
  \bibfield  {author} {\bibinfo {author} {\bibfnamefont {M.~M.}\ \bibnamefont
  {Asmar}}\ and\ \bibinfo {author} {\bibfnamefont {W.-K.}\ \bibnamefont
  {Tse}},\ }\bibfield  {title} {\bibinfo {title} {Photo-induced non-collinear
  interlayer rkky coupling in bulk rashba semiconductors},\ }\href
  {https://doi.org/10.1088/1367-2630/ad6b43} {\bibfield  {journal} {\bibinfo
  {journal} {New Journal of Physics}\ }\textbf {\bibinfo {volume} {26}},\
  \bibinfo {pages} {083016} (\bibinfo {year} {2024})}\BibitemShut {NoStop}%
\bibitem [{\citenamefont {Wang}\ \emph {et~al.}(2013)\citenamefont {Wang},
  \citenamefont {Steinberg}, \citenamefont {Jarillo-Herrero},\ and\
  \citenamefont {Gedik}}]{FloqExp2}%
  \BibitemOpen
  \bibfield  {author} {\bibinfo {author} {\bibfnamefont {Y.~H.}\ \bibnamefont
  {Wang}}, \bibinfo {author} {\bibfnamefont {H.}~\bibnamefont {Steinberg}},
  \bibinfo {author} {\bibfnamefont {P.}~\bibnamefont {Jarillo-Herrero}},\ and\
  \bibinfo {author} {\bibfnamefont {N.}~\bibnamefont {Gedik}},\ }\bibfield
  {title} {\bibinfo {title} {{Observation of Floquet-Bloch States on the
  Surface of a Topological Insulator}},\ }\href
  {https://doi.org/10.1126/science.1239834} {\bibfield  {journal} {\bibinfo
  {journal} {Science}\ }\textbf {\bibinfo {volume} {342}},\ \bibinfo {pages}
  {453} (\bibinfo {year} {2013})}\BibitemShut {NoStop}%
\bibitem [{\citenamefont {Mahmood}\ \emph {et~al.}(2016)\citenamefont
  {Mahmood}, \citenamefont {Chan}, \citenamefont {Alpichshev}, \citenamefont
  {Gardner}, \citenamefont {Lee}, \citenamefont {Lee},\ and\ \citenamefont
  {Gedik}}]{FloqExp1}%
  \BibitemOpen
  \bibfield  {author} {\bibinfo {author} {\bibfnamefont {F.}~\bibnamefont
  {Mahmood}}, \bibinfo {author} {\bibfnamefont {C.-K.}\ \bibnamefont {Chan}},
  \bibinfo {author} {\bibfnamefont {Z.}~\bibnamefont {Alpichshev}}, \bibinfo
  {author} {\bibfnamefont {D.}~\bibnamefont {Gardner}}, \bibinfo {author}
  {\bibfnamefont {Y.}~\bibnamefont {Lee}}, \bibinfo {author} {\bibfnamefont
  {P.~A.}\ \bibnamefont {Lee}},\ and\ \bibinfo {author} {\bibfnamefont
  {N.}~\bibnamefont {Gedik}},\ }\bibfield  {title} {\bibinfo {title}
  {{Selective scattering between Floquet--Bloch and Volkov states in a
  topological insulator}},\ }\href {https://doi.org/10.1038/nphys3609}
  {\bibfield  {journal} {\bibinfo  {journal} {Nat. Phys.}\ }\textbf {\bibinfo
  {volume} {12}},\ \bibinfo {pages} {306} (\bibinfo {year} {2016})}\BibitemShut
  {NoStop}%
\bibitem [{\citenamefont {Zhou}\ \emph {et~al.}(2023)\citenamefont {Zhou},
  \citenamefont {Bao}, \citenamefont {Fan}, \citenamefont {Zhou}, \citenamefont
  {Gao}, \citenamefont {Zhong}, \citenamefont {Lin}, \citenamefont {Liu},
  \citenamefont {Yu}, \citenamefont {Tang}, \citenamefont {Meng}, \citenamefont
  {Duan},\ and\ \citenamefont {Zhou}}]{BlackPhosphorus}%
  \BibitemOpen
  \bibfield  {author} {\bibinfo {author} {\bibfnamefont {S.}~\bibnamefont
  {Zhou}}, \bibinfo {author} {\bibfnamefont {C.}~\bibnamefont {Bao}}, \bibinfo
  {author} {\bibfnamefont {B.}~\bibnamefont {Fan}}, \bibinfo {author}
  {\bibfnamefont {H.}~\bibnamefont {Zhou}}, \bibinfo {author} {\bibfnamefont
  {Q.}~\bibnamefont {Gao}}, \bibinfo {author} {\bibfnamefont {H.}~\bibnamefont
  {Zhong}}, \bibinfo {author} {\bibfnamefont {T.}~\bibnamefont {Lin}}, \bibinfo
  {author} {\bibfnamefont {H.}~\bibnamefont {Liu}}, \bibinfo {author}
  {\bibfnamefont {P.}~\bibnamefont {Yu}}, \bibinfo {author} {\bibfnamefont
  {P.}~\bibnamefont {Tang}}, \bibinfo {author} {\bibfnamefont {S.}~\bibnamefont
  {Meng}}, \bibinfo {author} {\bibfnamefont {W.}~\bibnamefont {Duan}},\ and\
  \bibinfo {author} {\bibfnamefont {S.}~\bibnamefont {Zhou}},\ }\bibfield
  {title} {\bibinfo {title} {Pseudospin-selective floquet band engineering in
  black phosphorus},\ }\href {https://doi.org/10.1038/s41586-022-05610-3}
  {\bibfield  {journal} {\bibinfo  {journal} {Nature}\ }\textbf {\bibinfo
  {volume} {614}},\ \bibinfo {pages} {75–80} (\bibinfo {year}
  {2023})}\BibitemShut {NoStop}%
\bibitem [{\citenamefont {Ito}\ \emph {et~al.}(2023)\citenamefont {Ito},
  \citenamefont {Sch{\"u}ler}, \citenamefont {Meierhofer}, \citenamefont
  {Schlauderer}, \citenamefont {Reudenstein}, \citenamefont {Reimann},
  \citenamefont {Afanasiev}, \citenamefont {Kokh}, \citenamefont
  {Tereshchenko}, \citenamefont {G{\"u}dde},\ and\ \citenamefont
  {et~al.}}]{FloqExp4}%
  \BibitemOpen
  \bibfield  {author} {\bibinfo {author} {\bibfnamefont {S.}~\bibnamefont
  {Ito}}, \bibinfo {author} {\bibfnamefont {M.}~\bibnamefont {Sch{\"u}ler}},
  \bibinfo {author} {\bibfnamefont {M.}~\bibnamefont {Meierhofer}}, \bibinfo
  {author} {\bibfnamefont {S.}~\bibnamefont {Schlauderer}}, \bibinfo {author}
  {\bibfnamefont {J.}~\bibnamefont {Reudenstein}}, \bibinfo {author}
  {\bibfnamefont {J.}~\bibnamefont {Reimann}}, \bibinfo {author} {\bibfnamefont
  {D.}~\bibnamefont {Afanasiev}}, \bibinfo {author} {\bibfnamefont {K.~A.}\
  \bibnamefont {Kokh}}, \bibinfo {author} {\bibfnamefont {O.~E.}\ \bibnamefont
  {Tereshchenko}}, \bibinfo {author} {\bibfnamefont {J.}~\bibnamefont
  {G{\"u}dde}},\ and\ \bibinfo {author} {\bibnamefont {et~al.}},\ }\bibfield
  {title} {\bibinfo {title} {{Build-up and dephasing of Floquet--Bloch bands on
  subcycle timescales}},\ }\href
  {https://www.nature.com/articles/s41586-023-05850-x} {\bibfield  {journal}
  {\bibinfo  {journal} {Nature}\ }\textbf {\bibinfo {volume} {616}},\ \bibinfo
  {pages} {696} (\bibinfo {year} {2023})}\BibitemShut {NoStop}%
\bibitem [{\citenamefont {Merboldt}\ \emph {et~al.}(2025)\citenamefont
  {Merboldt}, \citenamefont {Schüler}, \citenamefont {Schmitt}, \citenamefont
  {Bange}, \citenamefont {Bennecke}, \citenamefont {Gadge}, \citenamefont
  {Pierz}, \citenamefont {Schumacher}, \citenamefont {Momeni}, \citenamefont
  {Steil}, \citenamefont {Manmana~R.}, \citenamefont {Sentef}, \citenamefont
  {Reutzel},\ and\ \citenamefont {Mathias}}]{TrARPSGraph}%
  \BibitemOpen
  \bibfield  {author} {\bibinfo {author} {\bibfnamefont {M.}~\bibnamefont
  {Merboldt}}, \bibinfo {author} {\bibfnamefont {M.}~\bibnamefont {Schüler}},
  \bibinfo {author} {\bibfnamefont {D.}~\bibnamefont {Schmitt}}, \bibinfo
  {author} {\bibfnamefont {J.}~\bibnamefont {Bange}}, \bibinfo {author}
  {\bibfnamefont {W.}~\bibnamefont {Bennecke}}, \bibinfo {author}
  {\bibfnamefont {K.}~\bibnamefont {Gadge}}, \bibinfo {author} {\bibfnamefont
  {K.}~\bibnamefont {Pierz}}, \bibinfo {author} {\bibfnamefont
  {H.}~\bibnamefont {Schumacher}}, \bibinfo {author} {\bibfnamefont
  {D.}~\bibnamefont {Momeni}}, \bibinfo {author} {\bibfnamefont
  {D.}~\bibnamefont {Steil}}, \bibinfo {author} {\bibfnamefont
  {S.}~\bibnamefont {Manmana~R.}}, \bibinfo {author} {\bibfnamefont {M.~A.}\
  \bibnamefont {Sentef}}, \bibinfo {author} {\bibfnamefont {M.}~\bibnamefont
  {Reutzel}},\ and\ \bibinfo {author} {\bibfnamefont {S.}~\bibnamefont
  {Mathias}},\ }\bibfield  {title} {\bibinfo {title} {Observation of floquet
  states in graphene},\ }\href {https://doi.org/10.1038/s41567-025-02889-7}
  {\bibfield  {journal} {\bibinfo  {journal} {Nature Physics}\ }\textbf
  {\bibinfo {volume} {21}},\ \bibinfo {pages} {1093–1099} (\bibinfo {year}
  {2025})}\BibitemShut {NoStop}%
\bibitem [{\citenamefont {Sie}\ \emph {et~al.}(2015)\citenamefont {Sie},
  \citenamefont {McIver}, \citenamefont {Lee}, \citenamefont {Fu},
  \citenamefont {Kong},\ and\ \citenamefont {Gedik}}]{Floq_WS2}%
  \BibitemOpen
  \bibfield  {author} {\bibinfo {author} {\bibfnamefont {E.~J.}\ \bibnamefont
  {Sie}}, \bibinfo {author} {\bibfnamefont {J.~W.}\ \bibnamefont {McIver}},
  \bibinfo {author} {\bibfnamefont {Y.-H.}\ \bibnamefont {Lee}}, \bibinfo
  {author} {\bibfnamefont {L.}~\bibnamefont {Fu}}, \bibinfo {author}
  {\bibfnamefont {J.}~\bibnamefont {Kong}},\ and\ \bibinfo {author}
  {\bibfnamefont {N.}~\bibnamefont {Gedik}},\ }\bibfield  {title} {\bibinfo
  {title} {Valley-selective optical {Stark} effect in monolayer {WS2}},\
  }\href {https://doi.org/10.1038/nmat4156} {\bibfield  {journal} {\bibinfo
  {journal} {Nature Materials}\ }\textbf {\bibinfo {volume} {14}},\ \bibinfo
  {pages} {290} (\bibinfo {year} {2015})}\BibitemShut {NoStop}%
\bibitem [{\citenamefont {Shan}\ \emph {et~al.}(2021)\citenamefont {Shan},
  \citenamefont {Ye}, \citenamefont {Chu}, \citenamefont {Lee}, \citenamefont
  {Park}, \citenamefont {Balents},\ and\ \citenamefont
  {Hsieh}}]{Floquet_Modualation}%
  \BibitemOpen
  \bibfield  {author} {\bibinfo {author} {\bibfnamefont {J.-Y.}\ \bibnamefont
  {Shan}}, \bibinfo {author} {\bibfnamefont {M.}~\bibnamefont {Ye}}, \bibinfo
  {author} {\bibfnamefont {H.}~\bibnamefont {Chu}}, \bibinfo {author}
  {\bibfnamefont {S.}~\bibnamefont {Lee}}, \bibinfo {author} {\bibfnamefont
  {J.-G.}\ \bibnamefont {Park}}, \bibinfo {author} {\bibfnamefont
  {L.}~\bibnamefont {Balents}},\ and\ \bibinfo {author} {\bibfnamefont
  {D.}~\bibnamefont {Hsieh}},\ }\bibfield  {title} {\bibinfo {title} {Giant
  modulation of optical nonlinearity by {Floquet} engineering},\ }\href
  {https://doi.org/10.1038/s41586-021-04051-8} {\bibfield  {journal} {\bibinfo
  {journal} {Nature}\ }\textbf {\bibinfo {volume} {600}},\ \bibinfo {pages}
  {235} (\bibinfo {year} {2021})}\BibitemShut {NoStop}%
\bibitem [{\citenamefont {Kobayashi}\ \emph {et~al.}(2023)\citenamefont
  {Kobayashi}, \citenamefont {Heide}, \citenamefont {Johnson}, \citenamefont
  {Tiwari}, \citenamefont {Liu}, \citenamefont {Reis}, \citenamefont {Heinz},\
  and\ \citenamefont {Ghimire}}]{Floq_WS22}%
  \BibitemOpen
  \bibfield  {author} {\bibinfo {author} {\bibfnamefont {Y.}~\bibnamefont
  {Kobayashi}}, \bibinfo {author} {\bibfnamefont {C.}~\bibnamefont {Heide}},
  \bibinfo {author} {\bibfnamefont {A.~C.}\ \bibnamefont {Johnson}}, \bibinfo
  {author} {\bibfnamefont {V.}~\bibnamefont {Tiwari}}, \bibinfo {author}
  {\bibfnamefont {F.}~\bibnamefont {Liu}}, \bibinfo {author} {\bibfnamefont
  {D.~A.}\ \bibnamefont {Reis}}, \bibinfo {author} {\bibfnamefont {T.~F.}\
  \bibnamefont {Heinz}},\ and\ \bibinfo {author} {\bibfnamefont
  {S.}~\bibnamefont {Ghimire}},\ }\bibfield  {title} {\bibinfo {title} {Floquet
  engineering of strongly driven excitons in monolayer tungsten disulfide},\
  }\href {https://doi.org/10.1038/s41567-022-01849-9} {\bibfield  {journal}
  {\bibinfo  {journal} {Nature Physics}\ }\textbf {\bibinfo {volume} {19}},\
  \bibinfo {pages} {171} (\bibinfo {year} {2023})}\BibitemShut {NoStop}%
\bibitem [{\citenamefont {McIver}\ \emph {et~al.}(2020)\citenamefont {McIver},
  \citenamefont {Schulte}, \citenamefont {Stein}, \citenamefont {Matsuyama},
  \citenamefont {Jotzu}, \citenamefont {Meier},\ and\ \citenamefont
  {Cavalleri}}]{FloqExp3}%
  \BibitemOpen
  \bibfield  {author} {\bibinfo {author} {\bibfnamefont {J.~W.}\ \bibnamefont
  {McIver}}, \bibinfo {author} {\bibfnamefont {B.}~\bibnamefont {Schulte}},
  \bibinfo {author} {\bibfnamefont {F.-U.}\ \bibnamefont {Stein}}, \bibinfo
  {author} {\bibfnamefont {T.}~\bibnamefont {Matsuyama}}, \bibinfo {author}
  {\bibfnamefont {G.}~\bibnamefont {Jotzu}}, \bibinfo {author} {\bibfnamefont
  {G.}~\bibnamefont {Meier}},\ and\ \bibinfo {author} {\bibfnamefont
  {A.}~\bibnamefont {Cavalleri}},\ }\bibfield  {title} {\bibinfo {title}
  {{Light-induced anomalous Hall effect in graphene}},\ }\href
  {https://doi.org/10.1038/s41567-019-0698-y} {\bibfield  {journal} {\bibinfo
  {journal} {Nat. Phys.}\ }\textbf {\bibinfo {volume} {16}},\ \bibinfo {pages}
  {38} (\bibinfo {year} {2020})}\BibitemShut {NoStop}%
\bibitem [{\citenamefont {Park}\ \emph {et~al.}(2022)\citenamefont {Park},
  \citenamefont {Lee}, \citenamefont {Jang}, \citenamefont {Choi},
  \citenamefont {Park}, \citenamefont {Jung}, \citenamefont {Watanabe},
  \citenamefont {Taniguchi}, \citenamefont {Cho},\ and\ \citenamefont
  {Lee}}]{Park2022FloquetAndreev}%
  \BibitemOpen
  \bibfield  {author} {\bibinfo {author} {\bibfnamefont {S.}~\bibnamefont
  {Park}}, \bibinfo {author} {\bibfnamefont {W.}~\bibnamefont {Lee}}, \bibinfo
  {author} {\bibfnamefont {S.}~\bibnamefont {Jang}}, \bibinfo {author}
  {\bibfnamefont {Y.-B.}\ \bibnamefont {Choi}}, \bibinfo {author}
  {\bibfnamefont {J.}~\bibnamefont {Park}}, \bibinfo {author} {\bibfnamefont
  {W.}~\bibnamefont {Jung}}, \bibinfo {author} {\bibfnamefont {K.}~\bibnamefont
  {Watanabe}}, \bibinfo {author} {\bibfnamefont {T.}~\bibnamefont {Taniguchi}},
  \bibinfo {author} {\bibfnamefont {G.~Y.}\ \bibnamefont {Cho}},\ and\ \bibinfo
  {author} {\bibfnamefont {G.-H.}\ \bibnamefont {Lee}},\ }\bibfield  {title}
  {\bibinfo {title} {Steady floquet--andreev states in graphene josephson
  junctions},\ }\href {https://doi.org/10.1038/s41586-021-04364-8} {\bibfield
  {journal} {\bibinfo  {journal} {Nature}\ }\textbf {\bibinfo {volume} {603}},\
  \bibinfo {pages} {421} (\bibinfo {year} {2022})}\BibitemShut {NoStop}%
\bibitem [{\citenamefont {Liu}\ \emph {et~al.}(2025)\citenamefont {Liu},
  \citenamefont {Yang}, \citenamefont {Gaertner}, \citenamefont {Huckabee},
  \citenamefont {Suslov}, \citenamefont {Refael}, \citenamefont {Nathan},
  \citenamefont {Lewandowski}, \citenamefont {Foa~Torres}, \citenamefont
  {Esin}, \citenamefont {Barbara},\ and\ \citenamefont
  {Kalugin}}]{FloqTraspGraphen}%
  \BibitemOpen
  \bibfield  {author} {\bibinfo {author} {\bibfnamefont {Y.}~\bibnamefont
  {Liu}}, \bibinfo {author} {\bibfnamefont {C.}~\bibnamefont {Yang}}, \bibinfo
  {author} {\bibfnamefont {G.}~\bibnamefont {Gaertner}}, \bibinfo {author}
  {\bibfnamefont {J.}~\bibnamefont {Huckabee}}, \bibinfo {author}
  {\bibfnamefont {A.~V.}\ \bibnamefont {Suslov}}, \bibinfo {author}
  {\bibfnamefont {G.}~\bibnamefont {Refael}}, \bibinfo {author} {\bibfnamefont
  {F.}~\bibnamefont {Nathan}}, \bibinfo {author} {\bibfnamefont
  {C.}~\bibnamefont {Lewandowski}}, \bibinfo {author} {\bibfnamefont
  {L.~E.~F.}\ \bibnamefont {Foa~Torres}}, \bibinfo {author} {\bibfnamefont
  {I.}~\bibnamefont {Esin}}, \bibinfo {author} {\bibfnamefont {P.}~\bibnamefont
  {Barbara}},\ and\ \bibinfo {author} {\bibfnamefont {N.~G.}\ \bibnamefont
  {Kalugin}},\ }\bibfield  {title} {\bibinfo {title} {Signatures of {Floquet}
  electronic steady states in graphene under continuous-wave mid-infrared
  irradiation},\ }\href {https://doi.org/10.1038/s41467-025-57335-2} {\bibfield
   {journal} {\bibinfo  {journal} {Nature Communications}\ }\textbf {\bibinfo
  {volume} {16}},\ \bibinfo {pages} {2057} (\bibinfo {year}
  {2025})}\BibitemShut {NoStop}%
\bibitem [{\citenamefont {Forbes}\ \emph {et~al.}(2021)\citenamefont {Forbes},
  \citenamefont {de~Oliveira},\ and\ \citenamefont {Dennis}}]{lightSt1}%
  \BibitemOpen
  \bibfield  {author} {\bibinfo {author} {\bibfnamefont {A.}~\bibnamefont
  {Forbes}}, \bibinfo {author} {\bibfnamefont {M.}~\bibnamefont
  {de~Oliveira}},\ and\ \bibinfo {author} {\bibfnamefont {M.~R.}\ \bibnamefont
  {Dennis}},\ }\bibfield  {title} {\bibinfo {title} {Structured light},\ }\href
  {https://doi.org/10.1038/s41566-021-00780-4} {\bibfield  {journal} {\bibinfo
  {journal} {Nature Photonics}\ }\textbf {\bibinfo {volume} {15}},\ \bibinfo
  {pages} {253} (\bibinfo {year} {2021})}\BibitemShut {NoStop}%
\bibitem [{\citenamefont {He}\ \emph {et~al.}(2022)\citenamefont {He},
  \citenamefont {Shen},\ and\ \citenamefont {Forbes}}]{lightSt}%
  \BibitemOpen
  \bibfield  {author} {\bibinfo {author} {\bibfnamefont {C.}~\bibnamefont
  {He}}, \bibinfo {author} {\bibfnamefont {Y.}~\bibnamefont {Shen}},\ and\
  \bibinfo {author} {\bibfnamefont {A.}~\bibnamefont {Forbes}},\ }\bibfield
  {title} {\bibinfo {title} {Towards higher-dimensional structured light},\
  }\href {https://www.nature.com/articles/s41377-022-00897-3} {\bibfield
  {journal} {\bibinfo  {journal} {Light: Science \& Applications}\ }\textbf
  {\bibinfo {volume} {11}},\ \bibinfo {pages} {205} (\bibinfo {year}
  {2022})}\BibitemShut {NoStop}%
\bibitem [{\citenamefont {Padgett}\ \emph {et~al.}(2004)\citenamefont
  {Padgett}, \citenamefont {Courtial},\ and\ \citenamefont {Allen}}]{lightV4}%
  \BibitemOpen
  \bibfield  {author} {\bibinfo {author} {\bibfnamefont {M.}~\bibnamefont
  {Padgett}}, \bibinfo {author} {\bibfnamefont {J.}~\bibnamefont {Courtial}},\
  and\ \bibinfo {author} {\bibfnamefont {L.}~\bibnamefont {Allen}},\ }\bibfield
   {title} {\bibinfo {title} {Light’s orbital angular momentum},\ }\href
  {https://doi.org/10.1063/1.1768672} {\bibfield  {journal} {\bibinfo
  {journal} {Physics Today}\ }\textbf {\bibinfo {volume} {57}},\ \bibinfo
  {pages} {35} (\bibinfo {year} {2004})},\ \Eprint
  {https://arxiv.org/abs/https://doi.org/10.1063/1.1768672}
  {https://doi.org/10.1063/1.1768672} \BibitemShut {NoStop}%
\bibitem [{\citenamefont {Franke-Arnold}\ \emph {et~al.}(2008)\citenamefont
  {Franke-Arnold}, \citenamefont {Allen},\ and\ \citenamefont
  {Padgett}}]{lightV3}%
  \BibitemOpen
  \bibfield  {author} {\bibinfo {author} {\bibfnamefont {S.}~\bibnamefont
  {Franke-Arnold}}, \bibinfo {author} {\bibfnamefont {L.}~\bibnamefont
  {Allen}},\ and\ \bibinfo {author} {\bibfnamefont {M.}~\bibnamefont
  {Padgett}},\ }\bibfield  {title} {\bibinfo {title} {{Advances in optical
  angular momentum}},\ }\href {https://doi.org/10.1002/lpor.200810007}
  {\bibfield  {journal} {\bibinfo  {journal} {Laser Photon. Rev.}\ }\textbf
  {\bibinfo {volume} {2}},\ \bibinfo {pages} {299} (\bibinfo {year}
  {2008})}\BibitemShut {NoStop}%
\bibitem [{\citenamefont {Shen}\ \emph {et~al.}(2019)\citenamefont {Shen},
  \citenamefont {Wang}, \citenamefont {Xie}, \citenamefont {Min}, \citenamefont
  {Fu}, \citenamefont {Liu}, \citenamefont {Gong},\ and\ \citenamefont
  {Yuan}}]{lightV1}%
  \BibitemOpen
  \bibfield  {author} {\bibinfo {author} {\bibfnamefont {Y.}~\bibnamefont
  {Shen}}, \bibinfo {author} {\bibfnamefont {X.}~\bibnamefont {Wang}}, \bibinfo
  {author} {\bibfnamefont {Z.}~\bibnamefont {Xie}}, \bibinfo {author}
  {\bibfnamefont {C.}~\bibnamefont {Min}}, \bibinfo {author} {\bibfnamefont
  {X.}~\bibnamefont {Fu}}, \bibinfo {author} {\bibfnamefont {Q.}~\bibnamefont
  {Liu}}, \bibinfo {author} {\bibfnamefont {M.}~\bibnamefont {Gong}},\ and\
  \bibinfo {author} {\bibfnamefont {X.}~\bibnamefont {Yuan}},\ }\bibfield
  {title} {\bibinfo {title} {{Optical vortices 30 years on: OAM manipulation
  from topological charge to multiple singularities}},\ }\href
  {https://doi.org/10.1038/s41377-019-0194-2} {\bibfield  {journal} {\bibinfo
  {journal} {Light: Science \& Applications}\ }\textbf {\bibinfo {volume}
  {8}},\ \bibinfo {pages} {1} (\bibinfo {year} {2019})}\BibitemShut {NoStop}%
\bibitem [{\citenamefont {Kim}\ \emph {et~al.}(2022)\citenamefont {Kim},
  \citenamefont {Dehghani}, \citenamefont {Ahmadabadi}, \citenamefont
  {Martin},\ and\ \citenamefont {Hafezi}}]{MD1}%
  \BibitemOpen
  \bibfield  {author} {\bibinfo {author} {\bibfnamefont {H.}~\bibnamefont
  {Kim}}, \bibinfo {author} {\bibfnamefont {H.}~\bibnamefont {Dehghani}},
  \bibinfo {author} {\bibfnamefont {I.}~\bibnamefont {Ahmadabadi}}, \bibinfo
  {author} {\bibfnamefont {I.}~\bibnamefont {Martin}},\ and\ \bibinfo {author}
  {\bibfnamefont {M.}~\bibnamefont {Hafezi}},\ }\bibfield  {title} {\bibinfo
  {title} {{Floquet vortex states induced by light carrying an orbital angular
  momentum}},\ }\href {https://doi.org/10.1103/PhysRevB.105.L081301} {\bibfield
   {journal} {\bibinfo  {journal} {Phys. Rev. B}\ }\textbf {\bibinfo {volume}
  {105}},\ \bibinfo {pages} {L081301} (\bibinfo {year} {2022})}\BibitemShut
  {NoStop}%
\bibitem [{\citenamefont {Aich}\ and\ \citenamefont
  {Seradjeh}(2024)}]{babakVLB}%
  \BibitemOpen
  \bibfield  {author} {\bibinfo {author} {\bibfnamefont {S.}~\bibnamefont
  {Aich}}\ and\ \bibinfo {author} {\bibfnamefont {B.}~\bibnamefont
  {Seradjeh}},\ }\bibfield  {title} {\bibinfo {title} {Multiple tunable
  real-space degeneracies in graphene irradiated by twisted light},\ }\href
  {https://doi.org/10.1103/PhysRevB.110.054314} {\bibfield  {journal} {\bibinfo
   {journal} {Phys. Rev. B}\ }\textbf {\bibinfo {volume} {110}},\ \bibinfo
  {pages} {054314} (\bibinfo {year} {2024})}\BibitemShut {NoStop}%
\bibitem [{\citenamefont {Massaro}\ \emph {et~al.}(2025)\citenamefont
  {Massaro}, \citenamefont {Meese}, \citenamefont {Sandler},\ and\
  \citenamefont {Asmar}}]{Lauren2025}%
  \BibitemOpen
  \bibfield  {author} {\bibinfo {author} {\bibfnamefont {L.~I.}\ \bibnamefont
  {Massaro}}, \bibinfo {author} {\bibfnamefont {C.}~\bibnamefont {Meese}},
  \bibinfo {author} {\bibfnamefont {N.~P.}\ \bibnamefont {Sandler}},\ and\
  \bibinfo {author} {\bibfnamefont {M.~M.}\ \bibnamefont {Asmar}},\ }\bibfield
  {title} {\bibinfo {title} {Photoinduced multiply quantized vortex states in
  dirac-like materials},\ }\href {https://doi.org/10.1103/PhysRevB.111.085402}
  {\bibfield  {journal} {\bibinfo  {journal} {Phys. Rev. B}\ }\textbf {\bibinfo
  {volume} {111}},\ \bibinfo {pages} {085402} (\bibinfo {year}
  {2025})}\BibitemShut {NoStop}%
\bibitem [{\citenamefont {Park}\ \emph
  {et~al.}(2008{\natexlab{a}})\citenamefont {Park}, \citenamefont {Yang},
  \citenamefont {Son}, \citenamefont {Cohen},\ and\ \citenamefont
  {Louie}}]{Park2008NP}%
  \BibitemOpen
  \bibfield  {author} {\bibinfo {author} {\bibfnamefont {C.-H.}\ \bibnamefont
  {Park}}, \bibinfo {author} {\bibfnamefont {L.}~\bibnamefont {Yang}}, \bibinfo
  {author} {\bibfnamefont {Y.-W.}\ \bibnamefont {Son}}, \bibinfo {author}
  {\bibfnamefont {M.~L.}\ \bibnamefont {Cohen}},\ and\ \bibinfo {author}
  {\bibfnamefont {S.~G.}\ \bibnamefont {Louie}},\ }\bibfield  {title} {\bibinfo
  {title} {Anisotropic behaviours of massless dirac fermions in graphene under
  periodic potentials},\ }\href {https://doi.org/10.1038/nphys890} {\bibfield
  {journal} {\bibinfo  {journal} {Nature Physics}\ }\textbf {\bibinfo {volume}
  {4}},\ \bibinfo {pages} {213} (\bibinfo {year}
  {2008}{\natexlab{a}})}\BibitemShut {NoStop}%
\bibitem [{\citenamefont {Park}\ \emph
  {et~al.}(2008{\natexlab{b}})\citenamefont {Park}, \citenamefont {Yang},
  \citenamefont {Son}, \citenamefont {Cohen},\ and\ \citenamefont
  {Louie}}]{Park2008PRL}%
  \BibitemOpen
  \bibfield  {author} {\bibinfo {author} {\bibfnamefont {C.-H.}\ \bibnamefont
  {Park}}, \bibinfo {author} {\bibfnamefont {L.}~\bibnamefont {Yang}}, \bibinfo
  {author} {\bibfnamefont {Y.-W.}\ \bibnamefont {Son}}, \bibinfo {author}
  {\bibfnamefont {M.~L.}\ \bibnamefont {Cohen}},\ and\ \bibinfo {author}
  {\bibfnamefont {S.~G.}\ \bibnamefont {Louie}},\ }\bibfield  {title} {\bibinfo
  {title} {New generation of massless dirac fermions in graphene under external
  periodic potentials},\ }\href
  {https://doi.org/10.1103/PhysRevLett.101.126804} {\bibfield  {journal}
  {\bibinfo  {journal} {Physical Review Letters}\ }\textbf {\bibinfo {volume}
  {101}},\ \bibinfo {pages} {126804} (\bibinfo {year}
  {2008}{\natexlab{b}})}\BibitemShut {NoStop}%
\bibitem [{\citenamefont {Brey}\ and\ \citenamefont
  {Fertig}(2009)}]{BreyFertig2009}%
  \BibitemOpen
  \bibfield  {author} {\bibinfo {author} {\bibfnamefont {L.}~\bibnamefont
  {Brey}}\ and\ \bibinfo {author} {\bibfnamefont {H.~A.}\ \bibnamefont
  {Fertig}},\ }\bibfield  {title} {\bibinfo {title} {Emerging zero modes for
  graphene in a periodic potential},\ }\href
  {https://doi.org/10.1103/PhysRevLett.103.046809} {\bibfield  {journal}
  {\bibinfo  {journal} {Physical Review Letters}\ }\textbf {\bibinfo {volume}
  {103}},\ \bibinfo {pages} {046809} (\bibinfo {year} {2009})}\BibitemShut
  {NoStop}%
\bibitem [{\citenamefont {Arovas}\ \emph {et~al.}(2010)\citenamefont {Arovas},
  \citenamefont {Brey}, \citenamefont {Fertig}, \citenamefont {Kim},\ and\
  \citenamefont {Ziegler}}]{Arovas2010}%
  \BibitemOpen
  \bibfield  {author} {\bibinfo {author} {\bibfnamefont {D.~P.}\ \bibnamefont
  {Arovas}}, \bibinfo {author} {\bibfnamefont {L.}~\bibnamefont {Brey}},
  \bibinfo {author} {\bibfnamefont {H.~A.}\ \bibnamefont {Fertig}}, \bibinfo
  {author} {\bibfnamefont {E.-A.}\ \bibnamefont {Kim}},\ and\ \bibinfo {author}
  {\bibfnamefont {K.}~\bibnamefont {Ziegler}},\ }\bibfield  {title} {\bibinfo
  {title} {Dirac spectrum in piecewise constant one-dimensional potentials},\
  }\href {https://doi.org/10.1088/1367-2630/12/12/123020} {\bibfield  {journal}
  {\bibinfo  {journal} {New Journal of Physics}\ }\textbf {\bibinfo {volume}
  {12}},\ \bibinfo {pages} {123020} (\bibinfo {year} {2010})}\BibitemShut
  {NoStop}%
\bibitem [{\citenamefont {Snyman}(2009)}]{Snyman2009GappedState}%
  \BibitemOpen
  \bibfield  {author} {\bibinfo {author} {\bibfnamefont {I.}~\bibnamefont
  {Snyman}},\ }\bibfield  {title} {\bibinfo {title} {Gapped state of a carbon
  monolayer in periodic magnetic and electric fields},\ }\href
  {https://doi.org/10.1103/PhysRevB.80.054303} {\bibfield  {journal} {\bibinfo
  {journal} {Physical Review B}\ }\textbf {\bibinfo {volume} {80}},\ \bibinfo
  {pages} {054303} (\bibinfo {year} {2009})}\BibitemShut {NoStop}%
\bibitem [{\citenamefont {Semenoff}\ \emph {et~al.}(2008)\citenamefont
  {Semenoff}, \citenamefont {Semenoff},\ and\ \citenamefont
  {Zhou}}]{Semenoff2008}%
  \BibitemOpen
  \bibfield  {author} {\bibinfo {author} {\bibfnamefont {G.~W.}\ \bibnamefont
  {Semenoff}}, \bibinfo {author} {\bibfnamefont {V.}~\bibnamefont {Semenoff}},\
  and\ \bibinfo {author} {\bibfnamefont {F.}~\bibnamefont {Zhou}},\ }\bibfield
  {title} {\bibinfo {title} {Domain walls in gapped graphene},\ }\href
  {https://doi.org/10.1103/PhysRevLett.101.087204} {\bibfield  {journal}
  {\bibinfo  {journal} {Physical Review Letters}\ }\textbf {\bibinfo {volume}
  {101}},\ \bibinfo {pages} {087204} (\bibinfo {year} {2008})}\BibitemShut
  {NoStop}%
\bibitem [{\citenamefont {Maksimova}\ \emph {et~al.}(2012)\citenamefont
  {Maksimova}, \citenamefont {Azarova}, \citenamefont {Telezhnikov},\ and\
  \citenamefont {Burdov}}]{Maksimova2012}%
  \BibitemOpen
  \bibfield  {author} {\bibinfo {author} {\bibfnamefont {G.~M.}\ \bibnamefont
  {Maksimova}}, \bibinfo {author} {\bibfnamefont {E.~S.}\ \bibnamefont
  {Azarova}}, \bibinfo {author} {\bibfnamefont {A.~V.}\ \bibnamefont
  {Telezhnikov}},\ and\ \bibinfo {author} {\bibfnamefont {V.~A.}\ \bibnamefont
  {Burdov}},\ }\bibfield  {title} {\bibinfo {title} {Graphene superlattice with
  periodically modulated dirac gap},\ }\href
  {https://doi.org/10.1103/PhysRevB.86.205422} {\bibfield  {journal} {\bibinfo
  {journal} {Physical Review B}\ }\textbf {\bibinfo {volume} {86}},\ \bibinfo
  {pages} {205422} (\bibinfo {year} {2012})}\BibitemShut {NoStop}%
\bibitem [{\citenamefont {De~Martino}\ \emph {et~al.}(2023)\citenamefont
  {De~Martino}, \citenamefont {Dell'Anna}, \citenamefont {Handt}, \citenamefont
  {Miserocchi},\ and\ \citenamefont {Egger}}]{DeMartino2023}%
  \BibitemOpen
  \bibfield  {author} {\bibinfo {author} {\bibfnamefont {A.}~\bibnamefont
  {De~Martino}}, \bibinfo {author} {\bibfnamefont {L.}~\bibnamefont
  {Dell'Anna}}, \bibinfo {author} {\bibfnamefont {L.}~\bibnamefont {Handt}},
  \bibinfo {author} {\bibfnamefont {A.}~\bibnamefont {Miserocchi}},\ and\
  \bibinfo {author} {\bibfnamefont {R.}~\bibnamefont {Egger}},\ }\bibfield
  {title} {\bibinfo {title} {Two-dimensional dirac fermions in a mass
  superlattice},\ }\href {https://doi.org/10.1103/PhysRevB.107.115420}
  {\bibfield  {journal} {\bibinfo  {journal} {Physical Review B}\ }\textbf
  {\bibinfo {volume} {107}},\ \bibinfo {pages} {115420} (\bibinfo {year}
  {2023})}\BibitemShut {NoStop}%
\bibitem [{\citenamefont {Fujita}\ \emph {et~al.}(1996)\citenamefont {Fujita},
  \citenamefont {Wakabayashi}, \citenamefont {Nakada},\ and\ \citenamefont
  {Kusakabe}}]{Fujita1996}%
  \BibitemOpen
  \bibfield  {author} {\bibinfo {author} {\bibfnamefont {M.}~\bibnamefont
  {Fujita}}, \bibinfo {author} {\bibfnamefont {K.}~\bibnamefont {Wakabayashi}},
  \bibinfo {author} {\bibfnamefont {K.}~\bibnamefont {Nakada}},\ and\ \bibinfo
  {author} {\bibfnamefont {K.}~\bibnamefont {Kusakabe}},\ }\bibfield  {title}
  {\bibinfo {title} {Peculiar localized state at zigzag graphite edge},\ }\href
  {https://doi.org/10.1143/JPSJ.65.1920} {\bibfield  {journal} {\bibinfo
  {journal} {Journal of the Physical Society of Japan}\ }\textbf {\bibinfo
  {volume} {65}},\ \bibinfo {pages} {1920} (\bibinfo {year}
  {1996})}\BibitemShut {NoStop}%
\bibitem [{\citenamefont {Nakada}\ \emph {et~al.}(1996)\citenamefont {Nakada},
  \citenamefont {Fujita}, \citenamefont {Dresselhaus},\ and\ \citenamefont
  {Dresselhaus}}]{Nakada1996}%
  \BibitemOpen
  \bibfield  {author} {\bibinfo {author} {\bibfnamefont {K.}~\bibnamefont
  {Nakada}}, \bibinfo {author} {\bibfnamefont {M.}~\bibnamefont {Fujita}},
  \bibinfo {author} {\bibfnamefont {G.}~\bibnamefont {Dresselhaus}},\ and\
  \bibinfo {author} {\bibfnamefont {M.~S.}\ \bibnamefont {Dresselhaus}},\
  }\bibfield  {title} {\bibinfo {title} {Edge state in graphene ribbons:
  Nanometer size effect and edge shape dependence},\ }\href
  {https://doi.org/10.1103/PhysRevB.54.17954} {\bibfield  {journal} {\bibinfo
  {journal} {Physical Review B}\ }\textbf {\bibinfo {volume} {54}},\ \bibinfo
  {pages} {17954} (\bibinfo {year} {1996})}\BibitemShut {NoStop}%
\bibitem [{\citenamefont {Brey}\ and\ \citenamefont
  {Fertig}(2006)}]{BreyFertig2006}%
  \BibitemOpen
  \bibfield  {author} {\bibinfo {author} {\bibfnamefont {L.}~\bibnamefont
  {Brey}}\ and\ \bibinfo {author} {\bibfnamefont {H.~A.}\ \bibnamefont
  {Fertig}},\ }\bibfield  {title} {\bibinfo {title} {Electronic states of
  graphene nanoribbons studied with the dirac equation},\ }\href
  {https://doi.org/10.1103/PhysRevB.73.235411} {\bibfield  {journal} {\bibinfo
  {journal} {Physical Review B}\ }\textbf {\bibinfo {volume} {73}},\ \bibinfo
  {pages} {235411} (\bibinfo {year} {2006})}\BibitemShut {NoStop}%
\bibitem [{\citenamefont {Son}\ \emph {et~al.}(2006)\citenamefont {Son},
  \citenamefont {Cohen},\ and\ \citenamefont {Louie}}]{Son2006}%
  \BibitemOpen
  \bibfield  {author} {\bibinfo {author} {\bibfnamefont {Y.-W.}\ \bibnamefont
  {Son}}, \bibinfo {author} {\bibfnamefont {M.~L.}\ \bibnamefont {Cohen}},\
  and\ \bibinfo {author} {\bibfnamefont {S.~G.}\ \bibnamefont {Louie}},\
  }\bibfield  {title} {\bibinfo {title} {Energy gaps in graphene nanoribbons},\
  }\href {https://doi.org/10.1103/PhysRevLett.97.216803} {\bibfield  {journal}
  {\bibinfo  {journal} {Physical Review Letters}\ }\textbf {\bibinfo {volume}
  {97}},\ \bibinfo {pages} {216803} (\bibinfo {year} {2006})}\BibitemShut
  {NoStop}%
\bibitem [{\citenamefont {Tao}\ \emph {et~al.}(2011)\citenamefont {Tao},
  \citenamefont {Jiao}, \citenamefont {Yazyev}, \citenamefont {Chen},
  \citenamefont {Feng}, \citenamefont {Zhang}, \citenamefont {Capaz},
  \citenamefont {Tour}, \citenamefont {Zettl}, \citenamefont {Louie},
  \citenamefont {Dai},\ and\ \citenamefont {Crommie}}]{Tao2011}%
  \BibitemOpen
  \bibfield  {author} {\bibinfo {author} {\bibfnamefont {C.}~\bibnamefont
  {Tao}}, \bibinfo {author} {\bibfnamefont {L.}~\bibnamefont {Jiao}}, \bibinfo
  {author} {\bibfnamefont {O.~V.}\ \bibnamefont {Yazyev}}, \bibinfo {author}
  {\bibfnamefont {Y.-C.}\ \bibnamefont {Chen}}, \bibinfo {author}
  {\bibfnamefont {J.}~\bibnamefont {Feng}}, \bibinfo {author} {\bibfnamefont
  {X.}~\bibnamefont {Zhang}}, \bibinfo {author} {\bibfnamefont {R.~B.}\
  \bibnamefont {Capaz}}, \bibinfo {author} {\bibfnamefont {J.~M.}\ \bibnamefont
  {Tour}}, \bibinfo {author} {\bibfnamefont {A.}~\bibnamefont {Zettl}},
  \bibinfo {author} {\bibfnamefont {S.~G.}\ \bibnamefont {Louie}}, \bibinfo
  {author} {\bibfnamefont {H.}~\bibnamefont {Dai}},\ and\ \bibinfo {author}
  {\bibfnamefont {M.~F.}\ \bibnamefont {Crommie}},\ }\bibfield  {title}
  {\bibinfo {title} {Spatially resolving edge states of chiral graphene
  nanoribbons},\ }\href {https://doi.org/10.1038/nphys1991} {\bibfield
  {journal} {\bibinfo  {journal} {Nature Physics}\ }\textbf {\bibinfo {volume}
  {7}},\ \bibinfo {pages} {616} (\bibinfo {year} {2011})}\BibitemShut {NoStop}%
\bibitem [{\citenamefont {Zhang}\ \emph {et~al.}(2013)\citenamefont {Zhang},
  \citenamefont {Yazyev}, \citenamefont {Feng}, \citenamefont {Xie},
  \citenamefont {Tao}, \citenamefont {Chen}, \citenamefont {Jiao},
  \citenamefont {Pedramrazi}, \citenamefont {Zettl}, \citenamefont {Louie},
  \citenamefont {Dai},\ and\ \citenamefont {Crommie}}]{Zhang2013}%
  \BibitemOpen
  \bibfield  {author} {\bibinfo {author} {\bibfnamefont {X.}~\bibnamefont
  {Zhang}}, \bibinfo {author} {\bibfnamefont {O.~V.}\ \bibnamefont {Yazyev}},
  \bibinfo {author} {\bibfnamefont {J.}~\bibnamefont {Feng}}, \bibinfo {author}
  {\bibfnamefont {L.}~\bibnamefont {Xie}}, \bibinfo {author} {\bibfnamefont
  {C.}~\bibnamefont {Tao}}, \bibinfo {author} {\bibfnamefont {Y.-C.}\
  \bibnamefont {Chen}}, \bibinfo {author} {\bibfnamefont {L.}~\bibnamefont
  {Jiao}}, \bibinfo {author} {\bibfnamefont {Z.}~\bibnamefont {Pedramrazi}},
  \bibinfo {author} {\bibfnamefont {A.}~\bibnamefont {Zettl}}, \bibinfo
  {author} {\bibfnamefont {S.~G.}\ \bibnamefont {Louie}}, \bibinfo {author}
  {\bibfnamefont {H.}~\bibnamefont {Dai}},\ and\ \bibinfo {author}
  {\bibfnamefont {M.~F.}\ \bibnamefont {Crommie}},\ }\bibfield  {title}
  {\bibinfo {title} {Experimentally engineering the edge termination of
  graphene nanoribbons},\ }\href {https://doi.org/10.1021/nn303730v} {\bibfield
   {journal} {\bibinfo  {journal} {ACS Nano}\ }\textbf {\bibinfo {volume}
  {7}},\ \bibinfo {pages} {198} (\bibinfo {year} {2013})}\BibitemShut {NoStop}%
\bibitem [{\citenamefont {Yao}\ \emph {et~al.}(2009)\citenamefont {Yao},
  \citenamefont {Yang},\ and\ \citenamefont {Niu}}]{Yao2009EdgeStates}%
  \BibitemOpen
  \bibfield  {author} {\bibinfo {author} {\bibfnamefont {W.}~\bibnamefont
  {Yao}}, \bibinfo {author} {\bibfnamefont {S.~A.}\ \bibnamefont {Yang}},\ and\
  \bibinfo {author} {\bibfnamefont {Q.}~\bibnamefont {Niu}},\ }\bibfield
  {title} {\bibinfo {title} {Edge states in graphene: From gapped flat-band to
  gapless chiral modes},\ }\href
  {https://doi.org/10.1103/PhysRevLett.102.096801} {\bibfield  {journal}
  {\bibinfo  {journal} {Physical Review Letters}\ }\textbf {\bibinfo {volume}
  {102}},\ \bibinfo {pages} {096801} (\bibinfo {year} {2009})}\BibitemShut
  {NoStop}%
\bibitem [{\citenamefont {Apel}\ \emph {et~al.}(2011)\citenamefont {Apel},
  \citenamefont {Pal},\ and\ \citenamefont
  {Schweitzer}}]{Apel2011GNRPotentials}%
  \BibitemOpen
  \bibfield  {author} {\bibinfo {author} {\bibfnamefont {W.}~\bibnamefont
  {Apel}}, \bibinfo {author} {\bibfnamefont {G.}~\bibnamefont {Pal}},\ and\
  \bibinfo {author} {\bibfnamefont {L.}~\bibnamefont {Schweitzer}},\ }\bibfield
   {title} {\bibinfo {title} {Energy gap in graphene nanoribbons with
  structured external electric potentials},\ }\href
  {https://doi.org/10.1103/PhysRevB.83.125431} {\bibfield  {journal} {\bibinfo
  {journal} {Physical Review B}\ }\textbf {\bibinfo {volume} {83}},\ \bibinfo
  {pages} {125431} (\bibinfo {year} {2011})}\BibitemShut {NoStop}%
\bibitem [{\citenamefont {Calvo}\ \emph {et~al.}(2025)\citenamefont {Calvo},
  \citenamefont {Foa~Torres},\ and\ \citenamefont {Berdakin}}]{Torres1}%
  \BibitemOpen
  \bibfield  {author} {\bibinfo {author} {\bibfnamefont {H.~L.}\ \bibnamefont
  {Calvo}}, \bibinfo {author} {\bibfnamefont {L.~E.~F.}\ \bibnamefont
  {Foa~Torres}},\ and\ \bibinfo {author} {\bibfnamefont {M.}~\bibnamefont
  {Berdakin}},\ }\bibfield  {title} {\bibinfo {title} {Engineering floquet
  moiré patterns for scalable photocurrents},\ }\href
  {https://doi.org/10.1021/acs.nanolett.4c05716} {\bibfield  {journal}
  {\bibinfo  {journal} {Nano Letters}\ }\textbf {\bibinfo {volume} {25}},\
  \bibinfo {pages} {1630} (\bibinfo {year} {2025})},\ \bibinfo {note} {pMID:
  39825842}\BibitemShut {NoStop}%
\bibitem [{\citenamefont {Akhmerov}\ and\ \citenamefont
  {Beenakker}(2008)}]{AkhmerovBeenakker2008}%
  \BibitemOpen
  \bibfield  {author} {\bibinfo {author} {\bibfnamefont {A.~R.}\ \bibnamefont
  {Akhmerov}}\ and\ \bibinfo {author} {\bibfnamefont {C.~W.~J.}\ \bibnamefont
  {Beenakker}},\ }\bibfield  {title} {\bibinfo {title} {Boundary conditions for
  dirac fermions on a terminated honeycomb lattice},\ }\href
  {https://doi.org/10.1103/PhysRevB.77.085423} {\bibfield  {journal} {\bibinfo
  {journal} {Phys. Rev. B}\ }\textbf {\bibinfo {volume} {77}},\ \bibinfo
  {pages} {085423} (\bibinfo {year} {2008})}\BibitemShut {NoStop}%
\bibitem [{\citenamefont {Jask{\'o}lski}\ \emph {et~al.}(2011)\citenamefont
  {Jask{\'o}lski}, \citenamefont {Ayuela}, \citenamefont {Pelc}, \citenamefont
  {Santos},\ and\ \citenamefont {Chico}}]{Jaskolski2011}%
  \BibitemOpen
  \bibfield  {author} {\bibinfo {author} {\bibfnamefont {W.}~\bibnamefont
  {Jask{\'o}lski}}, \bibinfo {author} {\bibfnamefont {A.}~\bibnamefont
  {Ayuela}}, \bibinfo {author} {\bibfnamefont {M.}~\bibnamefont {Pelc}},
  \bibinfo {author} {\bibfnamefont {H.}~\bibnamefont {Santos}},\ and\ \bibinfo
  {author} {\bibfnamefont {L.}~\bibnamefont {Chico}},\ }\bibfield  {title}
  {\bibinfo {title} {Edge states and flat bands in graphene nanoribbons with
  arbitrary geometries},\ }\href {https://doi.org/10.1103/PhysRevB.83.235424}
  {\bibfield  {journal} {\bibinfo  {journal} {Phys. Rev. B}\ }\textbf {\bibinfo
  {volume} {83}},\ \bibinfo {pages} {235424} (\bibinfo {year}
  {2011})}\BibitemShut {NoStop}%
\bibitem [{\citenamefont {Fefferman}\ \emph {et~al.}(2022)\citenamefont
  {Fefferman}, \citenamefont {Fliss},\ and\ \citenamefont
  {Weinstein}}]{Fefferman2022}%
  \BibitemOpen
  \bibfield  {author} {\bibinfo {author} {\bibfnamefont {C.~L.}\ \bibnamefont
  {Fefferman}}, \bibinfo {author} {\bibfnamefont {S.}~\bibnamefont {Fliss}},\
  and\ \bibinfo {author} {\bibfnamefont {M.~I.}\ \bibnamefont {Weinstein}},\
  }\bibfield  {title} {\bibinfo {title} {Edge states in rationally terminated
  honeycomb structures},\ }\href {https://doi.org/10.1073/pnas.2212310119}
  {\bibfield  {journal} {\bibinfo  {journal} {Proc. Natl. Acad. Sci. U.S.A.}\
  }\textbf {\bibinfo {volume} {119}},\ \bibinfo {pages} {e2212310119} (\bibinfo
  {year} {2022})}\BibitemShut {NoStop}%
\bibitem [{\citenamefont {Castro~Neto}\ \emph {et~al.}(2009)\citenamefont
  {Castro~Neto}, \citenamefont {Guinea}, \citenamefont {Peres}, \citenamefont
  {Novoselov},\ and\ \citenamefont {Geim}}]{graphrev}%
  \BibitemOpen
  \bibfield  {author} {\bibinfo {author} {\bibfnamefont {A.~H.}\ \bibnamefont
  {Castro~Neto}}, \bibinfo {author} {\bibfnamefont {F.}~\bibnamefont {Guinea}},
  \bibinfo {author} {\bibfnamefont {N.~M.~R.}\ \bibnamefont {Peres}}, \bibinfo
  {author} {\bibfnamefont {K.~S.}\ \bibnamefont {Novoselov}},\ and\ \bibinfo
  {author} {\bibfnamefont {A.~K.}\ \bibnamefont {Geim}},\ }\bibfield  {title}
  {\bibinfo {title} {The electronic properties of graphene},\ }\href
  {https://doi.org/10.1103/RevModPhys.81.109} {\bibfield  {journal} {\bibinfo
  {journal} {Rev. Mod. Phys.}\ }\textbf {\bibinfo {volume} {81}},\ \bibinfo
  {pages} {109} (\bibinfo {year} {2009})}\BibitemShut {NoStop}%
\bibitem [{\citenamefont {Shirley}(1965)}]{Floq-Shirley}%
  \BibitemOpen
  \bibfield  {author} {\bibinfo {author} {\bibfnamefont {J.~H.}\ \bibnamefont
  {Shirley}},\ }\bibfield  {title} {\bibinfo {title} {\textit{Solution of the
  Schr{\"o}dinger Equation with a Hamiltonian Periodic in Time}},\ }\href
  {https://doi.org/10.1103/PhysRev.138.B979} {\bibfield  {journal} {\bibinfo
  {journal} {Phys. Rev.}\ }\textbf {\bibinfo {volume} {138}},\ \bibinfo {pages}
  {B979} (\bibinfo {year} {1965})}\BibitemShut {NoStop}%
\bibitem [{\citenamefont {Sambe}(1973)}]{Floq-Sambe}%
  \BibitemOpen
  \bibfield  {author} {\bibinfo {author} {\bibfnamefont {H.}~\bibnamefont
  {Sambe}},\ }\bibfield  {title} {\bibinfo {title} {{Steady States and
  Quasienergies of a Quantum-Mechanical System in an Oscillating Field}},\
  }\href {https://doi.org/10.1103/PhysRevA.7.2203} {\bibfield  {journal}
  {\bibinfo  {journal} {Phys. Rev. A}\ }\textbf {\bibinfo {volume} {7}},\
  \bibinfo {pages} {2203} (\bibinfo {year} {1973})}\BibitemShut {NoStop}%
\bibitem [{\citenamefont {Eckardt}\ and\ \citenamefont
  {Anisimovas}(2015)}]{VanVleck}%
  \BibitemOpen
  \bibfield  {author} {\bibinfo {author} {\bibfnamefont {A.}~\bibnamefont
  {Eckardt}}\ and\ \bibinfo {author} {\bibfnamefont {E.}~\bibnamefont
  {Anisimovas}},\ }\bibfield  {title} {\bibinfo {title} {{High-frequency
  approximation for periodically driven quantum systems from a Floquet-space
  perspective}},\ }\href {https://doi.org/10.1088/1367-2630/17/9/093039}
  {\bibfield  {journal} {\bibinfo  {journal} {New J. Phys.}\ }\textbf {\bibinfo
  {volume} {17}},\ \bibinfo {pages} {093039} (\bibinfo {year}
  {2015})}\BibitemShut {NoStop}%
\bibitem [{\citenamefont {Stacey}(1982)}]{Stacey1982}%
  \BibitemOpen
  \bibfield  {author} {\bibinfo {author} {\bibfnamefont {R.}~\bibnamefont
  {Stacey}},\ }\bibfield  {title} {\bibinfo {title} {Eliminating lattice
  fermion doubling},\ }\href {https://doi.org/10.1103/PhysRevD.26.468}
  {\bibfield  {journal} {\bibinfo  {journal} {Phys. Rev. D}\ }\textbf {\bibinfo
  {volume} {26}},\ \bibinfo {pages} {468} (\bibinfo {year} {1982})}\BibitemShut
  {NoStop}%
\bibitem [{\citenamefont {Szafran}\ \emph {et~al.}(2019)\citenamefont
  {Szafran}, \citenamefont {Mre\ifmmode \acute{n}\else
  \'{n}\fi{}ca-Kolasi\ifmmode~\acute{n}\else \'{n}\fi{}ska},\ and\
  \citenamefont {\ifmmode~\dot{Z}\else \.{Z}\fi{}ebrowski}}]{szafran2019fd}%
  \BibitemOpen
  \bibfield  {author} {\bibinfo {author} {\bibfnamefont {B.}~\bibnamefont
  {Szafran}}, \bibinfo {author} {\bibfnamefont {A.}~\bibnamefont {Mre\ifmmode
  \acute{n}\else \'{n}\fi{}ca-Kolasi\ifmmode~\acute{n}\else \'{n}\fi{}ska}},\
  and\ \bibinfo {author} {\bibfnamefont {D.}~\bibnamefont
  {\ifmmode~\dot{Z}\else \.{Z}\fi{}ebrowski}},\ }\bibfield  {title} {\bibinfo
  {title} {Finite-difference method for dirac electrons in circular quantum
  dots},\ }\href {https://doi.org/10.1103/PhysRevB.99.195406} {\bibfield
  {journal} {\bibinfo  {journal} {Phys. Rev. B}\ }\textbf {\bibinfo {volume}
  {99}},\ \bibinfo {pages} {195406} (\bibinfo {year} {2019})}\BibitemShut
  {NoStop}%
\bibitem [{\citenamefont {Zhang}\ \emph {et~al.}(2022)\citenamefont {Zhang},
  \citenamefont {Bao}, \citenamefont {Shen},\ and\ \citenamefont
  {Hu}}]{PhysRevC.106.L051303}%
  \BibitemOpen
  \bibfield  {author} {\bibinfo {author} {\bibfnamefont {Y.}~\bibnamefont
  {Zhang}}, \bibinfo {author} {\bibfnamefont {Y.}~\bibnamefont {Bao}}, \bibinfo
  {author} {\bibfnamefont {H.}~\bibnamefont {Shen}},\ and\ \bibinfo {author}
  {\bibfnamefont {J.}~\bibnamefont {Hu}},\ }\bibfield  {title} {\bibinfo
  {title} {Resolving the spurious-state problem in the dirac equation with the
  finite-difference method},\ }\href
  {https://doi.org/10.1103/PhysRevC.106.L051303} {\bibfield  {journal}
  {\bibinfo  {journal} {Phys. Rev. C}\ }\textbf {\bibinfo {volume} {106}},\
  \bibinfo {pages} {L051303} (\bibinfo {year} {2022})}\BibitemShut {NoStop}%
\bibitem [{\citenamefont {Beenakker}\ \emph {et~al.}()\citenamefont
  {Beenakker}, \citenamefont {Donís~Vela}, \citenamefont {Lemut},
  \citenamefont {Pacholski},\ and\ \citenamefont
  {Tworzydło}}]{Beenakker2023TangentFermions}%
  \BibitemOpen
  \bibfield  {author} {\bibinfo {author} {\bibfnamefont {C.~W.~J.}\
  \bibnamefont {Beenakker}}, \bibinfo {author} {\bibfnamefont {A.}~\bibnamefont
  {Donís~Vela}}, \bibinfo {author} {\bibfnamefont {G.}~\bibnamefont {Lemut}},
  \bibinfo {author} {\bibfnamefont {M.~J.}\ \bibnamefont {Pacholski}},\ and\
  \bibinfo {author} {\bibfnamefont {J.}~\bibnamefont {Tworzydło}},\ }\bibfield
   {title} {\bibinfo {title} {Tangent fermions: Dirac or majorana fermions on a
  lattice without fermion doubling},\ }\href
  {https://doi.org/https://doi.org/10.1002/andp.202300081} {\bibfield
  {journal} {\bibinfo  {journal} {Annalen der Physik}\ }\textbf {\bibinfo
  {volume} {535}},\ \bibinfo {pages} {2300081}}\BibitemShut {NoStop}%
\bibitem [{\citenamefont {Walsh}\ \emph {et~al.}(2026)\citenamefont {Walsh},
  \citenamefont {Caldwell}, \citenamefont {Sandler},\ and\ \citenamefont
  {Asmar}}]{walsh2026}%
  \BibitemOpen
  \bibfield  {author} {\bibinfo {author} {\bibfnamefont {T.~W.}\ \bibnamefont
  {Walsh}}, \bibinfo {author} {\bibfnamefont {E.~E.}\ \bibnamefont {Caldwell}},
  \bibinfo {author} {\bibfnamefont {N.~P.}\ \bibnamefont {Sandler}},\ and\
  \bibinfo {author} {\bibfnamefont {M.~M.}\ \bibnamefont {Asmar}},\ }\href
  {https://arxiv.org/abs/2606.17341} {\bibinfo {title} {Vortex-beam-driven
  dirac materials: Impurity and polarization effects on light-induced vortex
  and edge states}} (\bibinfo {year} {2026}),\ \Eprint
  {https://arxiv.org/abs/2606.17341} {arXiv:2606.17341 [cond-mat.mes-hall]}
  \BibitemShut {NoStop}%
\bibitem [{\citenamefont {Wakabayashi}\ \emph {et~al.}(2010)\citenamefont
  {Wakabayashi}, \citenamefont {Sasaki}, \citenamefont {Nakanishi},\ and\
  \citenamefont {Enoki}}]{Wakabayashi2010}%
  \BibitemOpen
  \bibfield  {author} {\bibinfo {author} {\bibfnamefont {K.}~\bibnamefont
  {Wakabayashi}}, \bibinfo {author} {\bibfnamefont {K.}~\bibnamefont {Sasaki}},
  \bibinfo {author} {\bibfnamefont {T.}~\bibnamefont {Nakanishi}},\ and\
  \bibinfo {author} {\bibfnamefont {T.}~\bibnamefont {Enoki}},\ }\bibfield
  {title} {\bibinfo {title} {Electronic states of graphene nanoribbons and
  analytical solutions},\ }\href
  {https://doi.org/10.1088/1468-6996/11/5/054504} {\bibfield  {journal}
  {\bibinfo  {journal} {Sci. Technol. Adv. Mater.}\ }\textbf {\bibinfo {volume}
  {11}},\ \bibinfo {pages} {054504} (\bibinfo {year} {2010})}\BibitemShut
  {NoStop}%
\bibitem [{\citenamefont {Jackiw}\ and\ \citenamefont
  {Rebbi}(1976)}]{JackiwRebbi1976}%
  \BibitemOpen
  \bibfield  {author} {\bibinfo {author} {\bibfnamefont {R.}~\bibnamefont
  {Jackiw}}\ and\ \bibinfo {author} {\bibfnamefont {C.}~\bibnamefont {Rebbi}},\
  }\bibfield  {title} {\bibinfo {title} {Solitons with fermion number 1/2},\
  }\href {https://doi.org/10.1103/PhysRevD.13.3398} {\bibfield  {journal}
  {\bibinfo  {journal} {Phys. Rev. D}\ }\textbf {\bibinfo {volume} {13}},\
  \bibinfo {pages} {3398} (\bibinfo {year} {1976})}\BibitemShut {NoStop}%
\bibitem [{\citenamefont {Asmar}\ \emph {et~al.}(2018)\citenamefont {Asmar},
  \citenamefont {Sheehy},\ and\ \citenamefont {Vekhter}}]{thin4}%
  \BibitemOpen
  \bibfield  {author} {\bibinfo {author} {\bibfnamefont {M.~M.}\ \bibnamefont
  {Asmar}}, \bibinfo {author} {\bibfnamefont {D.~E.}\ \bibnamefont {Sheehy}},\
  and\ \bibinfo {author} {\bibfnamefont {I.}~\bibnamefont {Vekhter}},\
  }\bibfield  {title} {\bibinfo {title} {{Topological phases of
  topological-insulator thin films}},\ }\href
  {https://doi.org/10.1103/PhysRevB.97.075419} {\bibfield  {journal} {\bibinfo
  {journal} {Phys. Rev. B}\ }\textbf {\bibinfo {volume} {97}},\ \bibinfo
  {pages} {075419} (\bibinfo {year} {2018})}\BibitemShut {NoStop}%
\bibitem [{\citenamefont {Karakachian}\ \emph {et~al.}(2020)\citenamefont
  {Karakachian}, \citenamefont {Nguyen}, \citenamefont {Aprojanz},
  \citenamefont {Zakharov}, \citenamefont {Yakimova}, \citenamefont
  {Rosenzweig}, \citenamefont {Polley}, \citenamefont {Balasubramanian},
  \citenamefont {Tegenkamp}, \citenamefont {Power},\ and\ \citenamefont
  {Starke}}]{Karakachian2020GNRARPES}%
  \BibitemOpen
  \bibfield  {author} {\bibinfo {author} {\bibfnamefont {H.}~\bibnamefont
  {Karakachian}}, \bibinfo {author} {\bibfnamefont {T.~T.~N.}\ \bibnamefont
  {Nguyen}}, \bibinfo {author} {\bibfnamefont {J.}~\bibnamefont {Aprojanz}},
  \bibinfo {author} {\bibfnamefont {A.~A.}\ \bibnamefont {Zakharov}}, \bibinfo
  {author} {\bibfnamefont {R.}~\bibnamefont {Yakimova}}, \bibinfo {author}
  {\bibfnamefont {P.}~\bibnamefont {Rosenzweig}}, \bibinfo {author}
  {\bibfnamefont {C.~M.}\ \bibnamefont {Polley}}, \bibinfo {author}
  {\bibfnamefont {T.}~\bibnamefont {Balasubramanian}}, \bibinfo {author}
  {\bibfnamefont {C.}~\bibnamefont {Tegenkamp}}, \bibinfo {author}
  {\bibfnamefont {S.~R.}\ \bibnamefont {Power}},\ and\ \bibinfo {author}
  {\bibfnamefont {U.}~\bibnamefont {Starke}},\ }\bibfield  {title} {\bibinfo
  {title} {One-dimensional confinement and width-dependent bandgap formation in
  epitaxial graphene nanoribbons},\ }\href
  {https://doi.org/10.1038/s41467-020-19051-x} {\bibfield  {journal} {\bibinfo
  {journal} {Nature Communications}\ }\textbf {\bibinfo {volume} {11}},\
  \bibinfo {pages} {6380} (\bibinfo {year} {2020})}\BibitemShut {NoStop}%
\bibitem [{\citenamefont {Jiang}\ \emph {et~al.}(2023)\citenamefont {Jiang},
  \citenamefont {Hsieh}, \citenamefont {Jones}, \citenamefont {Majchrzak},
  \citenamefont {Sahoo}, \citenamefont {Watanabe}, \citenamefont {Taniguchi},
  \citenamefont {Miwa}, \citenamefont {Chen},\ and\ \citenamefont
  {Ulstrup}}]{Jiang2023}%
  \BibitemOpen
  \bibfield  {author} {\bibinfo {author} {\bibfnamefont {Z.}~\bibnamefont
  {Jiang}}, \bibinfo {author} {\bibfnamefont {K.}~\bibnamefont {Hsieh}},
  \bibinfo {author} {\bibfnamefont {A.~J.~H.}\ \bibnamefont {Jones}}, \bibinfo
  {author} {\bibfnamefont {P.}~\bibnamefont {Majchrzak}}, \bibinfo {author}
  {\bibfnamefont {C.}~\bibnamefont {Sahoo}}, \bibinfo {author} {\bibfnamefont
  {K.}~\bibnamefont {Watanabe}}, \bibinfo {author} {\bibfnamefont
  {T.}~\bibnamefont {Taniguchi}}, \bibinfo {author} {\bibfnamefont {J.~A.}\
  \bibnamefont {Miwa}}, \bibinfo {author} {\bibfnamefont {Y.~P.}\ \bibnamefont
  {Chen}},\ and\ \bibinfo {author} {\bibfnamefont {S.}~\bibnamefont
  {Ulstrup}},\ }\bibfield  {title} {\bibinfo {title} {{Revealing flat bands and
  hybridization gaps in a twisted bilayer graphene device with microARPES}},\
  }\href {https://doi.org/10.1088/2053-1583/acf775} {\bibfield  {journal}
  {\bibinfo  {journal} {2D Materials}\ }\textbf {\bibinfo {volume} {10}},\
  \bibinfo {pages} {045027} (\bibinfo {year} {2023})}\BibitemShut {NoStop}%
\bibitem [{\citenamefont {Ruffieux}\ \emph {et~al.}(2016)\citenamefont
  {Ruffieux}, \citenamefont {Wang}, \citenamefont {Yang}, \citenamefont
  {S{\'a}nchez-S{\'a}nchez}, \citenamefont {Liu}, \citenamefont {Dienel},
  \citenamefont {Talirz}, \citenamefont {Shinde}, \citenamefont {Pignedoli},
  \citenamefont {Passerone}, \citenamefont {Dumslaff}, \citenamefont {Feng},
  \citenamefont {M{\"u}llen},\ and\ \citenamefont {Fasel}}]{Ruffieux2016}%
  \BibitemOpen
  \bibfield  {author} {\bibinfo {author} {\bibfnamefont {P.}~\bibnamefont
  {Ruffieux}}, \bibinfo {author} {\bibfnamefont {S.}~\bibnamefont {Wang}},
  \bibinfo {author} {\bibfnamefont {B.}~\bibnamefont {Yang}}, \bibinfo {author}
  {\bibfnamefont {C.}~\bibnamefont {S{\'a}nchez-S{\'a}nchez}}, \bibinfo
  {author} {\bibfnamefont {J.}~\bibnamefont {Liu}}, \bibinfo {author}
  {\bibfnamefont {T.}~\bibnamefont {Dienel}}, \bibinfo {author} {\bibfnamefont
  {L.}~\bibnamefont {Talirz}}, \bibinfo {author} {\bibfnamefont
  {P.}~\bibnamefont {Shinde}}, \bibinfo {author} {\bibfnamefont {C.~A.}\
  \bibnamefont {Pignedoli}}, \bibinfo {author} {\bibfnamefont {D.}~\bibnamefont
  {Passerone}}, \bibinfo {author} {\bibfnamefont {T.}~\bibnamefont {Dumslaff}},
  \bibinfo {author} {\bibfnamefont {X.}~\bibnamefont {Feng}}, \bibinfo {author}
  {\bibfnamefont {K.}~\bibnamefont {M{\"u}llen}},\ and\ \bibinfo {author}
  {\bibfnamefont {R.}~\bibnamefont {Fasel}},\ }\bibfield  {title} {\bibinfo
  {title} {On-surface synthesis of graphene nanoribbons with zigzag edge
  topology},\ }\href {https://doi.org/10.1038/nature17151} {\bibfield
  {journal} {\bibinfo  {journal} {Nature}\ }\textbf {\bibinfo {volume} {531}},\
  \bibinfo {pages} {489} (\bibinfo {year} {2016})}\BibitemShut {NoStop}%
\bibitem [{\citenamefont {Wang}\ \emph {et~al.}(2016)\citenamefont {Wang},
  \citenamefont {Talirz}, \citenamefont {Pignedoli}, \citenamefont {Feng},
  \citenamefont {M{\"u}llen}, \citenamefont {Fasel},\ and\ \citenamefont
  {Ruffieux}}]{Wang2016}%
  \BibitemOpen
  \bibfield  {author} {\bibinfo {author} {\bibfnamefont {S.}~\bibnamefont
  {Wang}}, \bibinfo {author} {\bibfnamefont {L.}~\bibnamefont {Talirz}},
  \bibinfo {author} {\bibfnamefont {C.~A.}\ \bibnamefont {Pignedoli}}, \bibinfo
  {author} {\bibfnamefont {X.}~\bibnamefont {Feng}}, \bibinfo {author}
  {\bibfnamefont {K.}~\bibnamefont {M{\"u}llen}}, \bibinfo {author}
  {\bibfnamefont {R.}~\bibnamefont {Fasel}},\ and\ \bibinfo {author}
  {\bibfnamefont {P.}~\bibnamefont {Ruffieux}},\ }\bibfield  {title} {\bibinfo
  {title} {Giant edge state splitting at atomically precise graphene zigzag
  edges},\ }\href {https://doi.org/10.1038/ncomms11507} {\bibfield  {journal}
  {\bibinfo  {journal} {Nature Communications}\ }\textbf {\bibinfo {volume}
  {7}},\ \bibinfo {pages} {11507} (\bibinfo {year} {2016})}\BibitemShut
  {NoStop}%
\bibitem [{\citenamefont {Brede}\ \emph {et~al.}(2023)\citenamefont {Brede},
  \citenamefont {Merino-D{\'i}ez}, \citenamefont {Berdonces-Layunta},
  \citenamefont {Sanz}, \citenamefont {Dom{\'i}nguez-Celorrio}, \citenamefont
  {Lobo-Checa}, \citenamefont {Vilas-Varela}, \citenamefont {Pe{\~n}a},
  \citenamefont {Frederiksen}, \citenamefont {Pascual}, \citenamefont
  {de~Oteyza},\ and\ \citenamefont {Serrate}}]{Brede2023}%
  \BibitemOpen
  \bibfield  {author} {\bibinfo {author} {\bibfnamefont {J.}~\bibnamefont
  {Brede}}, \bibinfo {author} {\bibfnamefont {N.}~\bibnamefont
  {Merino-D{\'i}ez}}, \bibinfo {author} {\bibfnamefont {A.}~\bibnamefont
  {Berdonces-Layunta}}, \bibinfo {author} {\bibfnamefont {S.}~\bibnamefont
  {Sanz}}, \bibinfo {author} {\bibfnamefont {A.}~\bibnamefont
  {Dom{\'i}nguez-Celorrio}}, \bibinfo {author} {\bibfnamefont {J.}~\bibnamefont
  {Lobo-Checa}}, \bibinfo {author} {\bibfnamefont {M.}~\bibnamefont
  {Vilas-Varela}}, \bibinfo {author} {\bibfnamefont {D.}~\bibnamefont
  {Pe{\~n}a}}, \bibinfo {author} {\bibfnamefont {T.}~\bibnamefont
  {Frederiksen}}, \bibinfo {author} {\bibfnamefont {J.~I.}\ \bibnamefont
  {Pascual}}, \bibinfo {author} {\bibfnamefont {D.~G.}\ \bibnamefont
  {de~Oteyza}},\ and\ \bibinfo {author} {\bibfnamefont {D.}~\bibnamefont
  {Serrate}},\ }\bibfield  {title} {\bibinfo {title} {Detecting the
  spin-polarization of edge states in graphene nanoribbons},\ }\href
  {https://doi.org/10.1038/s41467-023-42436-7} {\bibfield  {journal} {\bibinfo
  {journal} {Nature Communications}\ }\textbf {\bibinfo {volume} {14}},\
  \bibinfo {pages} {6677} (\bibinfo {year} {2023})}\BibitemShut {NoStop}%
\bibitem [{\citenamefont {Roelcke}\ \emph {et~al.}(2024)\citenamefont
  {Roelcke}, \citenamefont {Kastner}, \citenamefont {Graml}, \citenamefont
  {Biereder}, \citenamefont {Wilhelm}, \citenamefont {Repp}, \citenamefont
  {Huber},\ and\ \citenamefont {Gerasimenko}}]{TimeSTM}%
  \BibitemOpen
  \bibfield  {author} {\bibinfo {author} {\bibfnamefont {C.}~\bibnamefont
  {Roelcke}}, \bibinfo {author} {\bibfnamefont {L.~Z.}\ \bibnamefont
  {Kastner}}, \bibinfo {author} {\bibfnamefont {M.}~\bibnamefont {Graml}},
  \bibinfo {author} {\bibfnamefont {A.}~\bibnamefont {Biereder}}, \bibinfo
  {author} {\bibfnamefont {J.}~\bibnamefont {Wilhelm}}, \bibinfo {author}
  {\bibfnamefont {J.}~\bibnamefont {Repp}}, \bibinfo {author} {\bibfnamefont
  {R.}~\bibnamefont {Huber}},\ and\ \bibinfo {author} {\bibfnamefont {Y.~A.}\
  \bibnamefont {Gerasimenko}},\ }\bibfield  {title} {\bibinfo {title}
  {{Ultrafast atomic-scale scanning tunnelling spectroscopy of a single vacancy
  in a monolayer crystal}},\ }\href
  {https://doi.org/https://doi.org/10.1038/s41566-024-01390-6} {\bibfield
  {journal} {\bibinfo  {journal} {Nat. Photonics}\ }\textbf {\bibinfo {volume}
  {18}},\ \bibinfo {pages} {1} (\bibinfo {year} {2024})}\BibitemShut {NoStop}%
\bibitem [{\citenamefont {Gorbachev}\ \emph {et~al.}(2014)\citenamefont
  {Gorbachev}, \citenamefont {Song}, \citenamefont {Yu}, \citenamefont
  {Kretinin}, \citenamefont {Withers}, \citenamefont {Cao}, \citenamefont
  {Mishchenko}, \citenamefont {Grigorieva}, \citenamefont {Novoselov},
  \citenamefont {Levitov},\ and\ \citenamefont
  {Geim}}]{Gorbachev2014ValleyHall}%
  \BibitemOpen
  \bibfield  {author} {\bibinfo {author} {\bibfnamefont {R.~V.}\ \bibnamefont
  {Gorbachev}}, \bibinfo {author} {\bibfnamefont {J.~C.~W.}\ \bibnamefont
  {Song}}, \bibinfo {author} {\bibfnamefont {G.~L.}\ \bibnamefont {Yu}},
  \bibinfo {author} {\bibfnamefont {A.~V.}\ \bibnamefont {Kretinin}}, \bibinfo
  {author} {\bibfnamefont {F.}~\bibnamefont {Withers}}, \bibinfo {author}
  {\bibfnamefont {Y.}~\bibnamefont {Cao}}, \bibinfo {author} {\bibfnamefont
  {A.}~\bibnamefont {Mishchenko}}, \bibinfo {author} {\bibfnamefont {I.~V.}\
  \bibnamefont {Grigorieva}}, \bibinfo {author} {\bibfnamefont {K.~S.}\
  \bibnamefont {Novoselov}}, \bibinfo {author} {\bibfnamefont {L.~S.}\
  \bibnamefont {Levitov}},\ and\ \bibinfo {author} {\bibfnamefont {A.~K.}\
  \bibnamefont {Geim}},\ }\bibfield  {title} {\bibinfo {title} {Detecting
  topological currents in graphene superlattices},\ }\href
  {https://doi.org/10.1126/science.1254966} {\bibfield  {journal} {\bibinfo
  {journal} {Science}\ }\textbf {\bibinfo {volume} {346}},\ \bibinfo {pages}
  {448} (\bibinfo {year} {2014})}\BibitemShut {NoStop}%
\bibitem [{\citenamefont {Sui}\ \emph {et~al.}(2015)\citenamefont {Sui},
  \citenamefont {Chen}, \citenamefont {Ma}, \citenamefont {Shan}, \citenamefont
  {Tian}, \citenamefont {Watanabe}, \citenamefont {Taniguchi}, \citenamefont
  {Jin}, \citenamefont {Yao}, \citenamefont {Xiao},\ and\ \citenamefont
  {Zhang}}]{Sui2015ValleyTransport}%
  \BibitemOpen
  \bibfield  {author} {\bibinfo {author} {\bibfnamefont {M.}~\bibnamefont
  {Sui}}, \bibinfo {author} {\bibfnamefont {G.}~\bibnamefont {Chen}}, \bibinfo
  {author} {\bibfnamefont {L.}~\bibnamefont {Ma}}, \bibinfo {author}
  {\bibfnamefont {W.}~\bibnamefont {Shan}}, \bibinfo {author} {\bibfnamefont
  {D.}~\bibnamefont {Tian}}, \bibinfo {author} {\bibfnamefont {K.}~\bibnamefont
  {Watanabe}}, \bibinfo {author} {\bibfnamefont {T.}~\bibnamefont {Taniguchi}},
  \bibinfo {author} {\bibfnamefont {X.}~\bibnamefont {Jin}}, \bibinfo {author}
  {\bibfnamefont {W.}~\bibnamefont {Yao}}, \bibinfo {author} {\bibfnamefont
  {D.}~\bibnamefont {Xiao}},\ and\ \bibinfo {author} {\bibfnamefont
  {Y.}~\bibnamefont {Zhang}},\ }\bibfield  {title} {\bibinfo {title}
  {Gate-tunable topological valley transport in bilayer graphene},\ }\href
  {https://doi.org/10.1038/nphys3485} {\bibfield  {journal} {\bibinfo
  {journal} {Nature Physics}\ }\textbf {\bibinfo {volume} {11}},\ \bibinfo
  {pages} {1027} (\bibinfo {year} {2015})}\BibitemShut {NoStop}%
\end{thebibliography}%
\end{document}